\documentclass[onecolumn,showpacs,nofootinbib, aps, prd,10pt,amssymb]{revtex4-2}

\usepackage[cp1251]{inputenc}
\usepackage[english]{babel}

\usepackage{latexsym}
\usepackage{graphicx}
\usepackage{amssymb,amsmath}
\usepackage{relsize}
\usepackage{dsfont}
\usepackage{textcomp}
\usepackage{mathrsfs}
\usepackage{bm}
\usepackage{braket}
\usepackage[normalem]{ulem}
\usepackage{array}
\usepackage{booktabs}
\usepackage{todonotes}

\usepackage{hyperref} 

\DeclareMathAlphabet\mathbfcal{OMS}{cmsy}{b}{n}

\def\ii{{\rm i}}

\newcommand{\aap}{Astron.\ Astrophys.}
\newcommand{\mnras}{Mon.\ Not.\ R.\ Astron.\ Soc.}

\newcommand{\apjl}{Astrophys.\ J.\ Lett.}
\newcommand{\apjs}{Astrophys.\ J.\ Suppl.\ Ser.}
\newcommand{\araa}{Ann.\ Rev.\ Astron.\ Astrophys.}
\newcommand{\apss}{Astrophys.\ Space Sci.}

\newcolumntype{C}[1]{>{\centering\arraybackslash}m{#1}} 

\DeclareRobustCommand{\mg}[1]{\ifmmode\bm{#1}\else\textbf{#1}\fi}
\DeclareRobustCommand{\ek}[1]{\ifmmode\bm{#1}\else\textbf{#1}\fi}

\newcommand{\m}[1]{\pmb{#1}}

\begin{document}

\title{Beyond the Tayler instability: A new global instability
of toroidal magnetic fields in stars}

\author{
Mikhail~E.~Gusakov$^1$, Laura~Becerra$^2$, Elena~M.~Kantor$^1$, Andreas~Reisenegger$^{3,4}$, and Juan~Alejandro~Valdivia$^5$}
\affiliation{$^1$Ioffe Institute, Polytekhnicheskaya 26, 194021 Saint-Petersburg, Russia \\
$^2$Centro Multidisciplinario de F\'{\i}sica, Vicerrector\'{\i}a de Investigaci\'on, Universidad
Mayor, 8580745 Santiago, Chile\\
$^3$Departamento de F{\'\i}sica, Facultad de Ciencias B\'asicas, Universidad Metropolitana de Ciencias de la Educaci\'on, Av. Jos\'e Pedro Alessandri 774, \~Nu\~noa, Santiago, Chile \\
$^4$Centro de Desarrollo de Investigaci\'on UMCE, Universidad Metropolitana de Ciencias de la Educaci\'on, Santiago, Chile \\
$^5$Departamento de F\'{\i}sica, Facultad de Ciencias, Universidad de Chile, Las Palmeras 3425, \~Nu\~noa, Santiago, Chile
}

\begin{abstract} 
Stellar toroidal magnetic fields are known to be unstable to the Tayler instability. 
Here we demonstrate the existence of a complementary current-driven instability 
of essentially {\it arbitrary} toroidal-field configurations in stably stratified nonrotating stars 
with the 
following properties: (i) in ideal magnetohydrodynamics, it grows on the Alfv\'{e}n timescale 
$\tau_{\rm A}$; 
(ii) under certain conditions, it may reveal itself by driving
shellular differential rotation about an arbitrary axis perpendicular to the magnetic-field symmetry axis; 
(iii) it is large-scale in the angular directions $\theta$ and $\varphi$, and develops 
at radial wavenumbers $k \lesssim \mathcal{N}\tau_{\rm A}/R$, where $\mathcal{N}$ is the 
Brunt-V\"ais\"al\"a frequency and $R$ is the stellar radius. 
Thus, unlike the Tayler instability, the proposed instability is intrinsically global. 
Consequently, 
it may be less susceptible to dissipative suppression than the Tayler instability and can 
prevail over it in some regimes.
This instability may have broad implications for magnetic field generation in stars and 
could modify scenarios of magnetic field amplification within the Tayler-Spruit dynamo, contributing 
to models of efficient angular-momentum transport and chemical mixing in stellar interiors.
\end{abstract}

\date{\today}


\maketitle

\section{Introduction}

Magnetic fields play an important role in the dynamics and evolution of stars \cite{dl09}. The nature of these fields and the processes responsible for their generation are of central interest in astrophysics. In particular, large-scale toroidal magnetic fields are expected to arise naturally due to stellar differential rotation. 
The instability of such fields was demonstrated by Tayler (1973) \cite{tayler73} (see also the 
earlier work by Vandakurov \cite{vandakurov72}).
In his study, Tayler identified a specific class of unstable displacements in toroidal magnetic fields; the resulting instability is now referred to as the Tayler instability.%
%
\footnote{Some authors (e.g., \cite{kr08,rk10,ar11}) use the term ``Tayler instability'' to denote, by definition, any $m=1$ current-driven (pinch-type) instability of toroidal fields. 
In this work, we do not adopt this terminology and, by the Tayler instability in stars, we mean the current-driven $m=1$ instability whose most characteristic feature is that the corresponding perturbations are small-scale (at least) in the radial direction and, in the limit of ideal magnetohydrodynamics, remain unstable even as this scale tends to zero (see Sec.~\ref{Tayler} for more details).}
%

Over the years, this instability has been studied using analytical and semi-analytical methods (e.g., \cite{gv78,ag78,goossens80,gt80, gbt81, pt85, kitchatinov08, kr08, rk10, bu12,armm13,sb24,sb25}), as well as through three-dimensional numerical simulations (e.g., \cite{braithwaite06,dbm10,kys11,lj11, ib15,guerrero19,jfl23,monteiro23,meduri24}).

Since the pioneering works of Spruit \cite{spruit99, Spruit2002}, who argued that the instability of toroidal magnetic fields could drive a dynamo process in stars, now known as the Tayler-Spruit dynamo, interest in the Tayler instability has increased substantially. In particular, the Tayler-Spruit dynamo is used to explain the magnetic field observed in magnetars \cite{barrere22, barrere23, barrere25} and low-field magnetars \cite{igoshev_etal25}; its influence is taken into account when modeling chemical mixing in nondegenerate stars (see \cite{Spruit2002, mm04,mm05} and, for illustration, two recent studies \cite{asatiani_etal25, buldgen_etal25} and references therein). Most importantly for stellar evolution theory, the magnetic stresses arising in the course of such a dynamo give rise to an effective viscosity (e.g., \cite{Spruit2002, mm04,mm05, fuller19}) and to an additional, efficient transport of angular momentum in stars.

At the same time, asteroseismic data indicate that existing angular-momentum transport mechanisms 
appear to be insufficient to account for the observed level of internal (core-to-envelope) 
differential rotation in certain classes of stars (in particular, in low-mass evolved stars and the 
Sun).
In this context, the mechanism of angular-momentum transport due to the Tayler-Spruit dynamo acquires special importance and is regarded as one of the most promising candidates (see \cite{amr19, fuller19} and references therein).

It should be noted that the original dynamo scenario proposed by Spruit \cite{Spruit2002} was criticized in \cite{dp07,zbm07} and was later modified by Fuller et al.\ in Ref.\ \cite{fuller19}. 
Studies aimed at detecting the Tayler-Spruit dynamo in numerical simulations have appeared only 
relatively recently \cite{barrere23, petitdemange23, jfl23, barrere25, petitdemange24, Igoshev25} 
(but see \cite{braithwaite06}). A discussion of these results and their consistency with linear 
theory can be found in Ref.\ \cite{sb24}.
Currently, 
there is no clear consensus  
on 
which version of the dynamo, as described in Ref.~\cite{Spruit2002} or \cite{fuller19}, is reproduced in simulations (cf.\ Refs.\ \cite{petitdemange24} and \cite{barrere23,barrere25,jfl23}).

In addition, a number of studies have shown that neither version of the Tayler-Spruit dynamo satisfactorily reproduces the observational constraints on the internal differential-rotation profiles and the surface abundances of certain chemical elements in stars (e.g., \cite{cmb14, eggenberger_etal19, eggenberger_etal19b, hartogh_etal20, fl22, me24, asatiani_etal25, buldgen_etal25}).

In light of the above, our goal in this paper is to revisit the stability of toroidal magnetic fields in stars.
Using the energy principle of Bernstein et al.\ \cite{bernstein_etal58}, we show analytically that, 
in addition to the well-known Tayler-type unstable displacements, there exists another class of 
unstable modes that differ qualitatively from those originally proposed by Tayler.
Both types of displacements have comparable growth rates, typically of the order of 
the 
Alfv\'en frequency. Depending on stellar parameters and magnetic-field configuration, either the 
Tayler instability or the newly found instability may dominate.
Our theoretical predictions are 
consistent with
numerical simulations performed with the \href{https://pencil-code.nordita.org}{\texttt{Pencil Code}} \cite{pencil21}, a high-order finite-difference code for compressible magnetohydrodynamics (MHD) simulations.

The newly identified instability suggests that the traditional Tayler-Spruit dynamo mechanism may need to be reconsidered. This, in turn, could have implications for the angular-momentum transport and the accompanying chemical mixing in stellar interiors, as well as for the observational properties of magnetic stars.

The structure of this paper is as follows.
Section \ref{MHD} presents the governing equations and the basic assumptions for a stably stratified, non-rotating magnetized star in equilibrium.
In Sec.~\ref{bern3}, we set up the energy-principle framework and notation, state the
assumptions on the equilibrium toroidal field, and introduce the displacement parametrization used in the rest of the paper.
Section \ref{pert} further transforms the energy functional and develops a perturbative framework that identifies the expected structure and key properties of the most unstable displacements.
In Sec.~\ref{timescale}, we use this framework to derive an upper bound for the characteristic growth time of the new instability and its scaling with the relevant small parameters.
Section \ref{example} applies the formalism to a concrete stellar model and a specified toroidal-field configuration, performs the minimization over trial displacements, and obtains numerical estimates for the growth time.
Section \ref{poloidal1} estimates the strength of a (weak) poloidal magnetic field required to suppress the instability.
In Sec.~\ref{Tayler}, we compare the Tayler instability with the global instability identified here, emphasizing the differences in their preferred lengthscales and their sensitivity to dissipation.
Section \ref{sim} presents numerical simulations that illustrate and support the theoretical results.
Finally, Sec.~\ref{summary} summarizes our conclusions.

The paper also includes four appendices.
Appendix~\ref{accurate} tests the accuracy of the minimization procedure discussed in the main text, in particular that of Eq.~\eqref{div3}.
Appendix~\ref{shear1} presents integral expressions for the dissipation timescale associated with shear viscosity, which affects the instability found in this work.
In Appendix \ref{proof1}, we discuss a formula quantifying the effect of viscous dissipation on the 
instability timescale.
Finally, Appendix~\ref{approx} examines the sensitivity of the new instability to various simplifying approximations that are often adopted in the literature when studying MHD instabilities in stars.

\section{Preliminary comments and governing equations}
\label{MHD}

In this work, for simplicity, we consider a non-relativistic magnetized star, neglecting the effects of general relativity.  
It is assumed that the stellar matter has a non-barotropic equation of state,  
in which the pressure $p$ depends not only on the density $\rho$, but also on at least one additional parameter $x$ that is conserved along the motion of each fluid element: $p = p(\rho, x)$.%
%
\footnote{
There may be several such additional parameters; this will not affect the results  
of the work. The only modification required in the case of multiple parameters concerns the definition of the frozen adiabatic index $\gamma_{\rm fr}$. See  
the explanation following equation \eqref{frozen} below.
}
%
This parameter may be, for example, the ratio of the entropy per unit volume $s$ to $\rho$,  
$x = s/\rho$, or, if the star contains two species of particles,  
the ratio of the density of one species $\rho_1$ to the total density, $x = \rho_1/\rho$.  
For neutron stars, it is convenient to use the baryon number density $n_b$ instead of $\rho$,  
so that the parameter can be written as $x = n_e / n_b$ ($n_e$ is the electron number density).

The complete system of magnetohydrodynamic equations describing a non-rotating magnetized star  
in the nondissipative limit has the form
\begin{align}
&
\rho \frac{D {\pmb u}}{Dt} = - {\pmb \nabla}p -\rho {\pmb \nabla}\Phi +\m F_{\rm L},
&
\label{euler}\\
&
\frac{\partial \rho}{\partial t}+{\pmb \nabla}\cdot (\rho {\pmb u})=0,
&
\label{cont1}\\
&
\frac{\partial x}{\partial t} + {\pmb u}\cdot {\pmb \nabla x}=0,
&
\label{cont2}\\
&
\frac{\partial {\pmb B}}{\partial t}= {\pmb \nabla}\times ({\pmb u}\times {\pmb B}),
&
\label{faraday}\\
&
{\pmb \nabla}\times {\pmb B}= \frac{4 \pi}{c} {\pmb j},
&
\label{ampere}\\
&
{\pmb \nabla}\cdot {\pmb B}=0,
&
\label{divB}\\
&
\Delta \Phi =4 \pi G \rho.
&
\label{grav}
\end{align} 
Here, ${\pmb u}$ is the fluid velocity;  
$D{\pmb u}/Dt \equiv \partial{\pmb u}/\partial t + ({\pmb u} \cdot {\pmb \nabla}) {\pmb u}$;  
${\pmb B}$ is the magnetic field of the star;  
${\pmb j}$ is the electric current density;  
$\m F_{\rm L} = \m j \times \m B / c$ is the Lorentz force per unit volume;  
$G$ is the gravitational constant;  
$c$ is the speed of light;  
$\Phi$ is the gravitational potential.

\subsection{Equilibrium and small deviations from equilibrium}
\label{Eq_and_Linear}

In an equilibrium star, equations \eqref{ampere}, \eqref{divB}, and \eqref{grav} formally remain unchanged,  
while the Euler equation \eqref{euler} takes the form
\begin{align}
&
 {\pmb \nabla}p +\rho {\pmb \nabla}\Phi = \m F_{\rm L}.
&
\label{force}
\end{align} 
Here and below, the quantities $p$, $\Phi$, $\rho$, $\m F_{\rm L}$, etc.\ refer to their equilibrium values, and
Eulerian perturbations from these equilibrium values will be denoted by $\delta p$, $\delta \Phi$,  
$\delta \rho$, $\delta \m F_{\rm L}$,  
etc.\ (with the exception of the fluid velocity, which will be denoted by ${\pmb u}$).

Let us consider a small perturbation of the system \eqref{euler}--\eqref{grav}  
relative to equilibrium.  
Introducing the Lagrangian displacement ${\pmb \xi}$ according to the relation
\begin{align}
&
{\pmb u}=\frac{\partial {\pmb \xi}}{\partial t},
&
\label{xi}
\end{align}
the linearized system of equations takes the form:
\begin{align}
&
\rho \frac{\partial^2 {\pmb \xi}}{\partial t^2} = - {\pmb \nabla}\delta p -\delta \rho {\pmb 
\nabla}\Phi 
-\rho {\pmb \nabla} \delta\Phi
+\delta \m F_{\rm L},
&
\label{euler3}\\
&
\delta \rho+{\pmb \nabla}\cdot (\rho {\pmb \xi})=0,
&
\label{cont13}\\
&
\delta x + {\pmb \xi}\cdot {\pmb \nabla x}=0,
&
\label{cont23}\\
&
\delta {\pmb B}= {\pmb \nabla}\times ({\pmb \xi}\times {\pmb B}),
&
\label{faraday3}\\
&
{\pmb \nabla}\times \delta {\pmb B}= \frac{4 \pi}{c} \delta {\pmb j},
&
\label{ampere3}\\
&
\Delta \delta\Phi =4 \pi G \delta\rho,
&
\label{grav3}
\end{align}
where 
\begin{align}
&
\delta \m F_{\rm L}=(\delta{\pmb j}\times {\pmb B}+{\pmb j}\times \delta {\pmb B})/c.
&
\label{dFL}
\end{align}
Equation ${\pmb \nabla}\cdot \delta {\pmb B}=0$ is not presented here since it is automatically 
satisfied in view of Eq.\ \eqref{faraday3}.
Keeping in mind that perturbations occur at frozen parameter $x$, we can write:
\begin{align}
&
\delta p + \m \xi \cdot \m\nabla p = \frac{\gamma_{\rm fr} p}{\rho} \left(\delta \rho +
\m \xi \cdot \m\nabla \rho 
\right),
&
\label{dp}
\end{align}
where $\gamma_{\rm fr}$ is the frozen adiabatic index,
\begin{align}
&
\gamma_{\rm fr}=\frac{\rho}{p}\frac{\partial p(\rho, x)}{\partial \rho}.
&
\label{frozen}
\end{align}
In the case where the pressure, in addition to the density, is described by $N$ parameters $x_1,\ldots,x_N$ instead of  
a single parameter $x$, the partial derivative in Eq.\ \eqref{frozen}  
should be taken at constant $x_1,\ldots,x_N$.

\section{Bernstein et al.\  energy principle and our assumptions}
\label{bern3}

\subsection{Bernstein et  al.\ energy principle}
\label{bernshtein}

The stability of the equilibrium state of a magnetized star can be conveniently studied  
using the energy principle \cite{bernstein_etal58},  
described in detail in the book \cite{gp04} (see also, e.g., Refs.\ \cite{aw08, armm13},  
where applications of this principle to stars are considered).  
Below, we will focus on the stability of the toroidal magnetic field,  
which must be completely localized inside the star 
(see Sec.\ \ref{assumpt} below for the boundary conditions that we assume such a field satisfies).  
In this case, the energy functional $W[{\pmb \xi}]$, describing the ``potential'' energy  
of the system, takes the form:
\begin{align}
	& 	W[{\pmb \xi}] = W^{\rm hydro}[\m\xi]+W^{\rm mag}[{\pmb \xi}],
	&
	\label{en11}
\end{align}
and consists of two parts -- the hydrodynamic part, $W^{\rm hydro}[{\pmb \xi}]$,  
and the magnetic part, $W^{\rm mag}[{\pmb \xi}]$,
\begin{align}
	& 	W^{\rm hydro}[{\pmb \xi}] = 
	\frac{1}{2} \int \left[ \gamma_{\rm fr} p (\m\nabla\cdot\m\xi)^2 +
	(\m\xi\cdot\m\nabla p) (\m\nabla\cdot\m\xi) - (\m\xi\cdot\m\nabla\Phi)
	\m\nabla\cdot (\rho\m\xi) + \rho\m\xi\cdot\m\nabla\delta\Phi \right] dV,&
	\label{Whydro}\\
	&
	W^{\rm mag}[{\pmb \xi}] = 
	\frac{1}{2} \int \left[
	\frac{\delta {\pmb B}\cdot\delta {\pmb B}}{4 \pi} + \frac{1}{c}\, 
	\m j \cdot (\m\xi \times  \delta {\pmb B})
	\right] d V. &
	\label{Wmag}
\end{align}
In expressions \eqref{Whydro}, \eqref{Wmag}, the integration is performed  
over the volume of the star $V$.  
Furthermore, in our analysis, we will not use the complex formalism with respect to the function $\m \xi$; instead we will  assume that the Lagrangian displacement $\m \xi$ is a real function.

The Bernstein et al.\ energy principle for stability states that \cite{gp04}:
``An equilibrium is stable if (sufficient) and only if
(necessary) $W[\m\xi]>0$ for all displacements $\m\xi(\m r)$
that are bound in norm and satisfy the boundary conditions.''
In the case of a toroidal magnetic field localized inside the star,  
the boundary conditions for $\m \xi(\m r)$ are irrelevant and will not be discussed further.

According to the energy principle, if we can find a Lagrangian displacement  
$\m \xi$ such that $W[\m \xi] < 0$, then the system is unstable.  
In fact, this energy principle follows from a more general  
statement known as the spectral variational principle.  
To formulate it, note that the system of linearized magnetohydrodynamic equations  
\eqref{euler3}--\eqref{grav3} is self-adjoint (e.g., \cite{gp04}).  
This means that any Lagrangian displacement can be expanded in terms of the eigenmodes of the magnetized  
star $\m \xi_n(\m r)$, with eigenfrequencies $\omega_n$, such that $\omega_n^2$ is real.  
The spectral variational principle states that the eigenfunctions $\m \xi_n(\m r)$ make the Rayleigh quotient
\begin{align}
&
\Lambda \equiv \frac{W[\m \xi]}{I[\m\xi]}
&
\label{ray}
\end{align}
stationary, i.e. the functional derivative $\delta(W[\m\xi]/I[\m\xi])/\delta \m\xi=0$ for 
$\m\xi=\m\xi_n$. 
The corresponding stationary values of $\Lambda$ are the eigenvalues $\omega_n^2$ \cite{gp04}.
The positive ``kinetic energy'' functional $I[\m\xi]$ in Eq.\ \eqref{ray} is defined as
\begin{align}
&
I[\m\xi] \equiv \frac{1}{2} \int \rho \, \m\xi\cdot\m\xi \, dV.
&
\label{III}
\end{align}
Thus, solving the linearized hydrodynamic equations is fully  
equivalent to finding the functions that make the ratio \eqref{ray} stationary.  
The energy principle of Bernstein et al.\ clearly follows from the spectral variational principle.  
In particular, if there exists at least one unstable mode with $\omega_n^2 < 0$  
(which corresponds to an exponential growth of this mode $\propto e^{|\omega_n| t}$),  
then one can always choose a Lagrangian displacement $\m \xi$ such that $W[\m \xi] < 0$.  
However, the spectral variational principle allows one to go further and obtain  
an estimate for the characteristic growth time of the instability.  
Indeed, the absolute minimum of the functional \eqref{ray}  
is equal to the square of the frequency of the most unstable mode, $\omega_{\rm min}^2$  
(if the system is unstable, then $\omega_{\rm min}^2 < 0$).  
Therefore, for any trial function $\m \xi$ one can say that
\begin{align}
	&
\omega_{\min}^2 \leq \frac{W[\m \xi]}{I[\m\xi]}.
	&
	\label{III2}
\end{align}
By choosing increasingly accurate trial functions,  
one can improve the upper bound on $\omega_{\rm min}^2$.

From the estimate of the (negative) value of $\omega_{\rm min}^2$,  
it is straightforward to obtain an upper bound on  
the characteristic growth time of the instability, $\tau_{\rm min}$.  
Defining $\tau_{\rm min}^2 \equiv -1/\omega_{\rm min}^2$, we obtain from \eqref{III2}
\begin{align}
	&
	\tau_{\min} \leq \sqrt{\frac{ I[\m \xi]}{|W[\m\xi]|  } }.
	&
	\label{III33}
\end{align}
%

\subsection{Our assumptions}
\label{assumpt}

In the present paper, we assume that the unperturbed, axially symmetric star hosts a toroidal 
magnetic field,  
${\pmb B} = B(r,\theta) {\pmb e}_{\varphi}$, where $r$, $\theta$, and $\varphi$ are the usual spherical  
coordinates with the origin at the  
center of the star in the absence of the magnetic field;  
${\pmb e}_{i}$ denotes the corresponding  
unit vector ($i = r$, $\theta$, $\varphi$).  
We emphasize that the same non-magnetized star is assumed to be stable.  
We define the radius of such a star as $R$.

Requiring the charge current density ${\pmb j}$ to remain finite at the poles and at the center of the  
star, we conclude that the toroidal field $B(r,\theta) {\pmb e}_{\varphi}$ must  
vanish as $\theta \rightarrow 0$ and $\theta \rightarrow \pi$ at least as $\sin\theta$, and as $r \rightarrow 0$ at least as $r$.

Furthermore, in order to avoid (unphysical) surface currents, we must require that $B = 0$  
at the stellar surface.  
In addition, since we are mainly interested in stellar models with  
vanishing density at the surface, we also assume that not only the toroidal magnetic field  
but also the charge current density ${\pmb j}$ vanishes at the surface. This leads to  
additional conditions at the stellar surface: $\partial_r B(r,\theta) = 0$,  
$\partial_\theta B(r,\theta) = 0$. Here and in what follows, we employ the notation  
$\partial_i \equiv \partial/\partial x_i$,  
where $x_i$ stands for $r$, $\theta$, or $\varphi$.

To demonstrate the instability of the toroidal field, we need to find a Lagrangian  
displacement $\m \xi$ such that  
$W[\m \xi] < 0$.  
In what follows, we shall always ignore the last term in expression (\ref{Whydro}),  
which depends on the gravitational field perturbation $\delta \Phi$.  
It is well known \cite{mt69} that this term is always negative, so  
if we prove that the system is unstable without this term, it will remain  
unstable when this term is taken into account.

Our final important assumption concerns the strength of the magnetic field.  
We assume that the magnetic energy density is much smaller than the pressure  
everywhere in the region where the magnetic field is localized.  
Introducing the (constant) ratio $\varkappa$ %
%
\footnote{Usually, in the literature, instead of $\varkappa$ one uses the parameter $\beta \equiv 1/\varkappa$, but for our purposes it is more convenient to work with $\varkappa$.}
%
of some typical magnetic energy density $B^2/8\pi$ to  
the pressure $p$,
\begin{align}
	&
	\varkappa \equiv \frac{B^2}{8 \pi p} \ll 1,
	&
	\label{BP}	
\end{align}
we assume that any quantity $X$, characterizing the unperturbed star [e.g., the gravitational 
potential, $\Phi(r,\theta)$, or the ``frozen'' adiabatic index, $\gamma_{\rm fr}$]
can be expanded in a series in $\varkappa$:
\begin{align}
	&
	X(r,\theta)=X_0(r)+ \varkappa X_1(r,\theta) + \varkappa^2 X_2(r,\theta)+\ldots,
	&
	\label{Adecomp}	
\end{align}
where $X_0(r)$ is the value of this quantity in the star without a magnetic field,
while the remaining terms represent small perturbations caused by the field.
Note that, with this definition of the functions $X_i(r,\theta)$ ($i = 1, 2, \ldots$),
the ratio $X_i/X_0 \sim \mathcal{O}(1)$.

The decomposition (\ref{Adecomp}) can be used to relate 
the partial derivatives $\partial_\theta \Phi_1(r,\theta)$ and $\partial_\theta \rho_1(r,\theta)$
using the force balance equation (\ref{force}).
Taking the curl of this equation and using (\ref{Adecomp}), one obtains, to linear accuracy in 
$\varkappa$,
\begin{align}
	&
	-\partial_r \Phi_0 \, \partial_\theta \rho_1 +\partial_r \rho_0 \, \partial_\theta \Phi_1= 
	\frac{r (\m \nabla \times \m F_{\rm L})_\varphi}{\varkappa},
	&
	\label{rho1}	
\end{align}
where $(\m \nabla \times \, \m F_{\rm L})_\varphi$ is the $\varphi$-component of the vector 
$\m \nabla\times \m F_{\rm L}$.
Note that the quantity $(\m \nabla \times \, \m F_{\rm L})_\varphi$ on the right-hand side of \eqref{rho1}
is quadratic in the small magnetic field.
It is convenient to explicitly account for this smallness by introducing a function $\mathcal{B}$
[such that $\mathcal{B}^2/p \sim \mathcal{O}(1)$] according to the relation
\begin{align}
&
\m B(r, \theta)=\sqrt{ 8 \pi \varkappa} \, 
\bm{\mathcal{B}} (r, \theta).
&
\label{Bmac}
\end{align}
In this case, equation \eqref{rho1} can be rewritten as:
\begin{align}
&
-\partial_r \Phi_0 \, \partial_\theta \rho_1 +\partial_r \rho_0 \, \partial_\theta \Phi_1= 
2 r \, \{
\m \nabla \times[(\m \nabla \times \bm{\mathcal{B}}) \times \bm{\mathcal{B}}]
\}_\varphi.
&
\label{rho11}	
\end{align}
The relation (\ref{rho11}) appears to be helpful in the subsequent derivations.

\subsection{General analysis}
\label{general}

Let us assume that the given magnetic field configuration renders our star unstable.
Particular unstable modes develop at their imaginary characteristic frequencies $\omega$ such that 
$\ii\omega<0$.
Substituting
$\m\xi(\m r,t) = \m\xi(\m r) e^{-\ii \omega t}$,
the Euler equation \eqref{euler3} can be written as
\begin{align}
	&
	-\rho \omega^2 \m\xi = - {\pmb \nabla}\delta p -\delta \rho {\pmb 
		\nabla}\Phi 
	+\delta \m F_{\rm L},
	&
	\label{euler4}
\end{align}
where, as already discussed in Sec.\ \ref{assumpt}, we omit the term involving the perturbation of the gravitational potential, $\delta \Phi$.

The quantity $\delta \m F_{\rm L}$ can be estimated from Eq.\ \eqref{dFL} as
$|\delta \m F_{\rm L}| \sim B \delta B/L \sim B^2 |\m\xi|/L^2$, where $L$ is the typical spatial scale of the problem 
(of the order of some fraction of the stellar radius $R$).
Each of the remaining two terms on the right-hand side of Eq.\ \eqref{euler4} should not exceed 
$\delta F_{\rm L}$ on the order of magnitude, since, by assumption, the instability is caused by 
the magnetic field.
Indeed, if either of these terms were much larger than $\delta F_{\rm L}$,
magnetic effects (i.e., dependence on the magnetic field) 
in Eq.\ \eqref{euler4} could be neglected, and this equation would reduce to the usual eigenmode problem for a nonrotating non-magnetized star,
which, by assumption, is stable (and thus uninteresting for us)
%
\footnote{We would like to emphasize that a situation in which each of the terms ${\pmb 
\nabla}\delta p$ and $\delta \rho\, {\pmb \nabla}\Phi$ is large, but they nearly cancel when 
summed, is not realistic. Indeed, ${\pmb \nabla}\Phi$ is nearly radial. If we imagine a situation 
in which these two terms compensate each other with high precision, then ${\pmb \nabla}\delta p$ 
would also have to be nearly radial. This corresponds to radial perturbations, which are 
characterized by high frequencies and are unlikely to develop as a result of magnetic instability.}.
%

Assuming that all terms on the right-hand side of \eqref{euler4} are of the same order as $\delta F_{\rm L}$ or less, we obtain from \eqref{euler4} an estimate for the frequency $\omega$:
$|\omega| \lesssim 1/\tau_{\rm A}$, where $\tau_{\rm A} \sim L \sqrt{4 \pi \rho}/B$ is the Alfv\'en timescale.
We see that, as expected,
the characteristic timescale of instability development turns out to be of the order of $\tau_{\rm A}$ or greater.

From the fact that $|\m \nabla \delta p| \lesssim |\delta \m F_{\rm L}|$ and $|\delta \rho \m\nabla \Phi| \lesssim |\delta \m F_{\rm L}|$
we obtain the estimates $|\delta \rho/\rho| \lesssim |\varkappa \m\xi /L|$ and $|\delta p/p| \lesssim |\varkappa \m\xi /L|$
[see Eq.\ \eqref{BP} for the definition of $\varkappa$].
Recalling Eqs.\ \eqref{cont13} and \eqref{dp}, we then obtain that
\begin{align}
&
|\m\nabla\cdot(\rho \m\xi)|\lesssim \rho \varkappa \frac{|\m\xi|}{L},
&
\label{div_rhoxi}\\
&
|\xi_r| \lesssim \frac{1}{\gamma_{{\rm fr}, 0}-\gamma_{{\rm eq},0} }\, \varkappa \, |\m\xi|.
&
\label{xirxir}
\end{align}
The second estimate here follows from Eq.\ \eqref{dp} if one notices that, up to (small) terms of order 
$\varkappa$, one can write 
$\gamma_{{\rm fr}} \approx \gamma_{{\rm fr}, 0}(r)$, 
$\m \nabla \rho \approx \m \nabla \rho_0(r)$, and 
$\m \nabla p \approx \m \nabla p_0(r)$, 
where $\rho_0(r)$ and $p_0(r)$ are the equilibrium density and pressure profiles 
in the star without a magnetic field,
and introduces the equilibrium index $\gamma_{{\rm eq},0} $ according to
\begin{align}
&
\gamma_{{\rm eq},0}  = \frac{\rho_0}{p_0}\frac{d p_0[\rho_0, x_0(\rho_0)]}{d \rho_0},
&
\label{Gamma1}
\end{align}
where all derivatives in Eq.\ \eqref{Gamma1} are taken in the equilibrium spherically symmetric star
in the absence of a magnetic field (see Sec.\ \ref{assumpt}); 
$x_0(\rho_0)$ is the dependence of $x$ on the density 
in the non-magnetized star.
In a non-barotropic star, 
the frozen adiabatic index, $\gamma_{{\rm fr}, 0}$, 
is not equal to the equilibrium index, $\gamma_{{\rm eq},0} $ (see, e.g., Ref.\ 
\cite{reisenegger09} for details).
Since we assume that the non-magnetized star is stable, the difference $\gamma_{{\rm 
fr},0}-\gamma_{{\rm eq},0}$ between these indices must be positive for the stellar matter to be 
stably stratified (see, e.g., Refs.~\cite{reisenegger09, armm13}). This condition is equivalent to 
requiring that the squared Brunt-V$\ddot{\rm a}$is$\ddot{\rm a}$l$\ddot{\rm a}$ frequency 
$\mathcal{N}^2$ be positive throughout the star, where
\begin{align}
&
\mathcal{N}=g\sqrt{\frac{\gamma_{{\rm fr},0}-\gamma_{{\rm eq},0}}{\gamma_{{\rm fr},0}\gamma_{{\rm eq},0}}\, \frac{\rho_0}{p_0}}
&
\label{brunt}
\end{align}
and $g$ is the gravitational acceleration.

From estimates \eqref{div_rhoxi} and \eqref{xirxir}, it follows that 
$\m \nabla \cdot(\rho \m \xi)$ and $\xi_r$ in a non-barotropic ($\gamma_{{\rm fr}, 0} \neq \gamma_{{\rm eq},0} $) 
star are suppressed and generally proportional to $\varkappa$.
This result was obtained under the assumption that the problem has a single lengthscale $L$ over which all quantities vary.
As will become clear in what follows, an additional lengthscale associated with the perturbation may arise; in that case, the estimates for
$\m \nabla \cdot (\rho \m \xi)$ and $\xi_r$ {\it will be modified}. However, the main conclusion, that $\m \nabla \cdot (\rho \m \xi)$ and $\xi_r$ are suppressed, remains intact.

Note that in a barotropic star, where $\gamma_{{\rm fr}, 0} = \gamma_{{\rm eq},0} $, estimate \eqref{xirxir} does not hold,
as is formally evident from the vanishing denominator in this estimate. 
In this case, $\xi_r$ is expected to be comparable to the other components, $\xi_\theta$ and $\xi_{\varphi}$,
of the Lagrangian displacement $\m \xi$.

\subsection{Result of exact minimization in the case of a toroidal magnetic field}
\label{tor}

As shown by Tayler \cite{tayler73}, in the case of a toroidal field and non-axisymmetric 
Lagrangian displacements (i.e., when the azimuthal 
number $m \neq 0$), 
the minimum of $W[\m\xi]$ is achieved when $\m\xi$
satisfies a certain condition.%
%
\footnote{
\label{WI}
The minimization is carried out with respect to the component $\xi_\varphi$ or, 
equivalently, with respect to the function $\m\nabla \cdot \m\xi$. We stress that it is the 
potential energy functional $W[\m\xi]$ that is being minimized here by the discussed condition, not 
the ratio $W[\m\xi]/I[\m\xi]$. As discussed in Sec.\ \ref{bernshtein}, this approach is fully 
justified if one is only interested in proving the very existence of the instability 
($W[\m\xi]<0$). However, when determining the characteristic growth time of the instability and the 
optimal unstable Lagrangian displacement, one should, strictly speaking, minimize the ratio 
$W[\m\xi]/I[\m\xi]$ [see Eq.\ \eqref{ray}]. 
The results of this more accurate procedure are presented in Appendix \ref{accurate}, where it is shown 
that for weak magnetic fields ($\varkappa\ll 1$) the result of minimizing $W[\m\xi]/I[\m\xi]$ does 
not differ noticeably from the expressions \eqref{div0} and \eqref{div3} used below.
}
%
In the notation of Ref.\ \cite{gv78}, this condition is written as [see their equation (17)]
\begin{align}
	&
	\delta p = -\m\xi \cdot\m F_{\rm L},
	&
	\label{div0}
\end{align}
where $\delta p$ is given by \eqref{dp}.
Using Eqs.\ \eqref{force}, \eqref{cont13}, and \eqref{dp},
Eq.\ \eqref{div0} reduces to the form (see, e.g., Ref.\ \cite{armm13})
\begin{align}
	&
	\m\nabla\cdot\m\xi=	\frac{\rho}{\gamma_{\rm fr} p} \m \nabla \Phi\cdot \m \xi.
	&
	\label{div3}
\end{align}
This equation is valid for a toroidal magnetic field of arbitrary strength.

\subsection{\texorpdfstring{The function $\m\xi$}{The function mxi}}
\label{xixi}

In the general case, the Lagrangian displacement $\m \xi$ can be expanded into a Fourier series in $\varphi$:  
\begin{align}
	&
	\m\xi(r,\theta, \varphi)=\sum_{m \geq0}
	\left\{
		[\tilde{\xi}_r^{(m)}(r,\theta) {\rm sin}(m\varphi),
	\tilde{\xi}_\theta^{(m)}(r,\theta){\rm sin}(m\varphi), \tilde{\xi}_\varphi^{(m)}(r,\theta){\rm 
	cos}(m\varphi)]
\right.
	&
	\nonumber\\
	&
	\left.
	\quad\quad\quad\quad\quad\quad\quad\,
	+[\hat{\xi}_r^{(m)}(r,\theta) {\rm cos}(m\varphi),
	\hat{\xi}_\theta^{(m)}(r,\theta){\rm cos}(m\varphi), \hat{\xi}_\varphi^{(m)}(r,\theta){\rm 
	sin}(m\varphi)]
	\right\},
	&
	\label{xiexpr}
\end{align}
where $\xi_{r,\theta,\varphi}^{(m)}$ are unknown functions.  
It can be shown that each individual term in this expansion  
does not interfere with the others  
when substituted into the functional \eqref{en11}  
and integrated over $\varphi$.
In this work, we consider only non-axisymmetric Lagrangian displacements  
($m \neq 0$),  
for which formula \eqref{div3} holds.  
Theory and numerical results  
(see, e.g., \cite{tayler73,gv78,braithwaite06,brvg22b})  
hint that such displacements are the most unstable  
for our problem.
For $m \neq 0$, the first and second terms in the sum \eqref{xiexpr},  
which differ only by the substitution  
${\rm cos}(m\varphi) \rightleftharpoons {\rm sin}(m\varphi)$,  
give identical contributions to \eqref{en11} after integration in $\varphi$.  
Therefore, without loss of generality,  
we can search for $\m\xi$ in the form:
\begin{align}
&
\m\xi(r,\theta, \varphi)=[\tilde{\xi}_r^{(m)}(r,\theta) {\rm sin}(m\varphi),
\tilde{\xi}_\theta^{(m)}(r,\theta){\rm sin}(m\varphi), \tilde{\xi}_\varphi^{(m)}(r,\theta){\rm 
cos}(m\varphi)].
&
\label{xiexpr2}
\end{align}

As the main independent parameters, we choose the $r$- and $\theta$-components  
of this Lagrangian displacement.  
The $\varphi$-component is not independent and can be expressed  
in terms of the other components using Eq.\ \eqref{div3}.
We now introduce new notations, whose physical meaning will become clearer below:
\begin{align}
&
\tilde{\xi}_r^{(m)}= \varepsilon \mathfrak{g}(r,\theta),
&
\label{xir3}\\
&
\tilde{\xi}_\theta^{(m)} = - m r \,  \mathfrak{a}(r,\theta).
&
\label{xitheta3}
\end{align}
Here, $\mathfrak{g}(r,\theta)$ and $r \mathfrak{a}(r,\theta)$ are functions of the same order of smallness  
[$\mathfrak{g}(r,\theta)/\mathfrak{a}(r,\theta) \sim \mathcal{O}(r)$],  
and the small constant $\varepsilon \ll 1$ is introduced to reflect the result  
of Section \ref{general}, which states that the radial component of the Lagrangian displacement  
should be suppressed compared to the other components  
during the development of magnetic field instability in a non-barotropic star.  
A specific estimate for the constant $\varepsilon$ will be obtained below;  
it is clear, however, that $\varepsilon$ becomes smaller as the ratio  
$\varkappa = B^2/8 \pi p$ decreases  
[see estimate \eqref{xirxir} and the discussion following it].

Using the notations \eqref{xir3}, \eqref{xitheta3},  
the function $\tilde{\xi}_\varphi^{(m)}$ in Eq.\ \eqref{xiexpr2}  
can be expressed in terms of $\m \nabla \cdot \m \xi$, $\mathfrak{g}(r,\theta)$, and  
$\mathfrak{a}(r,\theta)$ as
\begin{align}
&
\tilde{\xi}_{\varphi}^{(m)}= \frac{\varepsilon \, {\rm sin}\theta}{m r} \partial_r[r^2 
\mathfrak{g}(r,\theta)]
- r \partial_\theta[\mathfrak{a}(r,\theta) \, {\rm sin}\theta]- \frac{r  {\rm sin}\theta}{m} 
\mathcal{D}(r,\theta),
&
\label{xiphi}
\end{align}
where we have introduced $\mathcal{D}(r,\theta)\equiv \m\nabla\cdot\m\xi/{\rm sin}(m \varphi)$.
As follows from Eq.\ \eqref{div3}, the function $\mathcal{D}(r,\theta)$ equals
\begin{align}
&
\mathcal{D}(r,\theta) = \frac{\rho}{\gamma_{\rm fr} p} [\varepsilon \, \partial_r \Phi \, 
\mathfrak{g}(r,\theta) 
- m \,  \partial_\theta \Phi \, \mathfrak{a}(r,\theta)].
&
\label{DDD}
\end{align}
Because $\partial_\theta \Phi=\mathcal{O}(\varkappa)$,
we have $\mathcal{D}(r,\theta)=\mathcal{O}(\varepsilon+\varkappa)$.

\section{Further transforming the energy functional: development of the perturbation theory}
\label{pert}

\subsection{\texorpdfstring{Expansion of $W[\m\xi]$ in powers of $\varepsilon$ and $\varkappa$}{Expansion of W[m xi] in powers of epsilon and varkappa}}

Now, knowing how $\m\xi$ depends on the functions $\mathfrak{g}(r,\theta)$ and $\mathfrak{a}(r,\theta)$, we can express the functional \eqref{en11} in terms of these functions.  
Expressing the equilibrium pressure gradient $\m\nabla p$ in \eqref{en11} using the relation  
$\m\nabla p = \m F_{\rm L} - \rho \m\nabla \Phi$ [see Eq.\ \eqref{force}],  
and neglecting the perturbation of the gravitational potential $\delta \Phi$  
as discussed in Sec.\ \ref{bernshtein},  
we obtain an expression for $W[\m\xi]$.  
After integration over the angle $\varphi$,  
this expression will depend on the magnetic field $B(r,\theta)$ and its first derivatives with respect to $r$ and $\theta$,  
on the functions $\mathfrak{g}(r,\theta)$ and $\mathfrak{a}(r,\theta)$ and their first derivatives,  
on the equilibrium density $\rho(r,\theta)$ and its first derivatives,  
on the equilibrium gradient of the gravitational potential $\Phi(r,\theta)$, and on  
the equilibrium functions $\gamma_{\rm fr}(r,\theta)$ and $p(r,\theta)$.

We will not write out this rather cumbersome expression explicitly here,  
as its analysis would not provide us with much useful information.  
Instead, recall that we have not yet used the assumption of Sec.\ \ref{assumpt}  
about the smallness of the magnetic field.  
This assumption allows us to expand all equilibrium functions $p$, $\rho$, $\Phi$, and $\gamma_{{\rm fr}}$  in $W[\m\xi]$ in a series over the small parameter $\varkappa$,  
representing them in the form \eqref{Adecomp}.  
Introducing, instead of the small magnetic field $\m B(r,\theta)$, the function $\bm{\mathcal{B}}(r,\theta)$  
according to Eq.\ \eqref{Bmac}, the functional $W[\m\xi]$ can be schematically written as:
\begin{align}
&
W[\m\xi]= \varepsilon^0 (\underline{\varkappa \alpha_{01}} 
+\underline{\underline{\varkappa^2 \alpha_{02}}} +\varkappa^3 
\alpha_{03}+\ldots)
&
\nonumber\\
&
\quad\quad\,
+\varepsilon \, (\underline{\underline{\varkappa \alpha_{11}}} +\varkappa^2 \alpha_{12}+\varkappa^3 
\alpha_{13}+\ldots)
&
\nonumber\\
&
\quad\quad\,
+\varepsilon^2 (\underline{\underline{\alpha_{20}}}+ \uwave{\varkappa \alpha_{21}} +\varkappa^2 \alpha_{22} 
+\ldots),
&
\label{wxi}
\end{align}
where the coefficients 
$\alpha_{ij}$ with $i=0,1,2$ and $j=0,1,2,\ldots$ are ordinary numbers for the given specific  
stellar model, toroidal field, and the functions $\mathfrak{g}(r,\theta)$ and $\mathfrak{a}(r,\theta)$.%
%
\footnote{
	Note that the expansion \eqref{wxi} does not contain terms $\sim \varepsilon^3$ and higher powers,  
	since the functional $W[\m\xi]$ is quadratic in $\m\xi$.}
%
If these quantities are not specified explicitly,  
the $\alpha_{ij}$ represent integrals over $dr d\theta$ of certain functions,  
which will be discussed below.  
Note that always $\alpha_{00} = \alpha_{10} = 0$.

We will be primarily interested in the largest terms (underlined once and twice) in the expansion  
\eqref{wxi}.%
%
\footnote{A single underline denotes the first order of perturbation theory.  
	Terms doubly underlined are combined by us into the second order of perturbation theory.
    The term underlined with a wavy line is discussed in Sec.\ \ref{n_estimate}.
}
%
These terms, in particular, depend on the functions  
$\partial_\theta \rho_1(r,\theta)$, $\Phi_0'(r)$, and $\rho_0'(r)$ (here the prime $'$ denotes  
differentiation with respect to $r$).  
Expressing these functions using, respectively, the relations \eqref{rho11},  
$\Phi_0'(r) = - p_0'(r)/\rho_0$,  
and $\rho_0'(r) = \rho_0 p_0'(r) / (p_0 \gamma_{{\rm eq},0})$,  
we arrive at the final expressions for the coefficients  
$\alpha_{01}$, $\alpha_{02}$, $\alpha_{11}$, and $\alpha_{20}$,  
which we will analyze below.

\subsection{First order of perturbation theory}
\label{first}

From Eq.~\eqref{wxi} it follows that the dominant contribution to $W[\m\xi]$ 
is provided by
the term $\varkappa \alpha_{01}$, which 
originates from
the magnetic contribution $W^{\rm mag}[\m\xi]$ to the energy functional \eqref{en11}. The 
coefficient $\alpha_{01}$ is given by%
%
\footnote{Treating $\varkappa \alpha_{01}$ as the leading term implicitly assumes the ordering 
$\varkappa \gg \varepsilon^2$, which allows us to regard $\varepsilon^2 \alpha_{20}$ as a higher-order term. Under the opposite assumption, $\varepsilon^2 \gg \varkappa$, the leading-order term of $W[\m{\xi}]$ would coincide with that of a non-magnetized fluid, which is assumed to be stable.
It can be shown that constructing the perturbation theory under the assumption $\varepsilon^2 \gg \varkappa$ leads to $\mathfrak{g}(r,\theta)=0$ (no displacement in the $r$-direction) and precludes $W[\m\xi]$ from being negative [see Eqs.\ \eqref{a20} and \eqref{W1} below]. }
%
%
\begin{align}
	&
	\alpha_{01}= 
    4 \pi \int r^2 \sin \theta  \left[ C_1 \mathfrak{a}^2+C_2 
	\partial_\theta(\mathfrak{a}^2)+
	C_3 (\partial_\theta \mathfrak{a})^2
	\right] dr d\theta,
	&
	\label{Wmagn2}
\end{align}
where 
\begin{align}
	&
	C_1=\frac{1}{4} m^2 
	\left[
	\mathcal{B}^2 \left(1+\frac{m^2-1}{\sin^2 \theta}\right)-{\rm cot}\theta \, \partial_\theta 
	\left(\mathcal{B}^2 \right)
	\right],
	&
	\label{C1}\\
	&
	C_2=-\frac{1}{4} m^2 \mathcal{B}^2 {\rm cot}\theta,
	&
	\label{C2}\\
	&
	C_3=\frac{1}{4} m^2 \mathcal{B}^2.
	&
	\label{C3}
\end{align}
Here and below, the integration over the angle $\theta$ is performed from $0$ to $\pi$, and over $r$ from $0$ to the stellar radius $R$, i.e., we integrate over the unperturbed volume of a spherically symmetric star without a magnetic field.  
Taking into account the non-sphericity of the star's equilibrium configuration caused by the magnetic field does not contribute to integrals of the type \eqref{Wmagn2}, due to our assumptions that both $B(r,\theta)$ and the matter density vanish at the stellar surface (see Sec.\ \ref{assumpt}).

Integrating by parts the second term in the expression for $C_1$ depending on 
$\partial_\theta 
\left(\mathcal{B}^2 \right)$, one arrives at the following expression
\begin{align}
	&
	\alpha_{01}= 
    4 \pi  \int  r^2 \sin\theta  \, 
	\frac{m^2  \mathcal{B}^2}{4} \, \left[ \, (\partial_\theta \mathfrak{a})^2+
	\frac{(m^2-1)}{{\rm sin}^2 \theta} \mathfrak{a}^2
	\right]
	\, dr d\theta.
	&
	\label{Wmagn31}
\end{align}
To obtain this formula we make use of the boundary conditions discussed in Sec.~\ref{assumpt}.
It is easy to see from Eq.\ \eqref{Wmagn31} that to maximally  destabilize  the system 
(i.e., to make $W[\m\xi] \simeq \varkappa \alpha_{01}$ as small as possible)
one needs 
to choose 
the lowest possible value of $m$, in our case $m=1$.%
%
\footnote{It might be shown that the case $m=0$ is also correctly described by this expression
but we do not consider it in this work because of the reasons stated in Sec.~\ref{xixi}.}
%
In the remainder of the paper, we assume $m=1$ and omit the superscript $(m)$. Finally, one obtains
\begin{align}
	&
	\alpha_{01}= 
    \pi  \int  r^2 \sin\theta  \, \mathcal{B}^2 \, 
	(\partial_\theta \mathfrak{a})^2
	\, dr d\theta.
	&
	\label{Wmagn3}
\end{align}
From Eq.~(\ref{Wmagn3}), it follows that, to first order in perturbation theory, the lowest 
possible value of $W[\m \xi]$ for the $m=1$ mode is zero.
This value is attained when $\partial_\theta \mathfrak{a}=0$, i.e.
\begin{align}
\mathfrak{a}(r,\theta)=a(r) \quad \text{in the first order of perturbation theory,}
\label{dtheta_a}
\end{align}
where $a(r)$ is some function of $r$.
Regarding Eq.~\eqref{dtheta_a}, we emphasize the following:
(i) strictly speaking, it needs to hold only in the region where the magnetic field is localized; and
(ii) Eq.~\eqref{dtheta_a} needs to be satisfied only to leading order in the expansion in 
$\varkappa$ and $\varepsilon$.
At higher orders $\partial_\theta \mathfrak{a}\neq 0$ and the function $\mathfrak{a}$ may depend on 
$\theta$.
One can check that allowing for the $\theta$ dependence of $\mathfrak{a}$ and substituting the 
ansatz \eqref{xiexpr2} into \eqref{en11} introduces only two terms in the energy functional 
$W[\m\xi]$ that depend on $\partial_\theta \mathfrak{a}(r,\theta)$, namely
\begin{align}
&
\pi \varkappa \int r^2 \sin\theta \, \mathcal{B}^2
\left[
(\partial_\theta \mathfrak{a})^2
-2\varepsilon\,\partial_\theta \mathfrak{a}\,\partial_r\mathfrak{g}
\right]dr\,d\theta.
&
\label{www}
\end{align}
Here, the first term is non-negative, while the second one, which originates from $\alpha_{11}$, is 
not sign-definite. Thus, the contribution of $\partial_\theta \mathfrak{a}$ to $W[\m\xi]$ can be 
profitable only if
\begin{align}
&
	|\partial_\theta \mathfrak{a}|\lesssim |\varepsilon \partial_r\mathfrak{g} | 
&
\label{dthetaa}
\end{align} 
with $\partial_\theta \mathfrak{a}$ and $\varepsilon \partial_r\mathfrak{g}$ having the same sign.
In this case, the terms in the energy functional $W[\m\xi]$ that contain
$\partial_\theta \mathfrak{a}(r,\theta)$
are of order $\mathcal{O}(\varkappa \varepsilon^2)$, that is, they contribute to the (small) 
third-order term underlined with a wavy line in Eq.~\eqref{wxi}.

What does the 
optimal 
Lagrangian displacement with $\mathfrak{a(r,\theta)}$ given by Eq.\ \eqref{dtheta_a}
correspond to in the first-order perturbation theory?
First, from Eq.\ \eqref{xir3} it follows that  
$\tilde{\xi}_{r} = 0$  
(taking into account the term depending on $\varepsilon$ in this formula would lead to terms in $W[\m\xi]$ that depend on $\varepsilon$, which we neglect at this order of perturbation theory).  
Similarly, from Eq.\ \eqref{xitheta3} we have 
$\tilde{\xi}_\theta =-r\mathfrak{a}(r,\theta)= -r [a(r)+\mathcal{O}(\varkappa 
\varepsilon^2)]\approx -r a(r)$.  
Furthermore, from Eq.\ \eqref{xiphi}  
it follows that $\tilde{\xi}_{\varphi} \approx -r \partial_\theta[\mathfrak{a}(r,\theta) {\rm 
sin}\theta] = 
-r \partial_\theta\{[a(r)+\mathcal{O}(\varkappa \varepsilon^2)] {\rm sin}\theta\} 
\approx -r a(r) {\rm cos}\theta$
[to show 
the first equality here,
it is necessary to note that the function $\mathcal{D}$,  
defined by Eq.\ \eqref{DDD}, is of order $\mathcal{O}(\varepsilon + \varkappa)$,  
i.e., it can be neglected in the first approximation].  
Thus, as follows from Eq.\ \eqref{xiexpr2},  
the Lagrangian displacement in this order of perturbation theory  
is $\m\xi \equiv \m\xi_{\rm rot} = [0, -r a(r) {\rm sin} \varphi,  
-r a(r) {\rm cos}\theta \, {\rm cos}\varphi ]$.  
As is easy to see, this Lagrangian displacement corresponds to a differential (in $r$) 
{\it shellular}
rotation 
about the Cartesian $x$-axis, given by the formula
\begin{align}
&
\m\xi_{\rm rot}(r, \theta, \varphi)=a(r) \, \hat{\m x}\times \m r,
&
\label{lagx}
\end{align}
where $\hat{\m x}=({\sin \theta \cos\varphi, \cos\theta 
\cos\varphi, -\sin\varphi})$ and $\m r=r {\pmb e}_r$.
Solid-body rotation corresponds to the case $a(r)={\rm const}$.

One can verify that if we 
looked for the Lagrangian displacement 
$\m\xi(r,\theta,\varphi)$ in the form of Eq.\ \eqref{xiexpr2},
but with $\sin\varphi$ 
replaced with $\cos\varphi$ and vice versa
(as argued in Sec.~\ref{xixi}, this is an alternative description),
and repeated all the derivations in this section
we would find that 
the corresponding optimal Lagrangian displacement $\m\xi_{\rm rot}$ in the first approximation 
is a shellular differential rotation about the $y$-axis.
This means that 
a shellular differential rotation with respect to an {\it arbitrary} axis $\hat{\m O}$ in the $x-y$ plane,
\begin{align}
	&
	\m\xi_{\rm rot}(r,\theta)=
	a(r) \, \hat{\m O}\times \m r,
	&
	\label{lagany}
\end{align}
is optimal in the first order of the perturbation theory in $\varkappa$ and $\varepsilon$.%
\footnote{
\label{nullify}
	In fact, 
    the Lagrangian displacement of the form \eqref{lagany}
does not change the magnetic part \eqref{Wmag} of the energy functional \eqref{en11},
i.e., $W^{\rm mag}[\m\xi_{\rm rot}] = 0$, and this result is exact
(it does not depend on whether the toroidal magnetic field is small or not).
To demonstrate this more directly, we integrate Eq.\ \eqref{Wmag} by parts using the identity
$\m\nabla \cdot (\m A_1 \times \m A_2) = (\m\nabla \times \m A_1)\cdot \m A_2
- \m A_1 \cdot (\m\nabla \times \m A_2)$
to obtain
\begin{align}
&
W^{\rm mag}[\m\xi] = \frac{1}{2}\int dV \, (\m\xi \times \m B)\cdot
\underline{\m\nabla \times \left(\frac{\delta \m B}{4 \pi} + \frac{1}{c}\m j \times \m \xi\right)}.
&
\label{Wmagzero}
\end{align}
One may verify that, upon setting $\m\xi = \m\xi_{\rm rot}$, the vector $(\m\xi \times \m B)$ in \eqref{Wmagzero}
is directed along $r$, while the $r$-component of the underlined vector vanishes.
Hence, the integrand in \eqref{Wmagzero} vanishes at each point inside the star, implying that
$W^{\rm mag}[\m\xi_{\rm rot}] = 0$.

To explain this result qualitatively, consider an arbitrary spherical shell. Since the background 
field is purely toroidal, the magnetic field lines are everywhere tangent to the shell and do not 
cross it. Now suppose that this shell is rotated as a rigid body about an axis perpendicular to the 
symmetry axis of the magnetic field. Because the field is frozen into the fluid, each fluid element 
carries its magnetic field with it. Under such a rigid rotation, the field associated with a given 
fluid element is merely reoriented, but its magnitude is unchanged. In other words, such 
displacement changes only the direction of the magnetic field, not its absolute value. Since this 
argument applies independently to every spherical shell, the magnetic energy is unchanged by the 
displacement $\m\xi = \m\xi_{\rm rot}$.
}
%
Note in this context that differential rotation about the symmetry axis $z$ is trivial \cite{fs78} 
since
it does not 
perturb
the system and, in particular, does not change the energy $W[\m \xi]$. Thus, in principle, we could also add such a displacement to the displacement of the form \eqref{lagany}. 
However, any trivial displacement increases the kinetic-energy functional \eqref{III}, thereby 
increasing the characteristic growth time of the potential instability; see Eq.\ \eqref{III33}.
In other words, trivial displacements are always unfavorable, and the optimal rotation should occur about an axis in the $x-y$ plane.
Note also that
the displacements \eqref{lagx} and \eqref{lagany}
do not apply to the non-magnetized region of the star, which need not sustain shellular differential rotation and may instead respond differently to the rotation of the magnetized regions.
The optimal displacement for the non-magnetized stellar region cannot be determined from the first few orders of perturbation theory considered in this paper.
In what follows, whenever needed in specific examples, we will assume that the magnetic field 
permeates the entire star or a substantial part of it, so that the approximation $\m \xi \approx \m 
\xi_{\rm rot}$ remains valid throughout the star to leading order.

Summarizing this section, we have found that for a toroidal field,  
the most unstable mode is the one with azimuthal number $m = 1$.  
The optimal Lagrangian displacement for this mode in the first order of the perturbation 
theory corresponds to a shellular differential rotation of the magnetized regions of the star 
about an arbitrary axis
in the $x-y$ plane,
for example, without loss of generality, about the $x$-axis [see Eq.\ \eqref{lagx}]
Moreover, this Lagrangian displacement is ``neutral''  
in the sense that the change it induces in the energy functional  
$W[\m\xi_{\rm rot}]$ in the first-order perturbation theory is zero,  
$W[\m\xi_{\rm rot}] \simeq \varkappa \alpha_{01} = 0$.

\subsection{Second-order perturbation theory}
\label{second}

Let us now consider the second-order perturbation theory, which includes terms  
proportional to $\varkappa^2$, $\varepsilon \varkappa$, and $\varepsilon^2$  
in the expansion \eqref{wxi} of the energy functional $W[\m\xi]$.  
Although, according to the initial assumption (and as we will verify below),  
the quantities $\varkappa$ and $\varepsilon$  
may differ significantly from each other, we group these terms together,  
keeping in mind that all higher-order contributions in the expansion \eqref{wxi}  
(for example, terms $\propto \varkappa^3$, $\varepsilon \varkappa^2$, and $\varepsilon^2 \varkappa$)  
are much smaller than at least {\it some} of the terms we retain.  
We will return to the discussion of this assumption in Sec.~\ref{n_estimate}.  

The coefficients $\alpha_{ij}$ for the second-order terms in the expansion \eqref{wxi} are given by:
\begin{align}
&
\alpha_{20}= \frac{1}{2} \int r^2 \sin \theta \, W_1 \, \mathfrak{g}(r,\theta)^2 \, dr d\theta,
&	
\label{a20}\\
&
\alpha_{11}= \frac{1}{2} \int r^2 \sin \theta \, W_2 \, \mathfrak{g}(r,\theta) \, dr d\theta
+ \underline{\frac{1}{2} \int \partial_r 
	\left[
    4 \pi  r^2 \, \cos \theta \, \mathfrak{a}(r,\theta) \,  \mathcal{B}^2 
	\mathfrak{g}(r,\theta)
	\right] dr d\theta},
&
\label{a11}\\
&
\alpha_{02} = \frac{1}{2} \int r^2 \sin \theta \, W_3 \, dr d\theta,
&
\label{a02}
\end{align}
where the functions $W_1$, $W_2$, and $W_3$ are defined as
\begin{align}
	& 	
	W_1 =  \frac{
		\pi 
		\, (\gamma_{{\rm fr}, 0}-\gamma_{{\rm eq},0} ) \, 
		[p_0'(r)]^2}{p_0 
		\gamma_{{\rm fr}, 0} \gamma_{{\rm eq},0} },
	&
	\label{W1}\\
	&
	W_2= \frac{2\pi}{p_0 \gamma_{{\rm fr}, 0} \gamma_{{\rm eq},0} }\left[
	- 
	2 \mathcal{B}^2
	\,
	{\rm cot}\theta
	\, p_0 \, \gamma_{{\rm fr}, 0} \, \gamma_{{\rm eq},0}  \, \partial_r\mathfrak{a}(r,\theta)
	+
	  p_0'(r) \, \partial_\theta \Phi_1 \, 
	(\gamma_{{\rm fr}, 0}-\gamma_{{\rm eq},0} ) \, \rho_0 \, \mathfrak{a}(r,\theta)
	\right],
	&
	\label{W2}\\
	&
	W_3=\frac{\pi
		\partial_\theta \Phi_1 \, \rho_0 \, \mathfrak{a}(r,\theta)^2}{ r p_0'(r) \, p_0\, 
		\gamma_{{\rm fr}, 0}\, 
		\gamma_{{\rm eq},0} }
	\left[ 
	r \, p_0'(r)\, \partial_\theta \Phi_1 \, 
	(\gamma_{{\rm fr}, 0}-\gamma_{{\rm eq},0} )  \, 
	\rho_0 
	+ 4 \mathcal{B} \, p_0 \, \gamma_{{\rm fr}, 0} \gamma_{{\rm eq},0}  \left( -
	\partial_\theta 
	\mathcal{B} +r 
	\,
	{\rm cot}\theta
	\, \partial_r \mathcal{B}\right)
	\right].
	&
	\label{W3}
\end{align}
In these equations, it is legitimate to replace $\mathfrak{a}(r,\theta)$ with the function $a(r)$, 
since $\mathfrak{a}(r,\theta)-a(r)\sim\mathcal{O}(\varkappa \epsilon^2)$. We nevertheless prefer 
not to do so, thereby effectively retaining some higher-order terms within the second-order 
treatment considered here.
Note that the underlined term in Eq.\ \eqref{a11} is a total derivative  
and vanishes upon integration over $r$ due to the boundary conditions (see Sec.~\ref{assumpt}).  
The corresponding functional $W[\xi]$, taken up to (and including) second-order perturbation theory,  
is given by [see Eq.\ \eqref{wxi}]
\begin{align}
&
W[\m\xi]\simeq  \varkappa^2 \alpha_{02}
+\varepsilon \varkappa \alpha_{11}
+\varepsilon^2 \alpha_{20}
&
\nonumber\\
&
=\frac{1}{2} \int r^2 \sin \theta 
\left[
\varepsilon^2 \, W_1 \, \mathfrak{g}(r,\theta)^2 + \varepsilon \varkappa \, W_2 \, 
\mathfrak{g}(r,\theta) + \varkappa^2 \, W_3
\right] dr d\theta.
&
\label{wxi2}
\end{align}
As can be seen, this functional depends,
among other contributions, on terms that are quadratic in 
the radial function  
$\mathfrak{g}(r,\theta)$ and does not depend on its derivatives.  
Furthermore, note that the coefficient $W_1$ given by Eq.\ \eqref{W1}  
is always positive for stably stratified matter,  
in which $\gamma_{{\rm fr}, 0} - \gamma_{{\rm eq},0}  > 0$ (see Sec.~\ref{general}).  
This means that there exists a function $\mathfrak{g}(r,\theta)$  
which {\it minimizes} $W[\m\xi]$.  
Performing the minimization, one obtains
\begin{align}
&
\mathfrak{g}(r,\theta) = - \frac{\varkappa}{\varepsilon} \, \frac{W_2}{2 W_1}.
&
\label{gr}
\end{align}
Given the asymptotic behavior of the magnetic field at the poles and the center,
it is straightforward to verify that $\mathfrak{g}(r,\theta)$ is always finite within the star.

At this point, let us make a comment on barotropic matter, i.e., when $\gamma_{{\rm fr},0}=\gamma_{{\rm eq},0}$.
In this case, one cannot minimize the functional~\eqref{wxi2} and arrive at Eq.~\eqref{gr}, because $W_1=0$ and the functional~\eqref{wxi2} is linear in $\mathfrak{g}(r,\theta)$.
Still, one can always make the functional~\eqref{wxi2} negative by choosing an appropriate sign and amplitude of $\mathfrak{g}(r,\theta)$.
Thus, a Lagrangian displacement of the type considered in this section (a shellular differential rotation plus a small radial component) destabilizes a barotropic star as well.
Nevertheless, numerical simulations (e.g., \cite{brvg22b}) indicate that, in barotropic stars, this type of instability is not the dominant one. Instead, the system typically develops an instability in which the radial displacement $\xi_r$ is comparable to the $\theta$- and $\varphi$-components, $\xi_\theta$ and $\xi_\varphi$. This is not surprising, because the results of Sec.\ \ref{general} imply that, in barotropic matter, the displacement $\xi_r$ is not suppressed.
We will mostly focus on stratified (non-barotropic) matter in what follows.

Returning to non-barotropic matter, note that the function $W_2$ in Eq.~\eqref{gr} depends on the 
function $\mathfrak{a}(r,\theta)$ and its derivative $\partial_r\mathfrak{a}(r,\theta)$ [see 
Eq.~\eqref{W2}].
Accordingly, using \eqref{W1} and \eqref{W2}, the right-hand side of \eqref{gr} can be estimated as
\begin{align}
&
\mathfrak{g}(r,\theta) \sim \frac{\varkappa}{\varepsilon} \, R
\left[\mathfrak{a}(r,\theta)+\frac{\gamma_{{\rm fr}, 0}\gamma_{{\rm eq}, 0}}{\gamma_{{\rm fr}, 
0}-\gamma_{{\rm eq},0} 
}\, \partial_r\mathfrak{a}(r,\theta)R 
\right], 
&
\label{gg}
\end{align}
where $R$ is the characteristic lengthscale over which the pressure $p_0(r)$ varies (we identify 
it with the stellar radius).
If we introduce the characteristic scale 
$1/k \lesssim R$  
for the variation of the function $\mathfrak{a}(r,\theta)$ in the radial direction, we arrive at 
the estimate
\begin{align}
&
\mathfrak{g}(r,\theta) \sim \frac{\varkappa n}{\varepsilon} 
\frac{\gamma_{{\rm fr}, 0}\gamma_{{\rm eq}, 0}}{\gamma_{{\rm fr}, 0}-\gamma_{{\rm eq},0} } R 
\,\mathfrak{a}(r,\theta)
\label{gg2}
&
\end{align}
where $n$ is a (possibly large) numerical coefficient of order $k R$, i.e. $n \sim k R$.

Let us now recall that, according to the initial assumption, the functions  
$\mathfrak{g}(r,\theta)$ and $r \mathfrak{a}(r,\theta)$ 
should be of the same order of magnitude  
(see Eqs.\ \eqref{xir3} and \eqref{xitheta3} and the text following them).
This means that $\mathfrak{g}(r,\theta)/\mathfrak{a}(r,\theta) \sim \mathcal{O}(r)$ and hence
$\varepsilon$ must be
\begin{align}
&
\varepsilon \sim \frac{n \varkappa \, \gamma_{{\rm fr}, 0}\gamma_{{\rm eq}, 0}}{\gamma_{{\rm fr}, 0}-\gamma_{{\rm eq},0} } \sim \frac{n \varkappa }{\gamma_{{\rm fr}, 0}-\gamma_{{\rm eq},0} }, 
&
\label{eps}
\end{align}
where we set $\gamma_{{\rm fr}, 0}\gamma_{{\rm eq}, 0}\sim 1$
in the second estimate.
We see that $\varepsilon$ will be of the order of $\varkappa$  
only if the characteristic radial scale of variation of the function $\mathfrak{a}(r,\theta)$  
matches $R$, that is, is of the order of the stellar radius, and, additionally, the difference $\gamma_{{\rm fr}, 0} - \gamma_{{\rm eq},0} $  
is not too small.  
Otherwise, $\varepsilon$ can be much larger (but must remain much less than  
unity, $\varepsilon \ll 1$, by assumption).  
A precise upper bound estimate for $\varepsilon$  
[and, consequently, for $n$, see Eq.\ \eqref{eps}]  
will be obtained in Sec.~\ref{n_estimate}.

The parameter $\varepsilon$ in Eq.\ \eqref{eps} is expressed in terms of the quantities $\varkappa$ and $\gamma_{{\rm fr},0}-\gamma_{{\rm eq},0}$. Alternatively, one can express $\varepsilon$ in terms of the typical Alfv\'en frequency $\omega_{\rm A} \sim B/[(4 \pi \rho_0)^{1/2}R]$ and the Brunt-V$\ddot{\rm a}$is$\ddot{\rm a}$l$\ddot{\rm a}$ frequency $\mathcal{N}$ defined in Eq.~\eqref{brunt}:
\begin{align}
&
\varepsilon \sim n \frac{\omega_{\rm A}^2}{\mathcal{N}^2}. 
&
\label{eps11}
\end{align}

Plugging Eq.\ \eqref{gr} into (\ref{wxi2}), one gets
\begin{align}
	& 	
W[\m\xi] \simeq
\frac{\varkappa^2}{2} \int r^2 \sin\theta 
\left(-\frac{W_2^2}{4 W_1}+\underline{W_3} \right) \, 
dr d\theta.
&
\label{wxi3}
\end{align}
Substituting the expressions \eqref{W1}--\eqref{W3} for $W_1$, $W_2$, and $W_3$, it is easy to see 
that the integrand in this expression is nonzero only in the region where the magnetic field is 
localized, in which $\mathfrak{a}(r,\theta)= a(r)+\mathcal{O}(\varkappa \varepsilon^2)\approx 
a(r)$; see Eq.\ \eqref{dtheta_a}.

It is evident that the expression \eqref{wxi3} for $W[\m\xi]$ consists of two terms,  
the first of which is always non-positive, while the second  
can have an arbitrary sign.  
The entire expression is formally proportional to the small quantity $\varkappa^2$.  
However, such smallness of the functional will occur only if the function $\mathfrak{a}(r,\theta)$  
does not vary significantly on the scale of the stellar radius, i.e.,  
$R \,\partial_r\mathfrak{a}(r,\theta) \sim \mathfrak{a}(r,\theta)$ (and, consequently, $n \sim 
1$).  
In this case, both the first and second terms inside the parentheses of the integrand in Eq.\ \eqref{wxi3}  
will be on the order of $p_0 r^2 \mathfrak{a}(r,\theta)^2 / R^2$.  
Of course, it is not excluded that even in this case, for some models of the toroidal field,  
one can choose a function $\mathfrak{a}(r,\theta)$ such that $W[\m\xi]$ becomes negative (and thus 
prove the instability of the field).  
However, such a result will not constitute a general proof of the instability  
of {\it any} toroidal field.%
%
\footnote{Note that, in the commonly used approximation $\Phi_1=0$, 
the function $W_3$ vanishes [see Eq.\ \eqref{W3}], and the instability sets in already for $n\sim 1$ 
(in fact, it sets in for any shellular differential rotation, see Appendix \ref{approx} for details).
}
%

To prove instability in the most general case, let us assume that the function  
$\mathfrak{a}(r,\theta)$ varies on small spatial scales, that is, 
$n = k R \gg 1$.  
In this case, the first term in \eqref{wxi3}, which depends on $W_2$,  
becomes much larger than the second (underlined) term.  
Indeed, as we have already established above, $W_2$ contains a term proportional to 
$\partial_r\mathfrak{a}(r,\theta) \sim n  
\mathfrak{a}(r,\theta) / R$,%
%
\footnote{For example, if one chooses $\mathfrak{a}(r,\theta)$ in the form $\mathfrak{a}(r,\theta) 
= \sin (r n / R)$, then $R \,\partial_r\mathfrak{a}(r,\theta) = n \cos(r n / R)$. }
%
therefore, the first term  
in \eqref{wxi3} is larger than the underlined term by a factor of $n^2$, and the latter can be neglected.  
As a result, we obtain an expression for $W[\m\xi]$ that is manifestly negative,
\begin{align}
	& 	
	W[\m\xi] \simeq
	-\frac{\varkappa^2}{8} \int r^2 \sin\theta \,
	\frac{W_2^2}{W_1} \, 
	dr d\theta,
	&
	\label{wxi4}
\end{align}
guaranteeing the instability of any toroidal field.  
Note that the true order of smallness of the functional $W[\m\xi]$ will be  
$W[\m\xi] \sim \varkappa^2 n^2 / (\gamma_{{\rm fr}, 0} - \gamma_{{\rm eq},0} ) \sim  
(\gamma_{{\rm fr}, 0} - \gamma_{{\rm eq},0} ) \varepsilon^2$,  
rather than $\varkappa^2$.

\subsection{\texorpdfstring{Estimate of the permissible magnitude of $n$}{Estimate of the permissible magnitude of n}}
\label{n_estimate}

In principle, by choosing $n$ arbitrarily large, one can make the functional \eqref{wxi4}  
or \eqref{wxi3} arbitrarily large in magnitude and negative.  
However, let us recall that $n$ and $\varepsilon$ are related by the estimate \eqref{eps},  
and $\varepsilon$, by assumption, must satisfy $\varepsilon \ll 1$.  
From this inequality, we immediately find $n \ll (\gamma_{{\rm fr}, 0} - \gamma_{{\rm eq},0} ) / \varkappa$.  
In fact, there is a stricter bound on $n$ arising from the requirement that the terms of the next (third) order  
in the expansion \eqref{wxi} of the functional $W[\m\xi]$  
do not affect the results obtained above in the second-order perturbation theory.

Before proceeding to the analysis of the third-order terms, let us briefly return to the second-order terms  
and analyze them in light of the results obtained in Sec.~\ref{second}.  
As we found in Sec.~\ref{second}, $\varepsilon \sim n \varkappa / (\gamma_{{\rm fr}, 0} - \gamma_{{\rm eq},0} )$,  
so the second-order terms $\varkappa^2 \alpha_{02}$, $\varkappa \varepsilon \alpha_{11}$, and $\varepsilon^2 \alpha_{20}$  
actually have different orders of smallness in the limit of interest $n \gg 1$.  
It is also important to note that the coefficient $\alpha_{11}$ depends on the large derivative 
$\partial_r\mathfrak{a}(r,\theta) \sim n  
\mathfrak{a}(r,\theta)/R$ [this dependence arises through the function $W_2$, see Eqs.\ \eqref{a11} 
and \eqref{W2}].  
This means that the term $\varkappa \varepsilon \alpha_{11}$ actually has the order of smallness  
$\varkappa \varepsilon \alpha_{11} \sim \varkappa \varepsilon n \sim \varepsilon^2(\gamma_{{\rm fr}, 0} - \gamma_{{\rm eq},0} )$,  
the same as the term $\varepsilon^2 \alpha_{20} \sim \varepsilon^2 (\gamma_{{\rm fr}, 0} - \gamma_{{\rm eq},0} )$,  
while the term $\varkappa^2 \alpha_{02} \sim \varkappa^2$ can be neglected  
[it corresponds to the small and subsequently discarded underlined term in Eq.\ \eqref{wxi3}].%
%
\footnote{
Note that more accurately $\varkappa \varepsilon \alpha_{11}$ can be estimated as
$\varkappa \varepsilon \alpha_{11} \sim \varkappa \varepsilon n \int \mathcal{B}^2 \, 
\mathfrak{a}(r,\theta)^2 \,dV$. 
\label{est}
}
%

This example shows that the terms formally grouped into the same order of smallness  
may, in fact, differ significantly in magnitude from each other.  
Another important observation, relevant also to the analysis of higher-order terms,  
is that one must carefully track the radial derivatives  
$\partial_r\mathfrak{a}(r,\theta)$ and $\partial_r\mathfrak{g}(r,\theta)$ in the integrands  
for the expansion coefficients $\alpha_{ij}$, since these derivatives  
generate an additional large multiplicative factor $n$.%
%
\footnote{The fact that the derivative 
$\partial_r\mathfrak{g}(r,\theta)$
leads to the appearance of an additional factor $n$,
i.e., 
$\partial_r\mathfrak{g}(r,\theta) \sim n \mathfrak{g}(r,\theta)$,  
follows from the form of the solution \eqref{gr}, which depends on $\mathfrak{a}(r,\theta)$,  
or from the estimate \eqref{gg}.
}
%

Let us now apply the considerations discussed above to the analysis of the third-order contributions  
in \eqref{wxi}, which have the form: $\varkappa^3 \alpha_{03}$,  
$\varepsilon \varkappa^2 \alpha_{12}$, and $\varepsilon^2 \varkappa \alpha_{21}$.  
Formally, the largest among them is the last term $\propto \varepsilon^2 \varkappa$;
it is underlined with a wavy line in Eq.\ \eqref{wxi}.
The other contributions could become comparable to it  
only if they depended on derivatives like $\partial_r\mathfrak{a}(r,\theta) \sim n 
\mathfrak{a}(r,\theta)/R$  
or $\partial_r\mathfrak{g}(r,\theta) \sim n \mathfrak{g}(r,\theta)/R$.  
It turns out, however, that the only terms of this kind  
are indeed contained in $\varepsilon^2 \varkappa \alpha_{21}$.
Thus, the dominant third-order contribution to $W[\m\xi]$ is:
\begin{align}
&
{\rm 3rd \,\, order \,\,contribution \,\, into} \,\, W[\m\xi]\approx
\pi\,\varkappa 
\int r^2 \sin \theta \, 
\mathcal{B}^2 \,
\left[\partial_\theta \mathfrak{a}(r,\theta)-\varepsilon\partial_r \mathfrak{g}(r,\theta) \right]^2 
\, dr d\theta.
&	
\label{W5}
\end{align}
Note that this contribution is always non-negative, i.e., it stabilizes the instability.  
Using Eqs.\ \eqref{gg2}, \eqref{eps}, and \eqref{dthetaa}, one can estimate  
$\left[\partial_\theta \mathfrak{a}(r,\theta)-\varepsilon\partial_r \mathfrak{g}(r,\theta) 
\right]^2 \sim n^2 \varepsilon^2 \mathfrak{a}(r,\theta)^2$.
Accordingly, the expression \eqref{W5} is of the order  
$\sim \varkappa \varepsilon^2 n^2 \int \mathcal{B}^2 \, \mathfrak{a}(r,\theta)^2 \, dV$.  
In order for the arguments of Sec.~\ref{second} to remain valid,  
this third-order contribution must be much smaller than the second-order one.  
The latter can be estimated as:  
$\varkappa \varepsilon \alpha_{11} \sim \varkappa \varepsilon n \int \mathcal{B}^2 \, 
\mathfrak{a}(r,\theta)^2 \, dV$  
(see footnote \ref{est}). Combining these two estimates, we obtain: 
\begin{align}
&
n \ll \frac{1}{\varepsilon} \quad\quad \Rightarrow \quad\quad n \ll 
\frac{\sqrt{\gamma_{{\rm fr}, 0}-\gamma_{{\rm eq},0} }}{\varkappa^{1/2}},
\quad\quad {\rm or}, 
\,\,{\rm equivalently}, \quad\quad n\ll \frac{\mathcal{N}}{\omega_{\rm A}},
&
\label{nest}
\end{align}
where we made use of Eqs.\ \eqref{eps} and \eqref{eps11}. 
As can be seen, this condition turns out to be much more stringent than the naive criterion obtained at the beginning of this section.  
Note also that such a restriction on $n$ ensures that the second-order terms remain much smaller than the first-order terms.  
Indeed, $\varepsilon^2(\gamma_{{\rm fr}, 0}-\gamma_{{\rm eq},0} ) \sim n^2 \varkappa^2 / (\gamma_{{\rm fr}, 0} - \gamma_{{\rm eq},0} ) \ll \varkappa$.

\subsection{\texorpdfstring
{A more accurate estimate 
of the optimal $n$ 
}
{A more accurate estimate 
of the optimal $n$ 
}
}
\label{diffeq}
In the previous section, we found that in order to guarantee $W[\m\xi] < 0$, and thus demonstrate the instability of the toroidal field, it is necessary to take $n \gg 1$, which, at the same time, must not be too large: $n \ll \sqrt{\gamma_{{\rm fr}, 0} - \gamma_{{\rm eq},0} } / \varkappa^{1/2}\sim \mathcal{N}/\omega_{\rm A}$. 
In reality, however, one should expect that $n$ will grow up to values of the order $\sim \sqrt{\gamma_{{\rm fr}, 0} - \gamma_{{\rm eq},0} } / \varkappa^{1/2}
\sim \mathcal{N}/\omega_{\rm A}$, 
after which the contribution from the third-order terms \eqref{W5} will begin to stabilize the 
system. 
Therefore, to obtain a more accurate estimate for $n$, it is necessary to minimize the functional 
\eqref{wxi2} with respect to both $\mathfrak{g}(r,\theta)$ and $\mathfrak{a}(r,\theta)$, taking 
into account the third-order term \eqref{W5}%
%
\footnote{Note that among higher-order terms, there are none that could compete with the terms included in expression (\ref{wxi6}).}:
%
\begin{align}
	&
	W[\m\xi]\simeq  
	\frac{1}{2} \int r^2 \sin \theta 
	\left[
	\varepsilon^2 \, W_1 \, \mathfrak{g}(r,\theta)^2 + \varepsilon \varkappa \, W_2 \, 
	\mathfrak{g}(r,\theta) + \varkappa^2 \, W_3 
	+ \varkappa 
	\, 2 \pi 
	\mathcal{B}^2 [\partial_\theta \mathfrak{a}(r,\theta)-\varepsilon\partial_r 
	\mathfrak{g}(r,\theta)]^2
	\right] dr d\theta.
	&
	\label{wxi6}
\end{align}
The minimization problem for this functional can be simplified by treating the last term in 
\eqref{wxi6} as small 
and replacing the functions $\mathfrak{a}(r,\theta)$ and $\mathfrak{g}(r,\theta)$ with their 
lower-order perturbative approximations:
$\mathfrak{a}(r,\theta)=a(r)$
and $\mathfrak{g}(r,\theta)=-(\varkappa/\varepsilon)\cdot W_2/(2W_1)$; see Eqs.~\eqref{dtheta_a} 
and \eqref{gr}.%
%
\footnote{
One can check that substituting $\mathfrak{g}(r,\theta)=-(\varkappa/\varepsilon)\cdot 
W_2/(2W_1)$ into the functional \eqref{wxi6} yields an integrand that vanishes outside the 
magnetized regions, so that only magnetized regions contribute. In these regions, 
$\mathfrak{a}(r,\theta)\approx a(r)$ holds to leading order; see Eq.~\eqref{dtheta_a}, and the 
sentence following Eq.~\eqref{wxi3}.
}
%
Although this procedure is not entirely rigorous (the last term may become large for sufficiently 
large $n$), it significantly simplifies the problem of finding the optimal $n$.  
Substituting $\mathfrak{g}(r,\theta)$, $W_1$, $W_2$, and $W_3$ into \eqref{wxi6} using formulas 
\eqref{gr}, \eqref{W1}--\eqref{W3}, 
we obtain%
%
\footnote{
\label{footnoteCrit}	
It is important to emphasize that 
the procedure described here is not a mathematically rigorous minimization of the functional 
\eqref{wxi6} (since the last term in this functional is treated 
perturbatively
). 
However, if we prove that instability exists for this choice of $\mathfrak{a}(r,\theta)$ 
and $\mathfrak{g}(r,\theta)$, then by virtue of the spectral variational principle from Sec.~\ref{bernshtein}, it must also exist for the true $\mathfrak{a}(r,\theta)$ and
$\mathfrak{g}(r,\theta)$, and moreover, the growth rate of this instability must be greater than or 
equal to what can be found from the approximate analysis.
}
%
\begin{align}
	& 	W[\m\xi] \simeq 
	\frac{\varkappa^2}{2} \int_0^R
	\left\{-[a'(r)]^2\,|F_1| + [a(r)]^2 \, F_2 -a(r)a'(r) \, F_3 + \varkappa \, [a''(r)]^2 \, 
	|F_4| \right\}
	dr,
	&
	\label{wxi7}
\end{align}
where
\begin{align}
	& 	
	F_1(r) = \int_0^\pi \frac{4\pi \,r^2 \, \mathcal{B}^4 \, {\rm cos}^2\theta\, p_0 \, 
	\gamma_{{\rm fr}, 0} \, 
		\gamma_{{\rm eq},0} }
	{{\rm sin}\theta \, [p_0'(r)]^2 \, (\gamma_{{\rm fr}, 0}-\gamma_{{\rm eq},0} )} \, d\theta,
	&
	\label{F1}\\
	&
	F_2(r)= \int_0^\pi 
	\frac{ 4 \pi\,r \, \mathcal{B} \, \partial_\theta\Phi_1 \rho_0 \left(-{\rm sin}\theta \, 
	\partial_\theta 
		\mathcal{B} +r \,{\rm cos}\theta \, \partial_r \mathcal{B} \right)}{ p_0'(r)}
	d\theta,
	&
	\label{F2}\\
	&
	F_3(r)= -\int_0^\pi 
	\frac{ 4\pi\,r^2 \, \mathcal{B}^2\, {\rm cos}\theta \, \partial_\theta \Phi_1 \, 
	\rho_0}{ 
	p_0'(r)}
	d\theta,
	&
	\label{F3}\\
&
F_4(r) = \int_0^\pi \frac{8\pi\,r^2  \mathcal{B}^6 {\rm cos^2 \theta} p_0^2 \gamma_{{\rm fr}, 0}^2 
\gamma_{{\rm eq},0} ^2}{ 
{\rm sin \theta} (\gamma_{{\rm fr}, 0}-\gamma_{{\rm eq},0} )^2 [p_0'(r)]^4} d\theta,
\label{F4}
&
\end{align}
and $R$ is the radius of the unperturbed star in the absence of the magnetic field.
Note that the functions $F_1(r)$ and $F_4(r)$ never become negative.
When deriving the expression for $F_4(r)$ in Eq.\ \eqref{wxi6}, 
we retained only the largest term depending on $[a''(r)]^2$ 
and neglected all other terms. 
Due to the conditions on the magnetic field (see Sec.\ \ref{assumpt}), all functions 
$F_1(r), \ldots, F_4(r)$ are bounded and behave well inside the star.

If the function $a(r)$ oscillates strongly, 
then in the functional \eqref{wxi7} the dominant terms are the negative first term and the positive last term, proportional to $\varkappa^2 n^2$ and $\varkappa^3 n^4$, respectively. 
In this case, the second and third terms in \eqref{wxi7} can be neglected.
If, as a trial oscillating function $a(r)$, one chooses%
\begin{align}
&
a(r) = \cos \left(\frac{n r}{R} \right), 
&
\label{ar}
\end{align}
then one obtains from Eq.\ \eqref{wxi7}:
\begin{align}
&
W[\m\xi] \approx \frac{\varkappa^2}{4} 
\left[
- \left(\frac{ n}{R}\right)^2  \int_0^R |F_1| dr 
+ \varkappa \left(\frac{n}{R}\right)^4 \int_0^R |F_4| dr
\right],
&
\label{wxi8}
\end{align}
where we used that 
$\int_0^R \cos^2\left(\frac{n r}{R}\right) |F_i| \, dr \approx \int_0^R \sin^2\left(\frac{n 
r}{R}\right) |F_i| \, dr \approx \frac{1}{2} \int_0^R |F_i| \, dr$ 
(for $i=1,4$).  
Minimizing \eqref{wxi8} with respect to $n$, we find that the optimal value of $n$ within this 
approximate treatment is
\begin{align}
&
n = 
\frac{R}{\sqrt{2 \varkappa} } \sqrt{\frac{\int_0^R|F_1| dr}{\int_0^R|F_4| dr}}
\sim \frac{\sqrt{\gamma_{{\rm fr}, 0}-\gamma_{{\rm eq},0} }}{\varkappa^{1/2}} 
\sim \frac{\mathcal{N}}{\omega_{\rm A}}
&
\label{nopt}
\end{align}
in complete agreement with our expectations.%
%
\footnote{
\label{maxvalue}
Note that the optimal value of $n$ is smaller than the maximum value for which the instability can persist [$W[\m\xi]<0$, see Eq.~(\ref{wxi8})] 
by only a factor of $\sqrt{2}$.} 
%
This choice of $n$ corresponds to a negative value of $W[\m\xi]$:
\begin{align}
&
W[\m\xi] \approx - \frac{\varkappa}{16} \frac{\left[\int_0^R |F_1| dr \right]^2}{\int_0^R |F_4| dr} 
\sim -\frac{\pi \varkappa}{8} \mathcal{B}^2 R^3 = - \frac{B^2 R^3}{64},
&
\label{wxi9}
\end{align}
That is, we indeed have instability with such a choice of the trial function $a(r)$.  
Obviously, the exact minimization of the functional \eqref{wxi7} (with fixed normalization $I[\m\xi]$)  
may yield a more optimal (but not universal) function $a(r)$,  
which depends on the specific form of the toroidal magnetic field (see Sec.~\ref{example}).

Thus, we have found that in ideal MHD the most unstable mode corresponds to
$n \sim \mathcal{N}/\omega_{\rm A}$.
For such values of $n$, one has $\varepsilon n \sim 1$ [see Eq.~\eqref{nest}].
As a result, the last term in Eq.~\eqref{wxi6}, which scales as
$\mathcal{O}(\varkappa \varepsilon^2 n^2) \sim \mathcal{O}(\varkappa)$,
is no longer a small correction: it becomes of the same order as the first term,
which depends on $|W_1|$ and likewise scales as $\mathcal{O}(\varkappa)$
[since $\varepsilon^2 \sim \varkappa n \varepsilon \sim \varkappa$].
Moreover, both of these terms become of the same order as the leading term, $\varkappa \alpha_{01}$, in the expansion \eqref{wxi}.
Therefore, the perturbative treatment developed in the preceding sections is, strictly speaking,
no longer justified in this regime.
This means, in particular, that shellular rotation [Eq.~\eqref{lagany}]
may only be a rough approximation to the optimal displacement associated with the global 
instability considered here.

The situation changes, however, once dissipative effects are included in the MHD equations.
As discussed in Secs.~\ref{diss} and \ref{opt_diss}, viscosity and magnetic diffusivity can substantially slow down the development
of small-scale displacements with $n \sim \mathcal{N}/\omega_{\rm A}$, which would otherwise be optimal in ideal MHD.
As a consequence, larger-scale modes with smaller $n$ become more favorable.
Such modes can again be described within the perturbative framework, and for them shellular rotation provides a very good approximation.
As will be shown below, this is precisely what is observed in the numerical experiments
presented in Sec.~\ref{sim}.

\subsection{Summary of Sec.\ \ref{pert}}
\label{summary_sec4}

Let us summarize the main results of Sec.\ \ref{pert} (see also Table \ref{fig:params_table}, which lists the key parameters used in constructing the unstable Lagrangian displacement).
We found that a toroidal field is destabilized by a Lagrangian displacement that can,
within the range of validity of the perturbation theory and to leading order in each spherical 
component,
be written as
\begin{align}
&
\m\xi(r,\theta,\varphi)=\m\xi_{\rm rot}(r,\theta,\varphi)+\xi_r(r,\theta,\varphi) \m e_r,
&
\label{displ22}
\end{align}
where the vector $\m\xi_{\rm rot}$ gives the dominant $\theta$- and $\varphi$-components of the 
displacement and corresponds to a shellular differential rotation about some axis perpendicular to 
the symmetry axis $z$, while $\xi_r \m e_r$ gives its dominant radial component. 
The $\theta$- and $\varphi$-components also contain corrections of relative order 
$\mathcal{O}(\varepsilon n)$, which are not written 
explicitly here [see, in particular, Eq.\ \eqref{xiphi}], whereas the radial component itself is 
smaller than the tangential one by a factor of order~$\varepsilon$:
\begin{align}
&
\left|\frac{\xi_r}{\xi_{\rm rot}}\right|
\sim \left|\frac{\xi_r}{\xi_{\theta}}\right| \sim \mathcal{O}(\varepsilon),
&
\label{ratio22}
\end{align}
see Eqs.\ \eqref{xir3} and \eqref{xitheta3}.
Note that the expression \eqref{displ22} is valid only in the region of the star occupied by the 
magnetic field [see the discussion after Eq.\ \eqref{lagany}].
Without loss of generality, we take the rotation axis to be the $x$ axis.
Then $\m\xi_{\rm rot}(r,\theta,\varphi)$ is given by Eq.\ \eqref{lagx}:
\begin{align}
&
\m\xi_{\rm rot}(r,\theta,\varphi)=a(r) \, \hat{\m x}\times \m r.
&
\label{lagx22}
\end{align}

If the characteristic radial variation scale of the function $a(r)$, which quantifies the degree of differential rotation, is not too small (i.e., if $n$, $k$, and $\varepsilon$ are below their optimal values listed in Table \ref{fig:params_table}), the radial component $\xi_r$ is obtained by an exact minimization of the energy functional $W[\m\xi]$ and is given by Eqs.\ \eqref{xir3} and \eqref{gr}.
Such a displacement does destabilize the toroidal field, but the instability develops more slowly than in the optimal case.

To minimize the growth time, one must consider larger (optimal) 
values of $n$, $k$, and $\varepsilon$, as listed in the third (or fourth) column of Table \ref{fig:params_table}.%
%
\footnote{
For even larger values of these parameters, the instability is completely suppressed 
by the higher-order term 
$\varkappa \varepsilon^2 \alpha_{21}$ in $W[\m\xi]$, discussed in Sec.\ \ref{n_estimate}.}
In this limiting case, the perturbative expansion breaks down, so that the corrections to 
shellular differential rotation cease to be higher-order in $\varkappa$ and $\varepsilon$.
Furthermore, Eq.~\eqref{gr} for $\xi_r$ should be regarded as only approximate, while the function 
$a(r)$ can be determined by minimizing the approximate functional \eqref{wxi7}.
Note that, as announced at the end of Sec.~\ref{diffeq}, this limiting case need not be realized in 
practice once dissipative terms are included in the MHD equations (see Secs.~\ref{diss} and 
\ref{opt_diss}).

Finally, it is worth emphasizing, that since both $\xi_{\rm rot}$ and $\xi_r$ depend on the function $a(r)$, which may vary rapidly with $r$, the radial derivatives of $\xi_{\rm rot}$ and $\xi_r$ should be estimated, in order of magnitude, as
\begin{align}
&
\partial_r \xi_{\rm rot} \sim n \xi_{\rm rot}/R
\quad {\rm and } \quad
\partial_r \xi_r \sim n \xi_r/R,
&
\label{estest}
\end{align}
which (for large $n$) can differ substantially from the naive estimates $\xi_{\rm rot}/R$ and $\xi_r/R$.

\begin{table}[t]
\centering
\setlength{\tabcolsep}{16pt}      
\renewcommand{\arraystretch}{1.4} 
\setlength{\extrarowheight}{4pt}  
\begin{tabular}{|c|c|c|c|}
\hline
\textbf{Parameter} & \textbf{Definition} & \multicolumn{2}{c|}{\textbf{Optimal/maximum value}} \\
\hline
$\varkappa \ll 1$ &
$\dfrac{B^{2}}{8\pi p}$ &
$-$ &
$-$ \\
\hline
$n \gtrsim 1$ &
$\dfrac{\partial_r\mathfrak{a}(r,\theta)}{\mathfrak{a}(r,\theta)}\,R\approx 
\dfrac{{a'}(r)}{{a}(r)}\,R$ &
$\dfrac{\sqrt{\gamma_{{\rm fr},0}-\gamma_{{\rm eq},0}}}{\varkappa^{1/2}}$ &
$\dfrac{\mathcal{N}}{\omega_{\rm A}}$ \\
\hline
$k \gtrsim 1/R$ &
$\dfrac{n}{R}$ &
$\dfrac{\sqrt{\gamma_{{\rm fr},0}-\gamma_{{\rm eq},0}}}{\varkappa^{1/2}R}$ &
$\dfrac{\mathcal{N}}{\omega_{\rm A}R}$ \\
\hline
$\varepsilon \ll 1$ &
$\left|\dfrac{\xi_r}{\xi_\theta}\right| \sim \dfrac{n \varkappa }{\gamma_{{\rm fr}, 0}-\gamma_{{\rm eq},0} }\sim n\,\dfrac{\omega_{\rm A}^{2}}{\mathcal{N}^{2}}$ &
$\dfrac{\varkappa^{1/2}}{\sqrt{\gamma_{{\rm fr},0}-\gamma_{{\rm eq},0}}}$ &
$\dfrac{\omega_{\rm A}}{\mathcal{N}}$ \\
\hline
\end{tabular}
\caption{Key parameters controlling the unstable displacement found in Sec.\ \ref{pert}. The expansion parameter $\varkappa$ is introduced in Eq.\ \eqref{BP}. The quantities $n$ and the characteristic wave vector of the perturbation, $k$, are defined after Eq.\ \eqref{gg}. The small parameter $\varepsilon$ is defined in Eq.\ \eqref{xir3} [see also Eq.\ \eqref{eps}].
The second column gives the definitions of these quantities. The third and fourth columns provide order-of-magnitude estimates of their optimal values, for which the instability is particularly efficient (these values are close to the maximum values for which the instability still exists, see footnote \ref{maxvalue}). In the third column, these values are expressed in terms of $\varkappa$ and $\gamma_{{\rm fr},0}-\gamma_{{\rm eq},0}$; in the fourth column, the optimal values are expressed in terms of the typical Alfv\'en frequency $\omega_{\rm A}$ and the Brunt-V\"ais\"al\"a frequency $\mathcal{N}$.
}
\label{fig:params_table}
\end{table}

\section{Characteristic instability growth time}
\label{timescale}

Using the trial function \eqref{ar} for $a(r)$
allows us not only to calculate $W[\m\xi]$,
but also to obtain an estimate for the upper bound on the characteristic instability growth time
$\tau_{\rm min}$.
For this purpose, we will use the spectral variational principle from Sec.~\ref{bernshtein}. 
We need to minimize the ratio $W[\m\xi]/I[\m\xi]$. 
In this ratio, the functional $W[\m\xi]$ is known and given by formula \eqref{wxi8},
and for $I[\m\xi]$ it is necessary to use Eq.\ \eqref{III},
in which the Lagrangian displacement $\m\xi$ should be substituted.
Proceeding in the same spirit as in Sec.~\ref{diffeq}, we adopt as $\m\xi$
the shellular differential rotation $\m\xi_{\rm rot}$ [see Eq.~\eqref{lagx}],
which, as shown in Sec.~\ref{first}, provides a good approximation
[to within terms of order $\sim \mathcal{O}(\varepsilon+\varkappa)$]
to the Lagrangian displacement within the regime of validity of the perturbation theory.
As a result, we obtain:
\begin{align}
&
I[\m\xi] \simeq \frac{1}{2} \int \rho \, \m\xi_{\rm rot}\cdot \m\xi_{\rm rot} \, dV
\approx \frac{4 \pi}{3} \int_0^R r^4 \rho_0 \, a(r)^2 \, dr
= \frac{4 \pi}{3} \int_0^R r^4 \rho_0 \, \cos\left(\frac{n r}{R}\right)^2 \, dr
\approx \frac{2 \pi}{3} \int_0^R r^4 \rho_0 \, dr.
&
\label{III4}
\end{align}
By order of magnitude, this integral can be estimated as $I[\m\xi] \sim (2\pi/15) \rho_0 R^5$.
It is evident that $I[\m\xi]$ does not depend on $n$, therefore the minimization of $W[\m\xi]/I[\m\xi]$
reduces to the minimization of $W[\m\xi]$, which was already performed in Sec.~\ref{diffeq}.
As a result, using \eqref{III33}, \eqref{wxi9}, and \eqref{III4}, we obtain
\begin{align}
&
\tau_{\rm min} \leq \sqrt{\frac{I[\m\xi]}{|W[\m\xi]|}} =
\sqrt{\frac{32 \pi}{3 \varkappa}} 
\frac{\left( \int_0^R r^4 \rho_0 \, dr\right)^{1/2} 
\left( \int_0^R |F_4| \, dr\right)^{1/2}}{\int_0^R |F_1| \, dr}
\sim \tau_{\rm A},
&
	\label{omin}
\end{align}
where $\tau_{\rm A}=1/\omega_{\rm A} \sim (4 \pi \rho_0)^{1/2}R/B$ 
is the typical Alfv\'en timescale.
We see that the typical instability growth time for our trial function $a(r)=\cos(n r/R)$
is $\sim \tau_{\rm A}$. 
Of course, the actual time depends on the specific field model and may be smaller
for a more optimal function $a(r)$. 
In the next section, using the example of a simple magnetic field model,
we will show that the optimal
function $a(r)$ can, in fact, differ significantly from the trial function of the form \eqref{ar}.

\section{Numerical example}
\label{example}

We illustrate the general results of the previous sections with an example of a specific model
of a toroidal magnetic field of the form
\begin{align}
&
\m B(r,\theta)= \mathcal{L} \rho r \sin \theta\, \m e_{\varphi},
&
\label{magmodel}
\end{align}
and study the instability associated with it.
In formula \eqref{magmodel}, $\mathcal{L}$ sets the magnetic field strength.
We chose this field model specifically because
it allows a relatively simple calculation of the hydrostatic structure of a magnetized star
\cite{sinha68, st72}.
Note that the stability of this field has already been studied in the work \cite{gv78}
based on the approach developed by Tayler \cite{tayler73}.

\subsection{Hydrostatic structure}
\label{hydrostatic}

Before studying the instability,
it is necessary to construct a hydrostatic equilibrium model of a star with the field \eqref{magmodel}.
We will assume that the stellar matter obeys a polytropic equation of state, $P = K \rho^{1 + 1/\mathfrak{n}}$, with a polytropic index $\mathfrak{n}=3$ as an approximation for a radiative (stably stratified) star.
Accordingly, in this case, $\gamma_{{\rm eq},0}  = 1 + 1/\mathfrak{n} = 4/3$.

\begin{figure}
	\center{\includegraphics[width=0.4\linewidth]{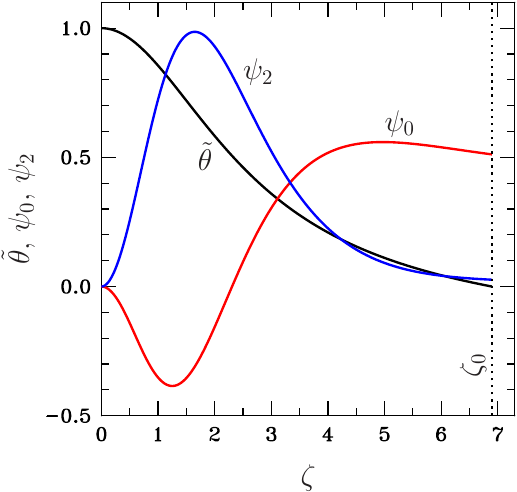}}
	\caption{Functions $\tilde{\theta}(\zeta)$, $\psi_0(\zeta)$, and $\psi_2(\zeta)$ versus dimensionless radial coordinate $\zeta$.}
	\label{Fig:thetapsi0psi2}
\end{figure}

We need to solve the equations of hydrostatic equilibrium \eqref{force}
and the Poisson equation \eqref{grav} simultaneously.
In the case of a weak field, this is done
by standard perturbation theory methods \cite{sinha68, st72};
the solution [see Eqs.\ \eqref{ThetaBig}--\eqref{Phigrav} below] is expressed in terms of the functions $\tilde{\theta}(\zeta)$, $\psi_0(\zeta)$, and $\psi_2(\zeta)$, 
which are shown in Fig.~\ref{Fig:thetapsi0psi2}.
The variable $\zeta$ is related to the radial coordinate by the equation
\begin{align}
&
r=\alpha \zeta, \quad\quad \alpha=\left(\frac{P_c (\mathfrak{n}+1)}{4 \pi G \, \rho_c^2} \right)^{1/2},
&
\label{alpha}
\end{align}
where $P_c$, $\rho_c$ are the central pressure and density of the star, respectively;
note that $K = P_c / \rho_c^{\gamma_{{\rm eq},0} }$.
The weakness of the field corresponds to the smallness of the dimensionless parameter $h^2 \equiv \mathcal{L}^2/(16 \pi^2 G) \ll 1$, 
which is equivalent to the condition $\varkappa \ll 1$.
Further in Sec.~\ref{example}, without loss of generality,
we will relate the parameters $\varkappa$ and $h^2$ through the equality
\begin{align}
&
\varkappa \equiv h^2.
&
\label{varkappa2}
\end{align}

Knowing the functions $\tilde{\theta}(\zeta)$, $\psi_0(\zeta)$, and $\psi_2(\zeta)$,
one can calculate the function $\Theta(\zeta,\mu)$,
\begin{align}
&
\Theta(\zeta, \mu) = \tilde{\theta}(\zeta)+ h^2 \left[ \psi_0(\zeta)+\psi_2(\zeta) P_2(\mu) \right],
&
\label{ThetaBig}
\end{align}
where $\mu=\cos \theta$ and $P_2(\mu)$ is the Legendre polynomial of order $2$.
Having $\Theta(\zeta,\mu)$,
one can, in turn, determine all other quantities,
in particular $\rho$ and $\Phi$:
\begin{align}
&
\rho= \rho_c \Theta(\zeta, \mu)^\mathfrak{n},
&
\label{rho}\\
&
\Phi = -D \left\{ 
\Theta(\zeta, \mu)+c_0 + h^2 \left[ c_1 + \frac{2}{3} \zeta^2 \, \tilde{\theta}(\zeta)^\mathfrak{n} \, 
(1- P_2(\mu))\right]
\right\},
&
\label{Phigrav}
\end{align}
where $D=(\mathfrak{n}+1)P_c/\rho_c$. For $\mathfrak{n}=3$ one has
$c_0 \approx 0.29263$, 
$c_1 \approx -0.31205$.
To obtain from Eqs.\ \eqref{ThetaBig}--\eqref{Phigrav} 
the corresponding expressions for a star without a magnetic field, 
it is sufficient to set $h = 0$.
The functions of the variables $\zeta$ and $\mu$ are easily converted into functions of $r$ and $\theta$
using the relations $\zeta = r/\alpha$ and $\mu = \cos\theta$.

The shape of the star, distorted by the magnetic field, follows from the condition $\Theta(\zeta, \mu) = 0$ and has the form 
$\zeta(\mu) = \zeta_0 + h^2 f(\mu)$, where $\zeta_0 \approx 6.897$ is the dimensionless radius of a spherically symmetric 
star without a magnetic field, and
\begin{align}
&
f(\mu)  = - \frac{[\psi_0(\zeta_0)+\psi_2(\zeta_0) P_2(\mu)]}
{
\frac{d 
\tilde{\theta}(\zeta)}{d\zeta}|_{\zeta=\zeta_0}
}.
&
\label{zeta}
\end{align}
Using this equation, one can write
\begin{align}
&
\frac{\zeta(1)-\zeta(0)}{\zeta(0)}=-\frac{3 h^2}{2 \zeta_0} 
 \frac{\psi_2(\zeta_0)}{\frac{d 
		\tilde{\theta}(\zeta)}{d\zeta}|_{\zeta=\zeta_0}}.
&
\label{ratio}
\end{align}
Keeping in mind that $\psi_2(\zeta_0) > 0$ and $d \tilde{\theta} / d\zeta < 0$,
we see that the polar radius is always longer than the equatorial one,
i.e., unlike the case of a rotating star, a magnetized star with toroidal field
is elongated along the poles.

\begin{figure}
	\center{\includegraphics[width=0.4\linewidth]{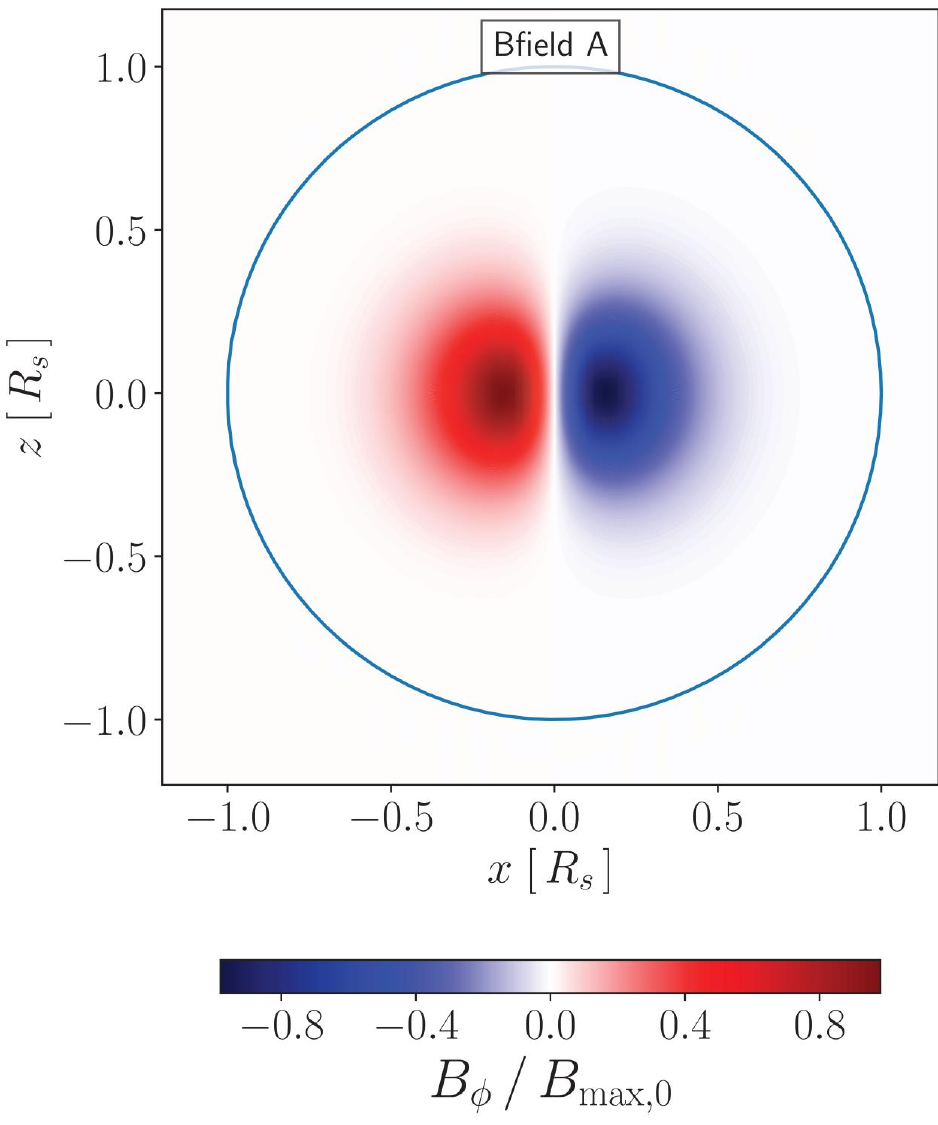}}
	\caption{Meridional cut of the star with the toroidal magnetic field used. The colors represent the toroidal-field magnitude (more precisely, its $y$-projection) normalized to its amplitude.
	}
	\label{Fig:MagmodelA}
\end{figure}

Below, as an illustration, we consider a model of a star
with $\rho_c = 25$~g~cm$^{-3}$ and $P_c = 5 \times 10^{16}$~dyn~cm$^{-2}$.
These parameters correspond to a star with mass $M \approx 2.38 M_\odot$ 
and radius $R \approx 1.93 R_\odot$.
The instability is studied for two magnetic field models with $h^2 = h_{500}^2 \approx 4.4035 
\times 10^{-3}$ (model I)
and $h^2 = h_{5000}^2 \approx 4.4035 \times 10^{-4}$ (model II) [the models differ only 
by normalization]. 
Model I corresponds to the ratio of gravitational to magnetic energy 
$|E_{\rm grav}/E_{\rm mag}| \approx 500$, while for model II this ratio is 5000.
Figure \ref{Fig:MagmodelA} shows a color plot of the magnetic field used, 
normalized to the maximum field strength in the star.

For further analysis, it is convenient to introduce the star-averaged Alfv\'en time according to the formula
\begin{align}
&
\tau_{\rm A} = \frac{R (4 \pi \overline{\rho})^{1/2}}{\overline{B}} \approx \frac{6745 \, {\rm 
s}}{h},
&
\label{tauA}
\end{align}
where $\overline{\rho}=M/V$ is the average stellar density ($V=4 \pi R^3/3$) and
$\overline{B}=(\int |B| dV)/V$ is the average absolute magnetic field value.

\subsection{Instability}
\label{instability1}

Now everything is prepared to determine the optimal function $a(r)$
that maximally destabilizes the star, and to estimate the characteristic instability growth time.
To do this, we need to minimize the ratio $\Lambda = W[\m\xi]/I[\m\xi]$
[see Sec.~\ref{bernshtein}], where $W[\m\xi]$ is given by Eq.\ \eqref{wxi7},
and $I[\m\xi]$ by Eq.\ \eqref{III4} (in which, however, the function $a(r)$
should be treated as unknown).
It is convenient to change the integration over the radial coordinate $r$ in the functionals $W[\m\xi]$ and $I[\m\xi]$ to integration over the dimensionless coordinate $\zeta = r/\alpha$. In this case, they can be rewritten as:
\begin{align}
&
W[\m\xi] \simeq
\frac{1}{2} \int_0^{\zeta_0}
\left\{-[a'(\zeta)]^2\,|\tilde{F}_1(\zeta)| + [a(\zeta)]^2 \, \tilde{F}_2(\zeta) 
-a(\zeta)a'(\zeta) \, \tilde{F}_3 + [a''(\zeta)]^2 \, 
|\tilde{F}_4| \right\}
d\zeta,
&
\label{W111}\\
&
I[\m\xi] \approx \int_0^{\zeta_0} \frac{4 \pi \alpha^5}{3}  \zeta^4 \rho_0 \, a(\zeta)^2 \, d\zeta,
&
\label{Ixi}
\end{align}
where we introduced new functions defined as $\tilde{F}_1 \equiv \varkappa^2 F_1 / \alpha$, 
$\tilde{F}_2 \equiv \varkappa^2 \alpha F_2$, 
$\tilde{F}_3 \equiv \varkappa^2 F_3$,
and $\tilde{F}_4 \equiv \varkappa^3 F_4 / \alpha^3$.
To compute them numerically using formulas \eqref{F1}--\eqref{F4},
one needs to take into account that $\varkappa = h^2$,
$8\pi \varkappa \mathcal{B}^2 = B^2$, and that $\Phi_1$, according to the definition 
\eqref{Adecomp} and Eqs.\ \eqref{ThetaBig} and \eqref{Phigrav},
is equal to (without the $h^2$ factor!):
\begin{align}
&
\Phi_1 = - D \left[\psi_0(\zeta)+\psi_2(\zeta) P_2(\mu)+ c_1 + \frac{2}{3} \zeta^2 \, 
\tilde{\theta}(\zeta)^\mathfrak{n} \, 
(1- P_2(\mu))\right].
&
\label{Phi11}
\end{align}
In addition, it is necessary to specify the ``frozen'' adiabatic index $\gamma_{{\rm fr}}$.
In all numerical calculations below, we set $\gamma_{\rm fr}= 5/3$.

\begin{figure}
	\center{\includegraphics[width=0.4\linewidth]{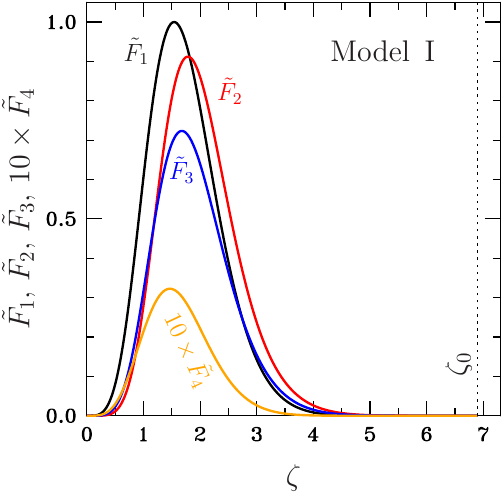}}
	\caption{Functions $\tilde{F}_i(\zeta)$ ($i=1,\ldots,4$) for model I of the magnetic field. 
		}
	\label{Fig:F1F2F3F4}
\end{figure}

Figure \ref{Fig:F1F2F3F4} shows the functions 
$\tilde{F}_i(\zeta)$ ($i = 1,\ldots,4$) for magnetic field model I. 
All functions are normalized so that the maximum value of $\tilde{F}_1(\zeta)$ is equal to 1.
As can be seen, the function $\tilde{F}_4$ is much smaller than the other functions. 
This is not surprising, since $\tilde{F}_4 \propto B^6 / p^3 \sim h^6$, whereas the other functions 
are proportional to $B^4 / p^2 \sim h^4$.
All functions vanish near $\zeta = 0$ and $\zeta = \zeta_0$, and
one can show that both $F_1$ and $F_4$ go as $\zeta^4$ when $\zeta \rightarrow 0$.
This behavior of the functions $\tilde{F}_i(\zeta)$ suggests the simplest approach 
to perform the numerical minimization of $\Lambda = W[\m\xi]/I[\m\xi]$.
Indeed, one could in principle vary this ratio with respect to $a(\zeta)$ and obtain a fourth-order
differential equation for $a(\zeta)$.%
%
\footnote{
\label{ortog}
Note that the differential operator corresponding to this equation is self-adjoint. 
Therefore, any two eigenfunctions $a_i(\zeta)$ and $a_j(\zeta)$, which are solutions 
to this equation and correspond to distinct eigenvalues $\Lambda_i$ and $\Lambda_j$,
are orthogonal with respect to the weight $\zeta^4 \, \rho_0(\zeta)$:
\begin{align}
\int_0^{\zeta_0}  \, \zeta^4\, \rho_0(\zeta) \, a_i(\zeta) \, a_j(\zeta)\,  d\zeta=0.
\nonumber
\end{align}
}
%
However, such an equation would be highly singular, since its coefficients vanish both at the center and at the boundary of the star. Solving such an equation numerically is difficult.
An alternative approach is to expand $a(\zeta)$ in some complete set of eigenfunctions
and minimize $W[\m\xi]/I[\m\xi]$ by adjusting the coefficients in such an expansion.
This is exactly what we did, expanding $a(\zeta)$ in Chebyshev polynomials $T_l(x(\zeta))$,
where $x(\zeta) = -1 + 2 \zeta / \zeta_0$, and retaining only the first 20 polynomials
(the solution is practically insensitive to further increases in the number of polynomials).

\begin{figure}
	\center{\includegraphics[width=0.4\linewidth]{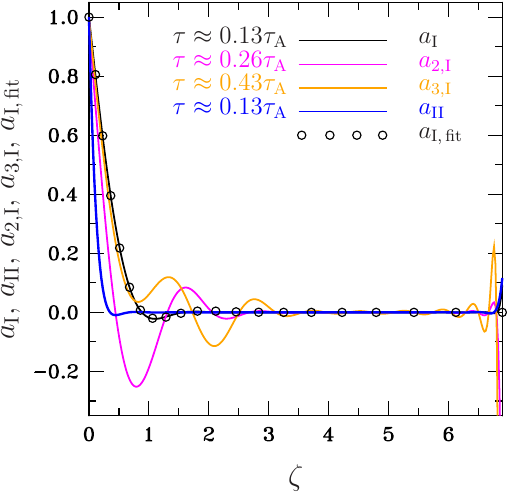}}
	\caption{Functions $a_{\rm I}$, $a_{2, {\rm I}}$, $a_{3, {\rm I}}$, $a_{\rm II}$, and $a_{\rm I,\, fit}$ versus $\zeta$. 
	}
	\label{Fig:a}
\end{figure}

The results of this standard minimization are presented in Fig.\ \ref{Fig:a}.
The figure shows the optimal functions $a(\zeta)$ (corresponding to the largest in absolute value and negative eigenvalues $\Lambda$) for models I ($E_{\rm grav}/E_{\rm mag} = 500$) and 
II ($E_{\rm grav}/E_{\rm mag} = 5000$).
These functions are denoted, respectively, by $a_{\rm I}(\zeta)$ and $a_{\rm II}(\zeta)$.
In addition, for model~I we also show the second and third most unstable eigenfunctions 
$a_{2, {\rm I}}(\zeta)$ and $a_{3, {\rm I}}(\zeta)$, which are orthogonal to the function $a_{\rm I}(\zeta)$ (see footnote \ref{ortog}).
For all the functions shown in the figure, the characteristic instability growth time 
$\tau = |\Lambda|^{-1/2}$ is also indicated.

Let us note the apparent (and unphysical) growth of these functions near the stellar surface.
This occurs because in the surface layers the density, pressure, and magnetic field all tend to zero, so the contributions of these regions to
$W[\m\xi]$ and $I[\m\xi]$ are negligible for essentially any choice of $a(\zeta)$, therefore, the minimization of $W[\m\xi]/I[\m\xi]$
 does not constrain $a(\zeta)$ in those layers. 
As a result, our growth-time estimates are insensitive to the
near-surface behavior.

The circles in Fig.\ \ref{Fig:a} show the result of fitting the most unstable mode $a_{\rm I}(\zeta)$ for model I:
\begin{align}
&
a_{\rm I, \, fit}(\zeta)= 
-22.73575\,e^{-1.034352 \zeta^2}
+
23.73575\,e^{-0.075038 \zeta - 1.004375 \zeta^2}.
&
\label{afit}
\end{align}
A similar fitting formula also exists for the function $a_{\rm II}(\zeta)$.
For both of these functions, the characteristic instability growth time,
measured in units of $\tau_{\rm A}$, is approximately
\begin{align}
&
\tau = |\Lambda|^{-1/2} \approx 0.13 \tau_{\rm A}.
&
\label{tauxxx}
\end{align}
This fact can be explained by noting that the result of the minimization is only weakly dependent 
on $\tilde{F}_2$ and $\tilde{F}_3$. Assuming $\tilde{F}_2 = \tilde{F}_3 = 0$ in Eq.\ \eqref{W111} 
and taking into account the asymptotic behavior $\tilde{F}_1 \propto \zeta^4$ and 
$\tilde{F}_4 \propto \zeta^4$ near $0$, one can show that $\Lambda$ scales with $h$ as 
$\Lambda \sim h^2$, i.e., the time $\tau$ measured in units of $\tau_{\rm A}$
should remain unchanged, which is consistent with the numerical results.
Moreover, with $\tilde{F}_2 = \tilde{F}_3 = 0$, the optimal function $a(\zeta)$ becomes 
self-similar for different values of $h$, that is, $a(\zeta/h)$ is independent of $h$ at small $h$.
This explains why the optimal solution localizes toward the center as 
$h \rightarrow 0$ [$a_{\rm II}(\zeta)$ is more localized near the center than 
$a_{\rm I}(\zeta)$].

The localization of the unstable mode as $h \rightarrow 0$ is a direct consequence of neglecting 
dissipative effects in our treatment based on ideal magnetohydrodynamics. In reality, 
accounting for dissipative effects
(e.g., introducing finite viscosity)
can significantly, by a factor of approximately $\tau / \tau_{\rm diss}$ (see, e.g., \cite{sb24}), slow down the growth of such a mode due to the large gradients involved (estimating the factor we have assumed that the dissipative timescale $\tau_{\rm diss}$ is much shorter than $\tau$). As a result, a different, less localized 
instability may have a better chance to develop: one that generates smaller gradients and is 
therefore less affected by dissipative processes (see also Sec.\ \ref{sim}).

Such instabilities, of course, do exist. For instance, if we take the less localized solution 
$a_{\rm I}(\zeta)$ obtained for model I and use it as the trial function $a(\zeta)$ in model II, 
the resulting displacement will still be unstable. Moreover, the corresponding characteristic 
growth time will be $\tau \approx 0.3\,\tau_{\rm A}$, that is, it will increase by less than a 
factor of 3 [cf.\ Eq.\ \eqref{tauxxx}].

\begin{figure}
	\center{\includegraphics[width=0.4\linewidth]{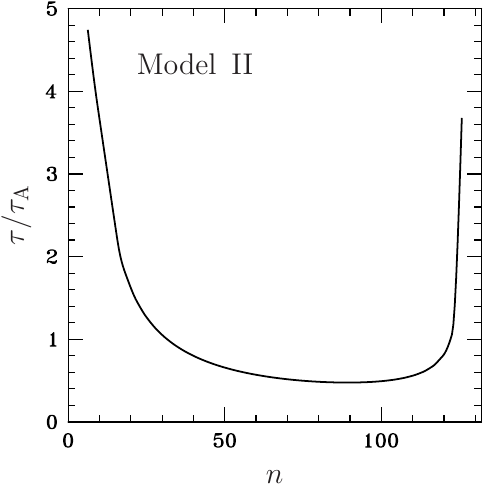}}
	\caption{Instability timescale for trial function $a(\zeta)= \cos(n \zeta/\zeta_0)$ versus 
	number $n$ for model II of the magnetic field. 
	}
	\label{Fig:tauvsn}
\end{figure}

Another interesting example is provided by the ``universally unstable'' trial function 
$a(\zeta) = \cos(n \zeta/\zeta_0)$ from Sec.\ \ref{diffeq} [see Eq.\ \eqref{ar}]. 
The instability timescale $\tau$ for this function is plotted in Fig.\ \ref{Fig:tauvsn} 
as a function of $n$ for model II. In agreement with Eq.\ \eqref{nopt}, 
the minimum of $\tau(n)$ is reached at 
$n \approx 89.2$
and equals $\tau \approx 0.477\,\tau_{\rm A}$. 

\subsection{A few subtle points}
\label{subtle}

In this section, we discuss several subtle points related to the results obtained above.

\begin{itemize}
	
\item Short instability timescale.
	
As follows from Eq.\ \eqref{tauxxx}, the characteristic instability growth time we obtained 
does not exceed $0.13\,\tau_{\rm A}$, i.e., it is much shorter than the average Alfv\'en time 
$\tau_{\rm A}$ [see Eq.\ \eqref{tauA}]. This fact can be easily explained by noting that 
the instability occurs in a small region near the center of the star. 
Accordingly, the magnetic field there is stronger than the average, and the characteristic 
lengthscale is much smaller than $R$. Therefore, the \textit{actual} Alfv\'en time 
characterizing this region must be much shorter than the mean stellar Alfv\'en time 
$\tau_{\rm A}$ defined by Eq.\ \eqref{tauA}.
For example, estimating the Alfv\'en time for the inner region $r \lesssim R/3$, where the 
magnetic field is mainly localized, using Eq.~(97), we obtain
$\tau_{\rm A}(R/3) \approx 0.115\,\tau_{\rm A}$,
which is nearly one order of magnitude smaller than the mean stellar Alfv\'en time.

This result is consistent with the findings of Ref.~\cite{gv78},  
which studied the local Tayler instability for the field model \eqref{magmodel}  
and found the minimum instability time [see their equation (66)]
\begin{equation}
\tau_{\rm GV} \simeq \left(8.5\times 10^{-5} \frac{E_{\rm mag}}{|E_{\rm grav}|} \frac{M \,R_\odot^3}{R^3\, M_\odot}\right)^{-1/2}.
\end{equation}
Substituting numbers, we find $\tau_{\rm GV} \approx 0.05 \tau_{\rm A}$,  
an even shorter time for the development of local instability.

\item 
Verification of the self-consistency of our approach and  
the ``optimality'' of the Lagrangian displacement we have found.

In deriving the functional \eqref{wxi7} from the original functional \eqref{en11},  
a kind of perturbation theory in the parameters $\varkappa$ and $\varepsilon$ was developed.  
This allowed us to significantly simplify the problem by discarding all small terms in Eq.~\eqref{en11}.  
It is of considerable interest to verify the self-consistency of our calculation  
by substituting the obtained Lagrangian displacement  
into the original {\it exact} expressions \eqref{en11}--\eqref{Wmag} and \eqref{III}  
for $W[\m\xi]$ and $I[\m\xi]$, and calculating  
$\Lambda = W[\m\xi]/I[\m\xi]$ and $\tau = |\Lambda|^{-1/2}$  
using these exact formulas.

Let us perform such a verification using the example of magnetic field model I.  
We have already defined the function $a(r)$ [see the fitting formula \eqref{afit},  
where $\zeta = r/\alpha$].  
Now we set $\mathfrak{a}(r,\theta) = a(r)$ and use Eq.~\eqref{gr} to find $\varepsilon 
\mathfrak{g}(r,\theta)$.  
Knowing $\mathfrak{a}(r,\theta)$ and $\varepsilon \mathfrak{g}(r,\theta)$,  
we determine $\tilde{\xi}_r^{(m)}$, $\tilde{\xi}_\theta^{(m)}$, and  
$\tilde{\xi}_\varphi^{(m)}$ (with $m = 1$) from Eqs.~\eqref{xir3}--\eqref{xiphi}.  
Thus, we obtain the full Lagrangian displacement $\m\xi(r,\theta,\varphi)$ given by Eq.~\eqref{xiexpr2}.  
Substituting it into the integrals \eqref{Whydro}, \eqref{Wmag}, and \eqref{III},  
we calculate the characteristic instability growth time  
corresponding to our displacement: $\tau \approx 0.12 \tau_{\rm A}$.  
The resulting time turns out to be close to, and even slightly less than,  
the value $\tau \approx 0.13 \tau_{\rm A}$ from the approximate calculation [see Eq.~\eqref{tauxxx}].  
This suggests that our approximations capture the essential physics and that our approach is 
self-consistent.
 
\item Angular momentum and its conservation.

Looking at Fig.~\ref{Fig:a}, one may get the impression that the star predominantly rotates  
in a certain direction, since $a(\zeta)$ in the figure is mostly positive.  
This raises the question: how is the conservation of angular momentum maintained in our star?  
Indeed, the correct Lagrangian displacement,  
which minimizes the ratio $\Lambda = W[\m\xi]/I[\m\xi]$ (see Sec.~\ref{bernshtein})  
and satisfies the equations of magnetohydrodynamics,  
should not change the angular momentum of the system,  
which is zero in our case.

Let us check how well angular momentum is conserved for our solution using the example of model I.  
Neglecting the small contribution from the electromagnetic field,  
the angular momentum of the system is proportional to the integral
\begin{align}
&
\m L \propto \int \rho \, \m r\times \m\xi \, dV
\approx
\int \rho_0(r) \, \m r \times \m\xi_{\rm rot}(r,\theta,\varphi) \, dV
= \int \rho_0(r) \, \m r \times [a(r) 
\hat{\m x}
\times \m r] \, dV
&
\nonumber\\
&
= \frac{8 \pi}{3} \, 
\hat{\m x}
\, \int_0^R  r^4 \, \rho_0(r) \, a(r) \, dr
=\frac{8 \pi \alpha^5 \rho_c}{3} \, 
\hat{\m x}\,
\int_0^{\zeta_0}  \zeta^4 \, 
\frac{\rho_0(\zeta)}{\rho_c} \, a(\zeta) \, 
d\zeta,
&
\label{angularmomentum}
\end{align}
where we used the expression \eqref{lagx} and changed the integration variable from $r$  
to $\zeta = r/\alpha$.  
If we substitute the fitting formula \eqref{afit} for $a(r)$  
into Eq.~\eqref{angularmomentum}, we obtain a nonzero but small value for the integral  
on the right-hand side of \eqref{angularmomentum}:  
$\int_0^{\zeta_0} \zeta^4 \, [\rho_0(\zeta)/\rho_c] \, a(\zeta) \, d\zeta \approx 0.0076$.  
To make this integral vanish exactly,  
a small solid-body rotation must be added to the system,  
i.e., $a(\zeta)$ should be replaced: $a(\zeta) \rightarrow a(\zeta) + \delta a$, where $\delta a$ is some constant.  
It turns out that in our case this constant is very small, $\delta a \approx -0.0007$  
[to be compared to the typical value of $a(\zeta) \sim 1$],  
and it has almost no effect on the potential energy of the system.%
%
\footnote{
In general, adding solid-body rotation to the system  
should not change its potential energy.  
The fact that replacing $a(\zeta) \rightarrow a(\zeta) + \delta a$  
does lead to a change in the potential energy,  
as approximately determined by the functional \eqref{wxi7} [or \eqref{W111}],  
is a direct consequence of the approximations we have made.  
In particular, following Ref.~\cite{mt69}, we have neglected from the very beginning  
the last term in expression (\ref{Whydro}),  
which depends on the gravitational field perturbation, $\delta \Phi$ (see Sec.~\ref{assumpt}).  
It can be shown that if this term is retained,  
then any solid-body rotation of the system indeed does not change its potential energy,  
since it is equivalent to a simple rotation of the coordinate system.
	}
%
Moreover, such a solid-body rotation contribution practically does not change the characteristic growth time of the instability.
This result shows that the obtained Lagrangian displacement satisfies  
the law of angular momentum conservation to a good accuracy.

Let us recall that, as already discussed, the procedure for obtaining the function $a(r)$  
in Sec.~\ref{instability1} is not absolutely rigorous.  
The fact that the angular momentum associated with the resulting Lagrangian displacement
remains practically zero (and can be made to vanish exactly without noticeably affecting the instability timescale) provides additional support for the claim that
the displacement we found is close to optimal.

\end{itemize}

\section{Stabilizing effect of the poloidal magnetic field}
\label{poloidal1}

Let us estimate the strength of the poloidal magnetic field needed to suppress 
the instability of the toroidal field.
To do that, we allow for the presence of the weak poloidal field $\m B_{\rm pol}$ 
of the form
\begin{align}
	&
	\m B_{\rm pol} = \m\nabla \Psi(r,\theta)\times \m\nabla\varphi.
	&
	\label{poloidal}
\end{align}
The idea is to determine the strength of the poloidal field  
at which the unstable Lagrangian displacement found in the previous sections  
becomes stabilized (i.e., ceases to be energetically favorable).  
As can be easily seen, the contribution of the poloidal field to $W[\m\xi]$  
appears already at first order in the perturbation theory in $\varkappa$ and $\varepsilon$  
(see Sec.~\ref{first}), when the Lagrangian displacement is taken to be pure shellular differential rotation  
$\m\xi_{\rm rot}(r,\theta,\varphi) = a(r) \hat{\m x} \times \m r$ [see Eq.~\eqref{lagx}].  
In this case, the main contribution to $W[\m\xi]$ comes from the magnetic part of the functional  
$W^{\rm mag}$, given by expression \eqref{Wmag}.  
Substituting $B(r,\theta)\, \m e_\varphi + \m B_{\rm pol}(r,\theta)$ and  
$\m\xi(r,\theta,\varphi) = \m\xi_{\rm rot}(r,\theta,\varphi)$  
into \eqref{Wmag}, we obtain a series of terms proportional to the functions  
$[a'(r)]^2$, $a'(r) a(r)$, and $a(r)^2$,  
and containing {\it no} mixed terms depending simultaneously  
on both the poloidal and toroidal fields.  
Taking into account that $R a'(r) \gg a(r)$,  
the dominant contribution from the poloidal field will come from the term $\propto [a'(r)]^2$:
\begin{align}
&
\frac{1}{2} \, \int_0^R |F_5| [a'(r)]^2 \, dr, \quad {\rm where}\quad
	F_5(r)= \int_0^{\pi} \frac{1}{4} \left( \frac{{\rm cos}^2 \theta +1}{{\rm sin} 
	\theta} 
	\right)\,(\partial_\theta \Psi)^2 \, d\theta.
	&
	\label{Wpoloidal}
\end{align}
This non-negative term will always act ``against'' the instability.  
Thus, taking into account the poloidal field, the problem of determining the optimal function $a(r)$  
reduces to minimizing the ratio $\widetilde{W}[\m\xi]/I[\m\xi]$,  
where $\widetilde{W}[\m\xi]$ is the potential energy functional  
including the contribution \eqref{Wpoloidal} from the poloidal field [cf.\ Eq.~\eqref{wxi7}]:
\begin{align}
	& 	\widetilde{W}[\m\xi] \simeq 
	\frac{1}{2} \int_0^R
	\left\{[a'(r)]^2\,(|F_5|-\varkappa^2\, |F_1|) + \varkappa^2 \, [a(r)]^2 \, F_2 -\varkappa^2\,  
	a(r)a'(r) \, F_3 + \varkappa^3 \, [a''(r)]^2 \, 
	|F_4| \right\}
	dr,
	&
	\label{wxiPol}
\end{align}
Even without performing an exact minimization of the ratio $\widetilde{W}[\m\xi]/I[\m\xi]$,  
it is clear that the poloidal field will completely suppress the instability  
once the poloidal contribution  
$(1/2) \int_0^R [a'(r)]^2 |F_5| \, dr$  
exceeds the main destabilizing term  
$-(1/2) \int_0^R \varkappa^2 \, [a'(r)]^2 |F_1| \, dr$ in the functional \eqref{wxiPol},  
that is (substituting $F_1$ and $F_5$)
\begin{align}
&
\int \frac{1}{4} \left( \frac{{\rm cos}^2 \theta +1}{{\rm sin} \theta} 
\right) \, (\partial_\theta \Psi)^2 \, [a'(r)]^2 \, dr d\theta
\gtrsim 
 \int \frac{r^2 \, B^4 \, \cos^2 \theta p_0 \gamma_{{\rm fr}, 0} \gamma_{{\rm eq},0} }{16 \pi \, 
 \sin \theta 
[p_0'(r)]^2 (\gamma_{{\rm fr}, 0}-\gamma_{{\rm eq},0} )} \, [a'(r)]^2 \, dr d\theta.
&
\label{ineq1}
\end{align}
If the regions of localization of the toroidal and poloidal magnetic fields coincide,  
then the integrals can be roughly estimated by order of magnitude,  
yielding the following lower bound for the stabilizing poloidal magnetic field $B_{\rm pol}$:
\begin{align}
	&
	B_{\rm pol}^2 \gtrsim 
    \frac{B^2}{4 \pi}
    \frac{B^2}{p_0}  \frac{\gamma_{{\rm fr}, 0} 
	\gamma_{{\rm eq},0} }{\gamma_{{\rm fr}, 0}-\gamma_{{\rm eq},0} } 
    \sim 2  B^2 \frac{\omega_{\rm A}^2}{\mathcal{N}^2}.
	&
	\label{Bpol_est}
\end{align}
This estimate can be rewritten by introducing the gravitational energy of the star, $E_{\rm grav}$,  
the energy of the poloidal field, $E_{\rm pol} = \int B_{\rm pol}^2/(8\pi) \, dV$,  
and the energy of the toroidal field, $E_{\rm tor} = \int B^2/(8\pi) \, dV$.
Estimating $B^2/(8 \pi p_0) \sim E_{\rm tor}/E_{\rm grav}$, one can rewrite \eqref{Bpol_est} as
\begin{align}
	&
	\frac{E_{\rm pol}}{E_{\rm tor}} \gtrsim 
	\beta \,  \frac{E_{\rm tor}}{E_{\rm grav}} 
	\, 
	\frac{\gamma_{{\rm fr}, 0} \gamma_{{\rm eq},0} }{\gamma_{{\rm fr}, 0}-\gamma_{{\rm eq},0} },
	&
	\label{Bpol_est2}
\end{align}
where $\beta$ is some model-dependent coefficient.  
Let us estimate $\beta$ for model I of the star with a toroidal magnetic field  
given by Eq.~\eqref{magmodel} and the optimal function $a(r)$ specified by Eq.~\eqref{afit}  
(recall that for model I, $E_{\rm grav}/E_{\rm tor} = 500$).  
To this end, we choose the poloidal magnetic field $\m B_{\rm pol}$ from Ref.~\cite{GKO17},  
for which $\Psi(r,\theta) = B_{\rm max} \, R^2 \, f(r/R) \, \sin^2 \theta$, where  
$f(x) = x^2/2 - 3x^4/5 + 3x^6/14$, and $B_{\rm max}$ is the normalization constant.%
%
\footnote{We emphasize that this particular choice of the poloidal magnetic field is made solely  
for the purpose of obtaining a more quantitative estimate of $\beta$,  
and it is not in equilibrium with the toroidal field of the form given by Eq.~\eqref{magmodel}.
}
%
Now substituting these data into inequality \eqref{ineq1},  
we obtain a lower bound estimate for the constant $B_{\rm max}$.  
This estimate can then be rewritten in the form of inequality \eqref{Bpol_est2},  
in which $\beta$ turns out to be approximately $\beta \approx 14.6$.  
According to this estimate, inequality \eqref{ineq1} for the chosen parameters  
begins to hold when $E_{\rm pol} \gtrsim 0.2 E_{\rm tor}$.  
It is worth emphasizing, however, that for a more realistic choice of the ratio $E_{\rm grav}/E_{\rm tor}$,  
the energy of the poloidal field required to suppress the instability  
will be significantly smaller.
For example, choosing $E_{\rm grav}/E_{\rm tor}\sim 10^6$ \cite{reisenegger09} reduces the ratio 
to $E_{\rm pol}/E_{\rm tor}\sim 10^{-4}$.

One sees that a very small poloidal field is sufficient to suppress the instability.
It remains to note that the scaling (\ref{Bpol_est2}) is well-known and was suggested from 
heuristic arguments in \cite{braithwaite09} and later confirmed in \cite{armm13, brvg22b}.

\section{Tayler instability versus the global instability identified in this work}
\label{Tayler}

\subsection{Ideal MHD}
\label{nodiss}

Here we compare Tayler's approach with ours.  
In Ref.\ \cite{tayler73}, Tayler analyzed the stability of a toroidal axisymmetric magnetic field using the energy functional method.  
In contrast to our approach, Tayler did not expand the energy functional in series in $\varepsilon$ and $\varkappa$.  
Instead, he reorganized the terms to explicitly identify a non-negative component.  
Although the original paper used cylindrical coordinates, spherical coordinates are more suitable  
for global problems in stars. Thus, Tayler's results were rederived in spherical coordinates \cite{gv78} by presenting the functional for azimuthal number $m\neq 0$ as
\begin{eqnarray}
	W[\m\xi] = \int \Biggl[
	\frac{B^{2}}{4\pi\,r^{2}}
	\Bigl(
	r\,\frac{\partial \xi_{r}}{\partial r}
	\;+\;
	\frac{\partial \xi_{\theta}}{\partial \theta}
	\;-\;
	\xi_{\theta}\,\cot\theta
	\Bigr)^{2} \nonumber \\
	\;+\;
	\left({a_{m}}\,\xi_{r}^{2}
	\;+\;
	{b_{m}}\,\xi_{r}\,\xi_{\theta}
	\;+\;
	{c_{m}}\,\xi_{\theta}^{2}
	\right)
	\Biggr] \, dV, \label{TaylerW}
\end{eqnarray}
where
\begin{align}
&
a_m = - \partial_r{\rho} \partial_r{\Phi}
- \frac{\rho^2}{\gamma_{\rm fr}\, p} \left( \partial_r{\Phi} \right)^2 
+ \frac{m^2\, B^2}{4\pi r^2 \sin^2\theta}
- \frac{B^2}{2\pi r^2}
- \frac{1}{2\pi r} \mathbf{B} \cdot \partial_r{\mathbf{B}},    \label{a}
&
\\
&
b_m = - \frac{1}{r} \partial_r{\Phi} \partial_\theta{\rho}
- \frac{1}{r} \partial_\theta{\Phi} \partial_r{\rho} 
- \frac{2 \rho^2}{\gamma_{\rm fr}\, p\, r}
\partial_r{\Phi} \partial_\theta{\Phi} 
- \frac{B^2}{\pi r^2} \cot\theta
- \frac{1}{2\pi r} \mathbf{B} \cdot \partial_r{\mathbf{B}} \cot\theta
- \frac{1}{2\pi r^2} \mathbf{B} \cdot \partial_\theta{\mathbf{B}},    \label{b}
&
\\
&
c_m = - \frac{1}{r^2} {\partial_\theta\rho} \partial_\theta{\Phi}
- \frac{\rho^2}{\gamma_{\rm fr}\, p\, r^2}
\left( \partial_\theta{\Phi} \right)^2 
+ \frac{m^2\, B^2}{4\pi r^2 \sin^2\theta}
- \frac{B^2}{2\pi r^2} \cot^2\theta
- \frac{1}{2\pi r^2} \mathbf{B} \cdot \partial_\theta{\mathbf{B}} \cot\theta.  \label{c}
&
\end{align}
The first term in Eq.~(\ref{TaylerW}) is non-negative,  
while the second term is a quadratic form of the functions  
$\xi_r$ and $\xi_{\theta}$, with coefficients $a_m$, $b_m$, and $c_m$ being certain functions of  
the thermodynamic quantities and the magnetic field [see Eqs.~\eqref{a}--\eqref{c}].  
In what follows, we discuss 
the Tayler instability using the energy functional $W[\m\xi]$ given by Eq.\ \eqref{TaylerW}.
Tayler \cite{tayler73} proved the following (see \cite{gv78} for details).
If at least one of the conditions,
\begin{align}
&
a_m < 0 \quad {\rm or} \quad  c_m < 0 \quad {\rm or} \quad b_m^2 > 4 a_m c_m 
&
\label{condx}
\end{align}
is fulfilled at  
some point of the star, then the field is unstable.%
%
\footnote{Subsequent work \cite{gbt81,armm13} has shown that such a point exists for any model of the stellar toroidal magnetic field.}
%
He demonstrated this by explicitly constructing a displacement that makes the functional negative. The displacement satisfies two properties. First, it perturbs the system only in the vicinity of a point where \eqref{condx} is fulfilled. Second, it eliminates the stabilizing non-negative term [the first term in Eq.~(\ref{TaylerW})] in the vicinity of that point.
To make the problem tractable, he adopted the WKB approximation  
with respect to the $r$ and $\theta$ coordinates,  
and substituted $\m \xi \propto \exp^{\imath(l \theta + k r + \phi)}$ into $W[\m \xi]$, 
assuming that $l \gg 1$ and $k \gg 1/R$.  
Interestingly, the condition $l\gg 1$ is not necessary for the onset of instability in many toroidal field configurations (see, e.g., \cite{gv78} and Sec.~\ref{sim} below for examples). Tayler (and we here) adopted it to make the argument universal, i.e., applicable to any model of the toroidal magnetic field.
In the azimuthal direction the perturbation was global with azimuthal number $m=1$ since, as Tayler showed, $m=1$ is the most unstable mode. Indeed, $m$ enters the energy functional $W[\m \xi]$ only as a factor $m^2$ in the non-negative terms of Eqs.~\eqref{a} and \eqref{c}.
In light of the above considerations,
the displacement should  
satisfy the following condition to neutralize the non-negative term:
\begin{align}
&
r k \xi_r+l \xi_\theta \approx 0, 
&
\label{WKB}
\end{align}
while the energy functional reduces to
\begin{eqnarray}
	W[\m\xi] = \int 
	\left({a_{m}}\,\xi_{r}^{2}
	\;+\;
	{b_{m}}\,\xi_{r}\,\xi_{\theta}
	\;+\;
	{c_{m}}\,\xi_{\theta}^{2}
	\right)
	 \, dV\approx\int 
	\left({a_{m}}\frac{l^2}{r^2k^2}
	\;-\;
	{b_{m}}\frac{l}{rk}
	\;+\;
	{c_{m}}
	\right)\xi_{\theta}^{2}
	 \, dV, \label{TaylerW1}
\end{eqnarray}
where $a_m$, $b_m$, and $c_m$ are evaluated at $m=1$.
Notably, the coefficients $b_m$ and $c_m$ contain only terms that depend 
on the magnetic field and vanish as the field vanishes. In the leading order in the magnetic field both coefficients scale quadratically with the field value: $b_m, c_m \sim \mathcal{O}(B^2)$. In contrast, the coefficient $a_m$ also includes magnetic field-independent contributions arising from the stellar matter stable stratification. 
These contributions are always positive and tend to stabilize the system, leading to $a_m>0$ for reasonable field values. 
It follows from this analysis that to make the quadratic form in Eq.~\eqref{TaylerW1} negative one must choose small values of $l/(kr)\sim\mathcal{O}(B)$. In that case the $b_m$-dependent term in \eqref{TaylerW1} can be neglected and the functional can be rewritten as:
\begin{align}
&
	W[\m\xi] \approx\int 
	\left({a_{m}}\frac{l^2}{r^2k^2}
	\;+\;
	{c_{m}}
	\right)\xi_{\theta}^{2}
	 \, dV, 
&     
\label{TaylerW2x}
\end{align}

Clearly, the necessary and sufficient condition for this form to become negative is 
\begin{align}
&
c_m<0. 
&
\label{cm}
\end{align}
Instability occurs when 
$kr$ exceeds the threshold value,
\begin{align}
&
 k r \gtrsim l \left(\frac{a_m}{|c_m|}\right)^{1/2}.
&
\label{thr}
\end{align}
Expressing $a_m$ and $|c_m|$ as [see Eqs.\ \eqref{a} and \eqref{c}]
\begin{align}
&
a_m \approx - \partial_r{\rho_0} \partial_r{\Phi_0}
- \frac{\rho_0^2}{\gamma_{{\rm fr}, 0}\, p_0} \left( \partial_r{\Phi_0} \right)^2  = \rho_0 \mathcal{N}^2,
&
\label{amx1}\\
&
|c_m| \sim \frac{B^2}{r^2},
&
\end{align}
where 
$\mathcal{N}$ is the Brunt-V$\ddot{\rm a}$is$\ddot{\rm a}$l$\ddot{\rm a}$ frequency \eqref{brunt},
one can represent \eqref{thr} as (see also, e.g., \cite{Spruit2002,sb24}):
\begin{align}
&
    k r \gtrsim \frac{\mathcal{N}}{\omega_{\rm A}} l,
&
\label{thr2}
\end{align}
where $\omega_{\rm A}=1/\tau_{\rm A}$ is the typical Alfv\'en frequency.
Obviously, the smaller the magnetic field, the larger $kr/l$ must be to fulfill the condition \eqref{thr2}. 
Near the threshold [at $k\sim \mathcal{N} l/(\omega_{\rm A}r)$], the growth rate of the 
instability is low but increases with increasing $k$ approaching an asymptote at $k\rightarrow 
\infty$ (see the blue solid curve in Fig.\ \ref{Fig:omega}).
It should be emphasized that Tayler did not claim that the proposed displacement was either optimal or unique.

In contrast to Tayler, we assume that the functional $W[\m\xi]$ can be expanded in the small parameters $\varepsilon$ and $\varkappa$. We then eliminate the non-negative leading-order term [not the same as the first term in Eq.~\eqref{TaylerW}], and consequently find that the unstable displacement is 
the shellular differential rotation in the leading order of the perturbation theory. 
Analysis of the next-order terms yields 
corrections to 
this differential rotation.
The shellular differential rotation is the only possible unstable displacement, but only for 
displacements $\m\xi$ that do not mix orders of the expansion (i.e., do not make some next-order 
terms comparable to the leading-order terms).

On the other hand, Tayler's displacement mixes orders, as seen from Eq.~\eqref{WKB}: small $\xi_r$ multiplied by $k$ becomes of order unity. This allows Tayler's displacement to be unstable as well, even though it is not a shellular differential rotation. While we find unstable displacements of the differential-rotation type that respect the ordering, the fastest-growing among them mixes orders [see Eqs.~\eqref{eps} and \eqref{nopt}, which imply $n\varepsilon\sim 1$]. 

In this sense our initial requirement that the leading-order term of $W[\m\xi]$ vanish is not compulsory for such ``mixing'' displacements. Similarly, Tayler's assumption that the first term in Eq.~\eqref{TaylerW} must vanish is not compulsory.
Here, Tayler's method and ours  
share a common strategy: both rely on an ansatz (distinct in each case) to derive unstable  
displacements, which themselves differ fundamentally.

One of the important features of the unstable displacements is the existence of limiting values of the typical radial wave numbers $k$ in both cases. However, these limits differ {\it qualitatively}. In the case of the global instability discussed in this paper, the typical value of $k$ is limited from above. 
According to the results of 
Sec.\ \ref{diffeq} [see, in particular, Eq.\ \eqref{nopt}], 
the displacement is unstable, 
i.e., it is not stabilized by
the effect of the third-order term, 
as long as
$n \sim k R \lesssim \sqrt{\gamma_{{\rm fr}, 0}-\gamma_{{\rm eq},0} }/\varkappa^{1/2} 
\sim \mathcal{N}/\omega_{\rm A}$, or%
%
\footnote{This is why this instability was not found, for example, in Refs.~\cite{kitchatinov08,kr08,rk10}, where the limit $k \rightarrow \infty$ was considered.}
%
\begin{align}
&
k \lesssim
\frac{\mathcal{N}}{\omega_{\rm A}R},
&
\end{align}
but $k$ practically is not limited from below 
[it should only be $\gtrsim 1/R$, see the comment after Eq.\ \eqref{wxi3}]. 
On the contrary, in the Tayler instability $k$ has the lower bound, 
$kr \gtrsim \mathcal{N} l/\omega_{\rm A}$, 
or equivalently, for $l\sim 1$ and $r\sim R$, 
$k \gtrsim \mathcal{N}/(\omega_{\rm A}R)
$
(at lower $k$ the displacement is stable), but at the same time it is not limited from above. Taken 
together, the relevant $k$ ranges read as
\begin{align}
    &
    \frac{\mathcal{N}}{\omega_{\rm A}R} \lesssim k \lesssim \infty \;\;\;\;\;\;  {\rm Tayler \;instability,}  \label{Taylerk}
    &
    \\
    &
    \frac{1}{R}\lesssim k \lesssim \frac{\mathcal{N}}{\omega_{\rm A}R} \quad \;\; {\rm global \;instability.}   \label{globalk}
    &
\end{align}
It is clear that the threshold value of $\mathcal{N}/(\omega_{\rm A}R)$ in these inequalities is 
determined only very approximately and may, in reality, differ by a factor of a few in either 
direction.

The properties of the global and Tayler instabilities in ideal MHD are summarized in the schematic 
Fig.~\ref{Fig:omega} (see the solid lines there).
The dashed curves account for dissipation and are discussed below in Sec.~\ref{diss}.
In Fig.~\ref{Fig:omega}, for both types of instability, the qualitative behavior of the ideal-MHD growth rate $1/\tau_0(k)$ 
(defined 
by $\m \xi \propto e^{t/\tau_0}$) is shown 
as a function of 
$\log(kR)$, where $k$ is the radial wavenumber. In accordance with the discussion above, the (red) solid curve 
for the 
global instability has a bell-shaped form with a maximum $1/\tau_0(k) \sim \omega_{\rm A}$ near $k 
R 
\sim \mathcal{N}/\omega_{\rm A}$. The Tayler instability (blue solid curve), in turn, comes into play only 
at $k R \gtrsim \mathcal{N}/\omega_{\rm A}$ and becomes most efficient at large $k R$, where 
$1/\tau_0(k)$ approaches its asymptotic maximal value, also of order $\omega_{\rm A}$. In the 
figure 
this maximum slightly exceeds the maximum of $1/\tau_0(k)$ for the global instability, which is not 
required in general but is consistent with our numerical experiments performed for a set of 
toroidal field models. 

\subsection{Dissipative effects}
\label{diss}

\begin{figure}
	\center{\includegraphics[width=0.5\linewidth]{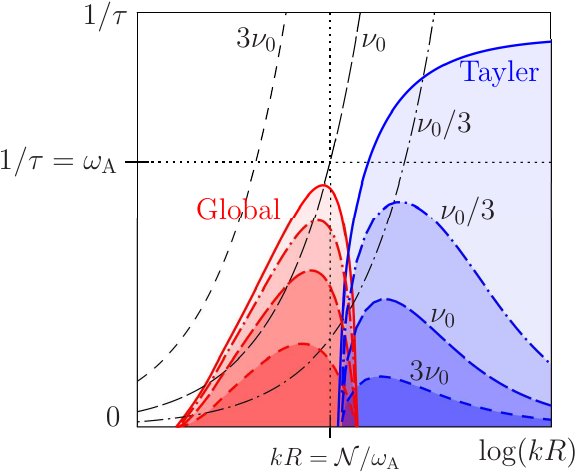}}
	\caption{A schematic plot of the growth rates for the global (red) and Tayler (blue) 
	instabilities as functions of the radial perturbation wavenumber $k$. Dots mark the threshold 
	value of $k$ [see Eqs.\ \eqref{Taylerk} and \eqref{globalk}] and the Alfv\'enic growth rate 
	$\omega_{\rm A}$. Solid lines show the ideal-MHD limit; colored dash-dotted and dashed lines 
	illustrate the effect of three representative shear-viscosity values: $\nu_0/3$ (dash-dotted), 
	$\nu_0$ (long-dashed), and $3\nu_0$ (short-dashed); see text for details. The corresponding 
	dissipative rates $1/\tau_{\nu}$ are shown by black curves, which scale as $k^2$.
	}
	\label{Fig:omega}
\end{figure}

Note that the weaker the magnetic field, the larger the value of $\mathcal{N}/\omega_{\rm A}$. 
Typically, $\mathcal{N}/\omega_{\rm A} \gtrsim 100$ for various stellar types and magnetic field 
strengths. 
For instance, $\mathcal{N}/\omega_{\rm A} \sim 100$ for magnetars, while for rotation-powered 
pulsars $\mathcal{N}/\omega_{\rm A} \sim 10^5$. 
Such high values of $\mathcal{N}/\omega_{\rm A}$, which imply large $k$ [see Eqs.\ \eqref{Taylerk}, \eqref{globalk}, and Fig.\ \ref{Fig:omega}], raise the question of whether dissipative 
processes such as, e.g., 
shear viscosity, magnetic diffusivity, thermal conductivity, and ambipolar diffusion, 
can significantly affect the development of these instabilities.
A comprehensive analysis of this problem is well beyond the scope of this paper and merits a 
separate study (see, e.g., Refs.\ \cite{ag78,sb24}, where such an analysis was performed 
within the short-wavelength approximation in studies of the Tayler instability).
Here, for illustration, we examine how shear viscosity, one of the effective dissipation mechanisms 
in neutron stars, could affect the instabilities under consideration.
We expect magnetic diffusivity (which is more relevant for ordinary non-degenerate stars) to 
influence the instability in a qualitatively similar way \cite{ag78, spruit99, Spruit2002}.

The characteristic dissipative timescale due to shear viscosity depends on the perturbation 
wavenumber and can be estimated as
\begin{align}
&
\tau_{\rm \nu} \approx 1/(\nu k^2), \label{taushear}
&
\end{align}
where
$\nu$ is the kinematic shear-viscosity coefficient.%
%
\footnote{In Appendix \ref{shear1} we present a more accurate expression for the dissipative timescale $\tau_{\rm \nu}$.}
%
Shear viscosity slows the growth of the instability (either global or Tayler) but cannot suppress 
it completely.
The corresponding instability timescale $\tau(k)$ can be written as (see Appendix \ref{proof1} for details):
\begin{align}
&
\tau(k) = \frac{\tau_0(k)^2}{2 \tau_{\rm \nu}(k)}+\tau_0(k) \, \sqrt{1+\frac{\tau_0(k)^2}{4 
\tau_\nu(k)^2}},
&
\label{tau}
\end{align}
where $\tau_0(k)$ is the instability timescale in the ideal MHD (see solid lines in Fig.\ 
\ref{Fig:omega}).
As it should be, $\tau(k)= \tau_0(k)$ for $\tau_0(k)\ll \tau_\nu(k)$ and
$\tau(k)= \tau_0^2(k)/\tau_\nu(k) \gg
\tau_0(k)$ for $\tau_0(k)\gg \tau_\nu(k)$ (e.g.,
\cite{sb24}).
As follows from Eq.\ \eqref{tau}, both instabilities are slowed down by shear viscosity, but
the Tayler instability is more sensitive to dissipation than the global one due to its shorter 
spatial scales. 
Since the (minimal) non-dissipative timescales $\tau_0 \sim 1/\omega_{\rm A}$ 
are similar for both instabilities (see Sec.~\ref{example}), the more efficient suppression of the 
Tayler instability by shear viscosity makes the global instability comparatively faster and allows 
it to develop.
Thus, if 
\begin{align}
&
    \tau_{\nu}\left(k = \frac{\mathcal{N}}{\omega_{\rm A}R}\right) \lesssim \tau_0 \left(k = 
    \frac{\mathcal{N}}{\omega_{\rm A}R} \right) \sim \omega_{\rm A}^{-1}, 
    \label{condtau}
&
\end{align}
we expect the global instability to dominate, while in the opposite limit, the Tayler instability 
may prevail. 
Note that a large dissipation, in a sense, guarantees the dominance of the global instability. However, depending on the specific field model, the global instability might still prevail even in the dissipationless limit or in the presence of only small dissipation.

For illustration, Fig.\ \ref{Fig:omega} qualitatively shows how viscosity modifies the ideal-MHD 
dependence $1/\tau_0(k)$ (solid lines) for the global (red lines) and Tayler (blue lines) 
instabilities.
The red and blue curves $1/\tau(k)$ are plotted for three viscosity values, $\nu_0/3$ (dash-dotted), $\nu_0$ 
(long-dashed), and $3\nu_0$ (short-dashed), where the reference viscosity $\nu_0$ is chosen such 
that $\tau_\nu(k=\mathcal{N}/(\omega_{\rm A}R))=1/\omega_{\rm A}$.
For each viscosity, $\tau(k)$ is computed from Eq.\ \eqref{tau}.
Additionally, Fig.~\ref{Fig:omega} shows the function $1/\tau_\nu(k)$, calculated for the same three viscosity values (see the black dash-dotted, long-dashed, and short-dashed curves, which scale as $k^2$).
The figure indicates that as viscosity increases, the global instability becomes the increasingly 
dominant destabilizing mechanism.
In summary, for systems with a relatively weak magnetic field [i.e., large $k=\mathcal{N}/(\omega_{\rm A}R)$] and high viscosity (or high magnetic diffusivity), the global magnetic-field instability is expected to develop in the star {\it earlier} than the Tayler instability.

Returning to Eq.\ \eqref{condtau}, it can be 
rewritten in the form:
\begin{align}
&
    \nu \gtrsim \frac{\omega_{\rm A}^2}{{\mathcal N}^2}\frac{R^2}{\tau_{\rm A}}\approx\frac{E_{\rm 
    mag}}{(\gamma_{{\rm fr}, 0}-\gamma_{{\rm eq},0} )E_{\rm grav}}\frac{R^2}{\tau_{\rm A}}.
&
\label{ineqmax}
\end{align}
Using values typical of neutron stars, we find that the global instability is expected to dominate 
over the Tayler instability when
\begin{align}
&
	\nu \gtrsim 
     10^{9}\,
  B_{15}^{3}\,
  \left(\frac{\rho}{7\times 10^{14}\,\mathrm{g\,cm^{-3}}}\right)^{-3/2}
  \left(\frac{\mathcal{N}}{10^{3}\, {\rm s}^{-1}}\right)^{-2}
  \left(\frac{R}{12\, \mathrm{km}}\right)^{-1} \,\rm cm^2 \,s^{-1},
&
    \label{nuNS}
\end{align}
where $B_{15}\equiv B/(10^{15}\,\rm G)$.
For nondegenerate stars, the magnetic diffusivity $\eta$ is a more relevant dissipative parameter, since it is usually larger than $\nu$. At the same time, as already mentioned above, we expect that the constraints obtained for $\nu$ in this section should be equally applicable to the coefficient $\eta$. With this in mind, for Sun-like stars the inequality \eqref{ineqmax} can be written as:
\begin{align}
&
	\eta \gtrsim 
     10^{3}\,
  B_{3}^{3}\,
  \left(\frac{\rho}{0.2\,\mathrm{g\,cm^{-3}}}\right)^{-3/2}
  \left(\frac{\mathcal{N}}{2 \times 10^{-3}\, {\rm s}^{-1}}\right)^{-2}
  \left(\frac{R}{0.7 R_\odot}\right)^{-1} \,\rm cm^2 \,s^{-1},
&
    \label{etaNS}
\end{align}
where $B_3 \equiv B/(10^3 \, {\rm G})$ and $R_\odot$ is the solar radius.
One finds that, for $B \sim 1$~kG, the threshold value of $\eta$ given by Eq.~\eqref{etaNS} is comparable to typical values of the Spitzer magnetic diffusivity $\eta$ in the radiative layers of solar-type stars (e.g., \cite{hkt04}).

Returning to the estimate \eqref{nuNS} for neutron stars, it is interesting to assess how effective the viscosity $\nu$ can be in practice.
Under typical neutron-star conditions, the kinematic shear viscosity is dominated by 
particle-particle collisions (see, e.g., \cite{ss18}).
As a rough estimate it can be approximated as
\begin{align}
&
	\nu \approx 3\times 10^{4} \left(\frac{T}{10^8\,\rm K}\right)^{-5/3} \, \rm cm^2 \,s^{-1}, \label{nupp}
&
\end{align}
where $T$ stands for the stellar temperature, and the dependence of $\nu$ on the matter density and composition is omitted.
Substituting now \eqref{nupp} into the inequality~(\ref{nuNS}), we find that particle shear viscosity effectively damps the Tayler instability for magnetic fields
\begin{align}
	&
	B_{15} \lesssim 0.03\left(\frac{T}{10^8\,\rm K}\right)^{-5/9}.
	\label{fieldvaluepp}
	&
\end{align}
By contrast, during the early stages of neutron-star evolution, when temperatures are still high 
enough that stellar matter is opaque to neutrinos, neutrino viscosity, associated with momentum 
transport by 
neutrinos, may play a significant role.
Assuming nondegenerate neutrinos and degenerate other particle species, Ref.\ \cite{td93} obtained the 
following expression for the neutrino viscosity (see also \cite{ip82} and the recent review \cite{rh24})
\begin{align}
&
	\nu \approx 5\times 10^7 \left(\frac{T}{3\times 10^{10}\,\rm K}\right)\left(\frac{\rho}{3\times 
	10^{14} \,\rm g\,cm^{-3}}\right)^{-4/3} \, \rm cm^2 \,s^{-1}, \label{nunu}
&
\end{align}
where we omit a factor of order unity in this formula that depends on the proton fraction.
Comparing \eqref{nunu} with the threshold in the inequality~(\ref{nuNS}), we find that neutrino viscosity effectively damps the Tayler instability for magnetic fields
\begin{align}
	&
	B_{15} \lesssim 0.4\left(\frac{T}{3\times 10^{10}\,\rm K}\right)^{1/3}.
	\label{fieldvalue}
	&
\end{align}
Summarizing, in nascent neutron stars, the global instability should dominate for magnetic-field strengths given by 
Eq.\ \eqref{fieldvalue} and possibly even higher, while at later evolutionary stages the corresponding threshold value is given by \eqref{fieldvaluepp}.
In addition to the viscosities discussed above, turbulent (eddy) viscosity driven by momentum 
transport by eddies may also play a role in nascent neutron stars (see, e.g., \cite{rh24}).
Its magnitude is uncertain, so we do not consider it here, but it may be efficient.

\subsection{Optimal wavenumbers in the presence of dissipation}
\label{opt_diss}

At the end of Sec.~\ref{diffeq}, we mentioned that including dissipation (viscosity or magnetic diffusivity) in the MHD equations can substantially slow down the development of small-scale perturbations with $n \sim kR \sim \mathcal{N}/\omega_{\rm A}$, which would be optimal in ideal MHD. Under these conditions, the most favorable displacements are those with smaller $n$ (that is, with smaller radial wavenumbers $k$), lying within the range of validity of the perturbation theory developed in Sec.~\ref{pert}.

In essence, this effect already follows from the discussion in the previous section. In particular, it is clearly visible in Fig.~\ref{Fig:omega}, where the maximum of $1/\tau(k)$ for the global instability shifts to the left, i.e., toward smaller $k$, as the viscosity $\nu$ increases. Below we illustrate this behavior with a simple analytic estimate.

Let us introduce the growth rates $\omega_0(k) \equiv 1/\tau_0(k)$ and $\omega(k) \equiv 
1/\tau(k)$. We denote the maximum value of $\omega_0(k)$ by $\omega_{0,{\rm opt}}$, and the 
corresponding value of $k$ by $k_{0,{\rm opt}}$. According to 
our previous results,
we have
$\omega_{0,{\rm opt}} \sim \omega_{\rm A}$ and $k_{0,{\rm opt}} \sim \mathcal{N}/(\omega_{\rm A}R)$.
Similarly, we denote the maximum of $\omega(k)$ by $\omega_{\rm opt}$, and the corresponding value of $k$ by $k_{\rm opt}$.

To estimate the effect of viscosity, we first need an approximate form of $\omega_0(k)$. 
Consider an unstable mode 
$\m \xi_k$ characterized by the radial 
wavenumber $k$. 
For such a mode,
the ideal-MHD growth rate is given by the Rayleigh quotient,
$\omega_0^2(k)= -W[\m \xi_k]/I[\m \xi_k]$
(see Sec.~\ref{bernshtein}).
Using the approximate functional $W[\m \xi]$ from Sec.~\ref{diffeq}, the functional $I[\m \xi]$ 
from Sec.~\ref{timescale}, and the estimate $a'(r)\sim k a(r)$ (Sec.~\ref{second}), we obtain 
the following simple $k$-dependence:
$\omega_0^2(k)=\alpha_1 k^2-\alpha_2 k^4$
[cf.\ Eq.~\eqref{wxi8}, where an analogous dependence on $n\sim kR$ arises for the same reason].
This form reflects the competition between the destabilizing contribution, which grows with $k$, 
and the stabilizing contribution associated with excessively rapid radial variation, which arises 
from the third-order term discussed in Sec.~\ref{diffeq} and scales as $k^4$.
Here $\alpha_1$ and $\alpha_2$ are some positive constants.
It is convenient to re-express them in terms of $\omega_{0,{\rm opt}}$ and $k_{0,{\rm opt}}$, which 
gives
\begin{align}
&
	\omega_0^2(k)=\omega_{0,{\rm opt}}^2 
	\left(2-\frac{k^2}{k_{0,{\rm opt}}^2}\right)\frac{k^2}{k_{0,{\rm opt}}^2}.
	&
    \label{w02}
\end{align}
This expression reaches its maximum value $\omega_{0,{\rm opt}}^2$ at $k=k_{0,{\rm opt}}$. 

Let us now include viscosity. Rewriting Eq.~\eqref{tau} in terms of $\omega(k)$, $\omega_0(k)$, and
$\omega_\nu(k)\equiv 1/\tau_\nu(k)=\nu k^2$,
we obtain
$\omega(k)=-\omega_\nu(k)/2+\sqrt{\omega_\nu(k)^2/4+\omega_0^2(k)}$.
Maximizing this expression with respect to $k$, we find that its maximum is
\begin{align}
&
\omega_{\rm opt}
=
\frac{2\omega_{0,{\rm opt}}^2}
{2\omega_{0,{\rm opt}}+\nu k_{0,{\rm opt}}^2},
&
\label{wopt}
\end{align}
and is attained at $k=k_{\rm opt}$, where
\begin{align}
&
k_{\rm opt}
=
k_{0,{\rm opt}}
\sqrt{
\frac{2\omega_{0,{\rm opt}}}
{2\omega_{0,{\rm opt}}+\nu k_{0,{\rm opt}}^2}
}.
&
\label{kopt}
\end{align}
Thus, as the viscosity increases, both the optimal growth rate $\omega_{\rm opt}$ and the optimal 
wavenumber $k_{\rm opt}$ decrease relative to their nondissipative counterparts,
$\omega_{0,{\rm opt}} \sim \omega_{\rm A}$ and
$k_{0,{\rm opt}} \sim \mathcal{N}/(\omega_{\rm A}R)$.
In particular, in the large-viscosity regime,
$\nu \gg 2\omega_{0,{\rm opt}}/k_{0,{\rm opt}}^2$,
which coincides with the condition for the global instability to dominate over the Tayler instability
[see Eq.~\eqref{ineqmax}],
we have
\begin{align}
&
\omega_{\rm opt}\approx\frac{2 \omega_{0,{\rm opt}}^2}{\nu k_{0,{\rm opt}}^2}\ll \omega_{0,{\rm opt}}, \quad\quad
k_{\rm opt}\approx \sqrt{\frac{2 \omega_{0,{\rm opt}}}{\nu}}\ll k_{0,{\rm opt}}.
&
\label{optest}
\end{align}
Note that the inequality $k_{\rm opt}\ll k_{0,{\rm opt}}$ implies that the stabilizing third-order term 
$\propto k^4$ in the expression \eqref{w02} for $\omega_0(k)$ can be neglected.
In this regime, the ratio of the viscous term to the Lorentz force perturbation $\delta\m F_{\rm L}$ in the Euler equation
for the optimal mode with $k=k_{\rm opt}$ can be estimated as
\begin{align}
&
\frac{\text{Viscous term}}{\text{Lorentz force perturbation}}\sim \frac{\nu \, \omega_{\rm opt} k_{\rm opt}^2}{\omega_{\rm A}^2}
\sim 2 \, \frac{2 \omega_{0,{\rm opt}}}{\nu k_{0,{\rm opt}}^2} \ll 1.
&
\label{ratiovisc}
\end{align}

In energy terms, the smallness of this ratio means that 
the magnetic energy functional $W^{\rm mag}[\m \xi]$ is much larger than the corresponding 
viscous dissipation term.%
%
\footnote{Indeed, the perturbed Lorentz force $\delta \m F_{\rm L}$ is related to the 
	magnetic energy 
	functional $W^{\rm mag}[\m \xi]$ by Eq.~\eqref{magapp} in Appendix \ref{approx}.}
%
Combined with the inequality $k_{\rm opt}\ll k_{0,{\rm opt}}$ [see Eq.\ \eqref{optest}],
this shows that, for the optimal ($k=k_{\rm opt}$) mode in the presence of viscosity,
perturbation theory in the parameters $\varkappa$ and $\varepsilon$ can be applied.
According to estimate \eqref{ratiovisc}, the viscous term does not contribute at first order in 
this perturbative expansion.
Thus, at this order, the energy functional $W[\m\xi]$ [see Eq.\ \eqref{wxi}] is determined solely 
by its magnetic part
$W^{\rm mag}[\m\xi]$, yielding the same term $\varkappa \alpha_{01}$
as in the first-order perturbation theory developed in Sec.~\ref{first}.
We therefore conclude that, although viscosity determines the optimal radial scale of the 
fastest-growing mode,
it does not modify the leading-order form of its Lagrangian displacement.
Accordingly, in the high-viscosity regime, this displacement should still be well described by the first-order perturbation theory of Sec.~\ref{pert}, that is, it should remain close to \textit{shellular differential rotation}.
This conclusion is further supported by the numerical simulations presented in the next section.

\section{Simulations}
\label{sim}

To illustrate our theoretical results presented in the previous sections, we carried out a series 
of numerical simulations using the 
\href{https://pencil-code.nordita.org}{\texttt{Pencil Code}}%
%
\footnote{\texttt{https://pencil-code.nordita.org}}
%
\cite{pencil21}, a high-order finite-difference code for compressible MHD flows. The code uses 
sixth-order centered spatial derivatives and a third-order Runge-Kutta time-stepping scheme to 
solve the MHD equations given in Appendix A of \cite{brvg22b}. The simulations are performed in a 
cubic box of side $L_{\rm box}=3 R$, centered on the star, with a Cartesian grid with 
equally spaced points and periodic boundary conditions. 
We verified that our results are insensitive to the box size; increasing the box size at fixed point density does not change them.

We describe stellar matter as an ideal monatomic gas, with the density and specific entropy
being the independent variables. The adiabatic index then equals $\gamma_{\rm fr}=5/3$. For the initial non-magnetized background configuration we choose a polytropic relation between the gas pressure
and density, $p\propto \rho^{\gamma_{\rm eq}}$, with $\gamma_{\rm eq}=4/3$. 
We define the stellar surface as the sphere at which the matter density $\rho$ equals $2 \times 
10^{-3}\rho_c$,
where $\rho_c$ is the central density.
The stellar radius $R$ then equals the radius of this sphere. We assume that the star is surrounded 
by a low-density ideal-gas atmosphere with a uniform temperature smoothly matching with its value 
at the stellar surface. The magnetic diffusivity is set to 0 inside the star and has a finite value 
outside, with a smooth transition between these two regions. 

Starting from a non-magnetized equilibrium stellar model, we introduce a magnetic field, which perturbs the star from equilibrium. To obtain the initial magnetic equilibrium, we evolve the system for several sound-crossing timescales with the magnetic field held fixed, so that the hydrodynamical forces relax against the Lorentz force. 
The perturbations of the gravitational potential by the magnetic field are neglected in all our simulations to reduce the numerical computation time (Cowling approximation). That is, the gravitational potential is calculated for the non-magnetized spherical background stellar configuration and fixed along the simulations. For more details on the initial setup, we refer the reader to Refs.~\cite{brvg22a,brvg22b}. 

\begin{figure}
	\center{\includegraphics[width=1\linewidth]{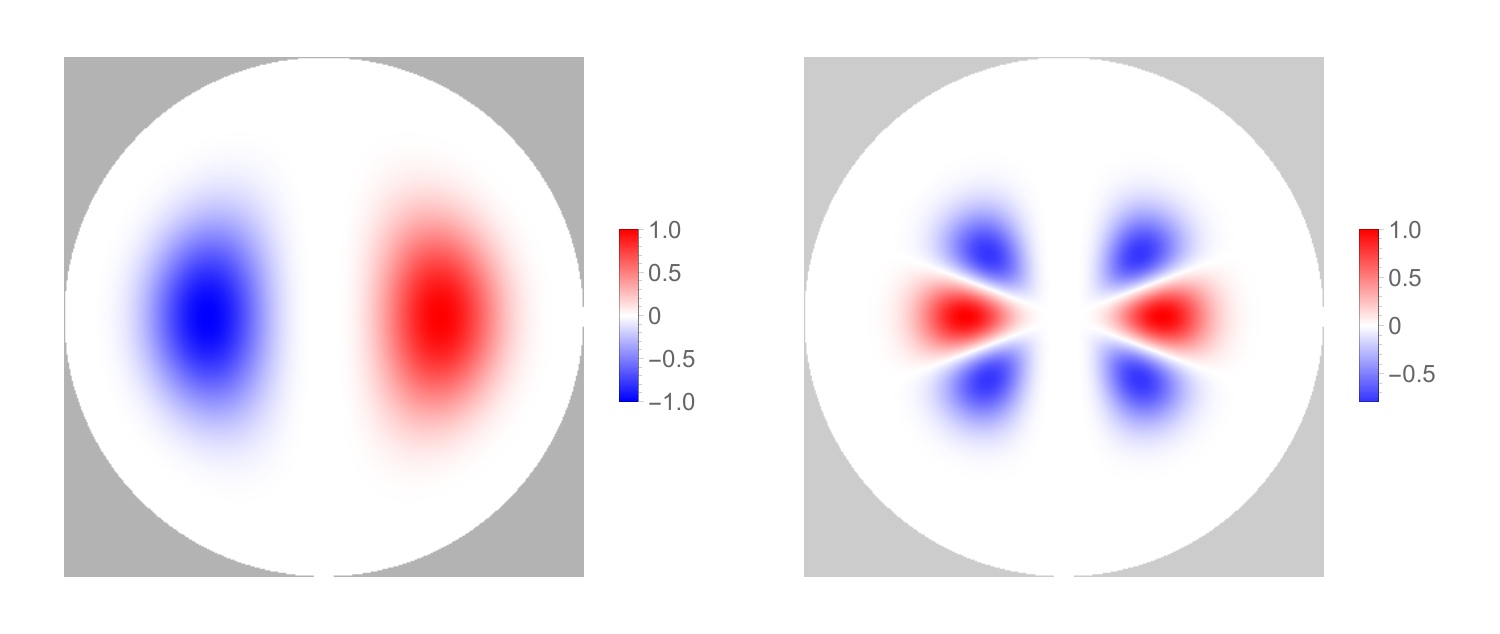}}
	\caption{Meridional cuts of the star showing color maps of the initial magnetic field used in our simulations (left) and the corresponding coefficient $c_m$ (right). Both quantities are normalized to their respective maxima. The Tayler instability is expected to develop in the blue regions of the right panel.
	}
	\label{Fig:ModelB}
\end{figure}

The evolution and stability of purely toroidal, purely poloidal, and mixed magnetic-field 
configurations were investigated in detail in \cite{brvg22b}. Here we use the magnetic-field model 
``Field I'' from \cite{brvg22b}, with the poloidal component set to zero (note that the model 
``Field I'' was originally proposed in \cite{armm13}).
The amplitude of the toroidal component is adjusted so that the ratio of gravitational to magnetic 
energy, $E_{\rm grav}/E_{\rm mag}$%
%
\footnote{For the definition of $E_{\rm grav}$ and $E_{\rm mag}$ we use equations (19) and (22) of 
\cite{brvg22a}, replacing the integration over the box by the integration over the stellar volume.},
%
equals $806$, $358$, $159$, or $89$ 
(see the third column of Table~\ref{param} and the corresponding values of $\mathcal{N}/\omega_{\rm A}$ in the fourth column). 
These values are clearly far too low to be representative of real stars.
However, adopting realistic magnetic field values is technically challenging for MHD simulations. A color map of the chosen field model is shown in the left panel of Fig.~\ref{Fig:ModelB}. The right panel of the figure shows the value of the $c_m$ coefficient [see Eq.~\eqref{c}]. According to \cite{tayler73, gv78} [see also Sec.\ \ref{nodiss}], 
in the regions of the star where $c_m<0$, the unstable 
(predominantly tangential to spherical surfaces)
displacements associated with the Tayler instability are expected to develop. 

\begin{figure}
	\center{\includegraphics[width=1\linewidth]{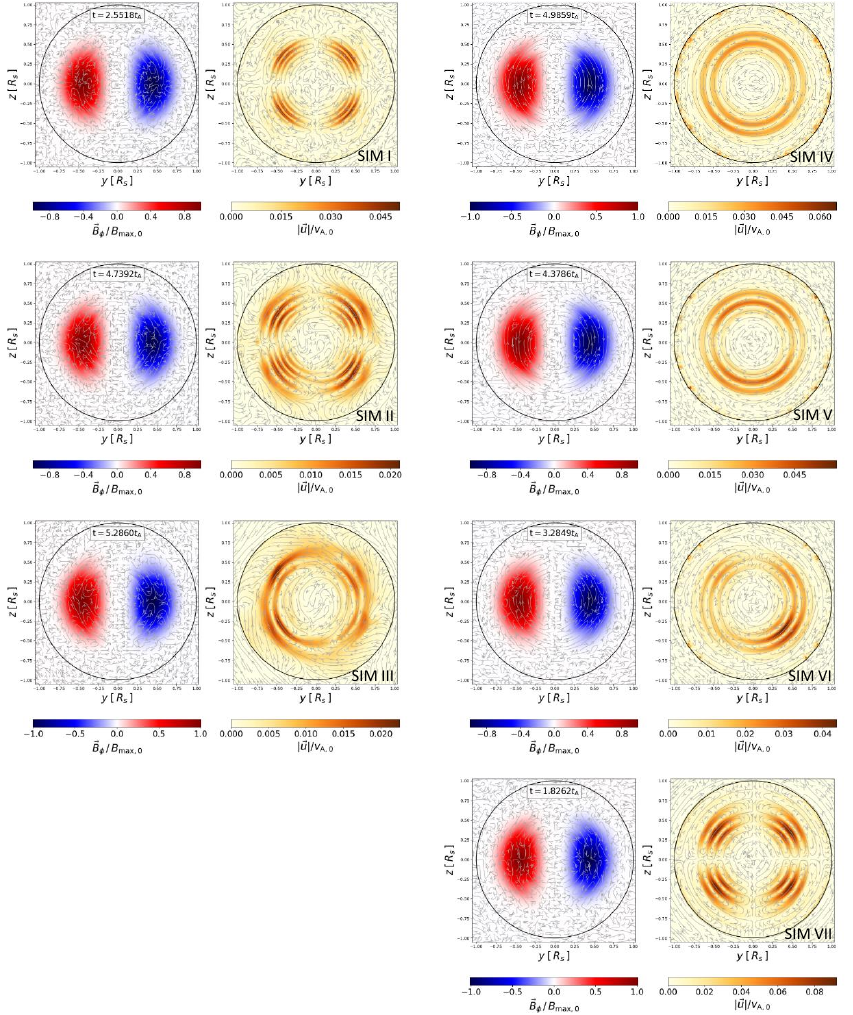}}
	\caption{
    Snapshots of the seven simulations. Simulations I, II, and III are displayed in the left column, and Simulations IV, V, VI, and VII in the right column, ordered from top to bottom. All panels show meridional cuts of the star. The $x$-axis, perpendicular to the plane of the figure, is selected to maximize in-plane displacements along the $\theta$ direction. The left sub-panel of each panel shows the magnetic field. The colors represent the toroidal-field magnitude (more precisely, its $x$-projection) normalized to its initial amplitude; arrows indicate the small, noisy poloidal component. The times for each snapshot are listed in the left sub-panels in units of the initial Alfv\'en time. The right sub-panel of  each panel shows the velocity field. Colors indicate its magnitude in units of the Alfv\'en speed; arrows indicate its direction.}
	\label{Fig:sim}
\end{figure}

In what follows, we discuss seven representative simulations 
(see the Supplemental Material~\cite{supplemental}
and the snapshots in Fig.~\ref{Fig:sim}). 
They differ in the numerical grid 
resolution, in the magnetic field amplitude, and in the values of the shear-viscosity coefficient $\nu$.
Namely, the first three simulations (Simulations I, II, and III) have high spatial resolution 
($288^3$) and differ only in the value of the shear viscosity, whereas the last four (Simulations 
IV, V, VI, and VII) have lower resolution ($128^3$) and differ in the magnetic-field amplitude.
The corresponding parameters are listed in Table~\ref{param} (see the columns 2-5). We define the 
Alfv\'en timescale $\tau_{\rm A}$, given in column six, as $\tau_{\rm A}=R \sqrt{4 \pi 
\bar\rho}/B_{\rm rms}$, where $\bar \rho$ is the average density (stellar mass over stellar volume) 
and $B_{\rm rms}$ is the root-mean square magnetic field in the star
\footnote{Note that this definition differs from that given by Eq.~\eqref{tauA}.}. 
In column seven, we give $\tau_s$, the sound crossing time across the distance $R$. It is 
calculated as $\tau_s=R/c_{\rm s,rms}$, where $c_{\rm s,rms}$ is the root-mean square value of the 
sound speed $c_{\rm s}=\sqrt{\gamma_{\rm fr}P/\rho}$, averaged over the stellar volume.

\begin{table}[t!]
    \centering{
    \begin{tabular}{|c| c| c |c| c| c| c|}
        \hline
      \;\;  Simulation \;& \; Resolution \;& \; $E_{\rm grav}/E_{\rm mag}$ \;& \; 
      $\mathcal{N}/\omega_{\rm A}$ \;& \; $\nu/[R^2 (G\rho_c)^{1/2}]$ \; & \; $\tau_{\rm 
      A}/(G\rho_c)^{-1/2}$ \; & \; $\tau_s/(G\rho_c)^{-1/2}$ \; \\ 
        \hline
       I & $288^3$ & 358 & 11 & $10^{-5}$ &  27  &  3.9 \\ 
       II & $288^3$ & 358 & 11 & $10^{-4}$ &  27  &  3.9 \\ 
       III & $288^3$ & 358 & 11 & $2\times 10^{-4}$ & 27  &  3.9 \\ 
       IV & $128^3$ & 806 & 16 & $10^{-5}$ & 41  &  3.9 \\ 
       V & $128^3$ & 358 & 11 & $10^{-5}$ & 27  &  3.9 \\ 
       VI & $128^3$ & 159 & 7.3 & $10^{-5}$ & 18  &  3.9 \\ 
       VII & $128^3$ & 89 & 5.4 & $10^{-5}$ & 13  &  3.9 \\ 
        \hline
    \end{tabular}}
    \caption{Parameters for the simulations presented in this paper. All simulations are performed in a computational box with side length $L_{\rm box}=3R$. 
    The adiabatic index is $\gamma_{\rm fr}=5/3$, and the equilibrium index is $\gamma_{\rm 
    eq}=4/3$. See text for details.
    }
    \label{param}
\end{table}

In addition to dissipation due to shear viscosity $\nu$, MHD motions also experience numerical 
dissipation (see, e.g., \cite{numdis} and references therein). By analyzing the evolution of the 
magnetic, kinetic, gravitational, and thermal energies during the simulations, we found that 
numerical resistivity can be neglected, whereas numerical viscosity may play a role. 
To gain qualitative insight into the contribution of numerical viscosity,
we ran sound-wave tests in nonmagnetized homogeneous 
matter in a box, with the shear-viscosity term removed from the equations. Varying the grid 
resolution and the sound wavenumber $k$, we found that the decrement $1/\tau_{\rm num}$ of the 
sound-wave velocity amplitude, $u = u_0 \exp(-t/\tau_{\rm num})$ (where $u_0$ is a small initial perturbation), for the adopted numerical 
scheme is given by
\begin{align}
&	
    \frac{1}{\tau_{\rm num}(k)}\approx 9.6\times 10^{-6}\left(\frac{N_{\rm gp}}{96}\right)^{-3}\left(\frac{k\,R}{\pi}\right)^4\frac{1}{\tau_s}. 
\label{taunum}
&
\end{align}
This is consistent with the third-order scheme of the \texttt{Pencil Code} \cite{pencil21}. Here, 
$N_{\rm gp}$ is the number of grid points {\it per length} $R$. 
In our magnetic simulations, $N_{\rm gp}$ is either $288/3=96$ (for the simulations I, II, and III) or $128/3$ (for the simulations IV, V, VI, and VII). 
Finally, $k\,R/\pi$ is the number of sound wave zeros per length $R$. 
In the remainder of this section, instead of the perturbation wave number $k$, we will use the 
typical lengthscale $d$.
For harmonic perturbations it is defined as $d=\pi/k$, and equals half the perturbation wavelength.
As follows from Eq.~\eqref{taunum}, smaller lengthscales $d$
and coarser grid resolutions result in faster numerical dissipation. 
Note, however, that Simulations VI and VII are likely subject to different numerical dissipation, since
a different numerical scheme (an upwind scheme)
was employed in these simulations for numerical stability when computing
spatial derivatives.

We estimate the typical value of $d$ for each simulation using the following formula, averaging the lengthscale over the meridional cuts shown in Fig.~\ref{Fig:sim}:
\begin{align}
    d\equiv \pi \,\frac{\int_{0.1}^{0.8}\int_0^\pi (|v_\theta|+|v_\phi|)\,r\, d\theta\, dr}{\int_{0.1}^{0.8}\int_0^\pi r \left( \left|\frac{\partial (v_\theta /r)}{\partial r}\right|+\left|\frac{\partial (v_\phi /r)}{\partial r}\right|\right) r\, d\theta\, dr}.
\end{align}
The radial integration is restricted to $0.1\le r\le 0.8$, excluding the central region, where the 
instability does not develop and the coordinate singularity may affect the estimate, as well as the 
outer layers, where residual post-relaxation noise is still present.

\begin{figure}
	\center{\includegraphics[width=0.7\linewidth]{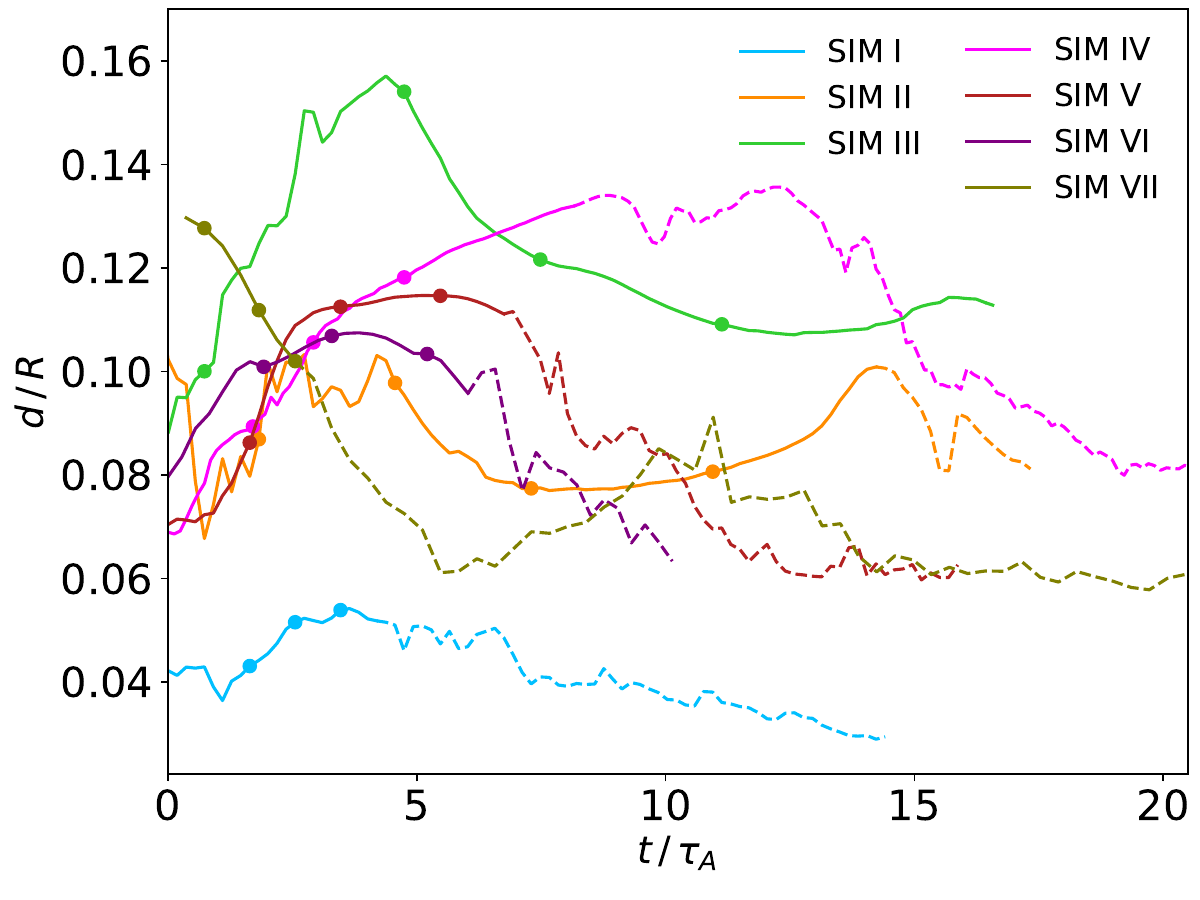}}
	\caption{Typical lengthscale $d$ as a function of time for Simulations I-VII. The solid lines correspond to the linear stages of the instability, while the dashed lines indicate the subsequent non-linear stages. The circles correspond to the snapshots shown in Figs.~\ref{Fig:diagnostics} and \ref{Fig:transition}.
 }
	\label{Fig:d}
\end{figure}

Figure~\ref{Fig:d} shows $d$ as a function of time for the Simulations I-VII. The value of $d$ is clearly rather noisy and can therefore serve only as a rough estimate of the lengthscale of the most unstable perturbation.
The least viscous Simulation I exhibits the instability on the smallest scale. 
Just as increasing the shear viscosity leads to an increase in the characteristic scale (see the curves for Simulations II and III), increasing the numerical viscosity (i.e., lowering $N_{\rm gp}$; Simulations IV--VII) also leads to a larger characteristic scale than in Simulation I.

Now, let us analyze the simulations. 
Simulation I
(see the top left panel of Fig.\ \ref{Fig:sim}) 
shows a clear onset of the Tayler instability in the stellar regions with $c_m<0$, implying that, in the low-viscosity regime (including both numerical and shear viscosity), the Tayler instability grows faster than the global one.

As discussed in Sec.~\ref{Tayler}, the typical stellar values of $\mathcal{N}/\omega_{\rm A}$ are large, usually exceeding $\sim 100$. Consequently, the Tayler instability in real stars develops on very small spatial scales in the radial direction, $d \sim\pi/k \lesssim \pi R \omega_{\rm A}/\mathcal{N}\lesssim 0.03R$. Under these conditions, there exists a value of shear viscosity that efficiently suppresses perturbations at $d \lesssim 0.03R$, while still allowing the global instability to grow at scales $0.03R \ll d \ll R$. Our estimates in Sec.~\ref{Tayler} indicate that such values of the shear viscosity can be reached in stars.
To illustrate this situation, we ran Simulations II and III with the same parameters as in Simulation I, but with an increased shear viscosity $\nu$. Unfortunately, for technical reasons, our simulations require unrealistically strong magnetic fields, corresponding to $\mathcal{N}/\omega_{\rm A}\sim 16,\, 11,\,7.3,\,{\rm or}\,5.4$, see Table \ref{param}. 
This allows the Tayler instability to develop on larger lengthscales, while narrowing the range of lengthscales available to the global instability [see inequalities~\eqref{Taylerk} and \eqref{globalk}].
Under these conditions, viscosities sufficient to damp the Tayler instability also suppress the 
global instability across nearly the entire range of accessible scales. 
Consequently, the snapshots of Simulations II and III (middle and lower left panels of 
Fig.~\ref{Fig:sim}) represent a mixture of the Tayler and global instabilities, with the latter 
contributing more prominently in the more viscous Simulation III. Indeed, the global 
instability manifests itself through the development of increasingly pronounced shellular rotation 
about the $x$ axis, which is perpendicular to the plane of the figure and passes through the center 
of the star. The presence of the global instability is also corroborated by the behavior of the 
characteristic parameter $\mathfrak{R}_x$, shown for these simulations in 
Fig.~\ref{Fig:diagnostics} below.

To artificially reproduce the suppression of the Tayler instability and isolate the global instability in its pure form, we carried out Simulation V, 
which had
the same $E_{\rm grav}/E_{\rm mag}=358$ as Simulations I--III, but a lower grid resolution ($128^3$). 
A snapshot from this simulation
is presented 
in the right-hand panel of the second row
of Fig.~\ref{Fig:sim}. 
The plane of the figure 
is aligned with the plane of the unstable displacement (shellular differential rotation). One clearly sees colored concentric circles in the right sub-panel, corresponding to the spherical shells participating in the shellular rotation, a typical characteristic feature of the global instability (see Sec.~\ref{opt_diss}). 
These circles are much more pronounced than similar structures in Simulations II and III.
According to Eq.~\eqref{taunum}, a lower grid resolution results in a larger numerical viscosity.
Therefore, for the adopted parameters, numerical viscosity suppresses the entire $k$ range relevant to the Tayler instability, while decreasing sharply toward lower $k$ values relevant to the global instability. This mimics the situation described above in this section, in which the Tayler instability is completely suppressed by shear viscosity, whereas the global instability is not.

The sensitivity of the instability simulations to the magnetic field strength is demonstrated by Simulations
IV-VII, carried out for $\mathcal{N}/\omega_{\rm A}$ equal to, respectively,
$16$, $11$, $7.3$, and $5.4$,
with all other parameters kept the same (see Table~\ref{param}).
The results are compared with those of the fiducial Simulation V, for which $\mathcal{N}/\omega_{\rm A}=11$.
Increasing the magnetic field relative to its value in Simulation V (Simulations VI and VII)
strengthens the Lorentz force and thus reduces the role of viscosity. Moreover, as the magnetic field increases, the threshold wave number, $\mathcal{N}/(\omega_{\rm A}R)$, decreases [see Eqs.~\eqref{Taylerk} and \eqref{globalk}]. Consequently, the range of $k$ relevant to the Tayler instability shifts to lower values, where dissipation is weaker. This makes the Tayler instability more likely to dominate over the global one, as seen in the simulations: Simulation VI exhibits the global instability with some admixture of Tayler instability, rather than the purely global instability found in Simulation V, while Simulation VII, with an even stronger magnetic field, shows a purely Tayler instability.
By contrast, reducing the magnetic field (Simulation IV) makes the onset of the Tayler instability even less favorable, and in this case we observe a purely global instability.

\begin{figure}
	\center{\includegraphics[width=0.8\linewidth]{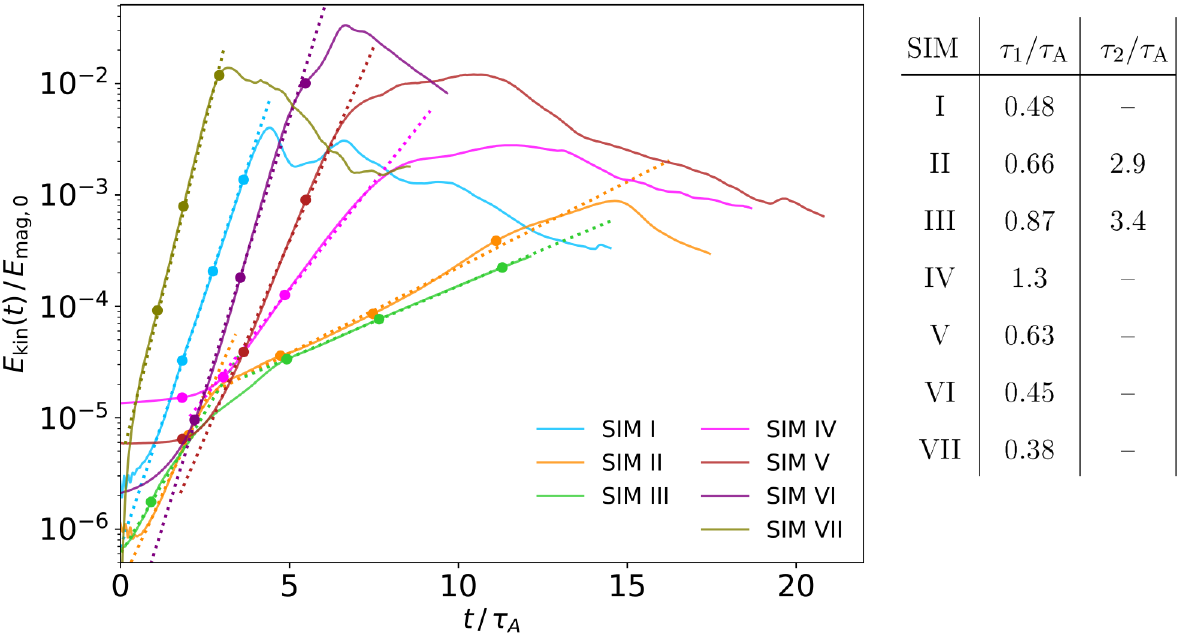}}
	\caption{
    Evolution of the kinetic energy, normalized to the initial magnetic energy, as a function of time, normalized to the Alfv\'en time. The seven solid lines correspond to the seven simulations. The dotted lines indicate exponential fits to the linear growth stages of the instability. For Simulations II and III, the two growth stages are fitted separately. The table lists the corresponding growth times inferred from these fits. The circles correspond to the snapshots shown in Figs.~\ref{Fig:diagnostics} and \ref{Fig:transition}.
	}
	\label{Fig:Energy}
\end{figure}

Figure \ref{Fig:Energy} shows the time evolution of the kinetic energy (in units of the initial magnetic energy) for the simulations discussed above (solid lines). The dotted lines indicate exponential fits to the linear stages. Simulations I, IV, V, VI, and VII exhibit a clear linear growth with a constant rate, corresponding, respectively, to the development of either Tayler or global instabilities. By contrast, in Simulations II and III the slope during the linear stage changes with time: the instability initially grows with a steep slope, which later becomes flatter. We interpret this behavior as follows. The initial steep slope corresponds to the more rapidly growing Tayler instability, which saturates at some point (probably because of the generated poloidal magnetic field that suppresses it). 
After that, the initially slower global instability, which may be further slowed down by the 
poloidal field, takes over, leading to the subsequent flattening of the slope.
The corresponding instability timescales inferred from the fits are listed in the table next to the figure. They are of the order of the Alfv\'en time. These timescales, however, are affected by both numerical and shear viscosity. When the relevant viscous timescales are known, Eq.~\eqref{tau} can be used to infer the corresponding ideal-MHD instability timescales. We apply this procedure to the initial stages of Simulations I-V, but not to the later stages of Simulations II and III, since those stages may already be affected by the generated poloidal field, nor to Simulations VI and VII, since $\tau_{\rm num}$ is unknown in these simulations.

Figure~\ref{Fig:d} allows us to (very roughly) estimate the typical lengthscale for a given simulation; the resulting values 
for Simulations I-V are listed in column three of Table~\ref{Table:tau}.
Using this lengthscale, we evaluate the numerical and shear viscous timescales from Eqs.~\eqref{taunum} and \eqref{taushear}, respectively. 
Their ratio, $\tau_\nu/\tau_{\rm num}$, is shown in column four, 
while the combined viscous timescale, $\tau_{\rm visc}=(1/\tau_\nu+1/\tau_{\rm num})^{-1}$, is given in column five.
The corresponding ideal-MHD timescale, $\tau_0$, inferred from Eq.~\eqref{tau} with $\tau_\nu$ replaced by $\tau_{\rm visc}$, is listed in column six.

It is clear that these estimates should be treated with considerable caution.
One of the main reasons is that the perturbations excited in the simulations span a range of lengthscales, whereas our procedure replaces this range by a single representative value. This simplification may affect the inferred timescales. Still, column four indicates that shear viscosity is unimportant for Simulations I, IV, and V, but contributes for Simulations II and III, as expected. 
Moreover, in all cases, it appears that $\tau_{0}(k=\pi/d) \sim \tau_{\rm visc}(k=\pi/d)$, 
suggesting that the perturbation lengthscale $d$ is likely set by viscosity: 
perturbations with smaller 
lengthscales
are viscously damped, whereas those with larger 
lengthscales
grow more slowly even without dissipation (see Sec.\ \ref{diss}). 
In the case of Simulation I, we expect that 
increasing the grid resolution will increase $\tau_{\rm num}$ and shift the Tayler instability toward smaller lengthscales $d$ (larger $k$).

\begin{table}[t!]
    \begin{tabular}{c|c|c|c|c|c|c}
        SIM \, &\, $\tau_1/\tau_{\rm A}$ \, &\, $d=\pi/k$ \, &\, $\tau_{\nu}/\tau_{\rm num}$ \, &\, $\tau_{\rm visc}/\tau_{\rm A}$ \, &\, $\tau_{0}(k=\pi/d)/\tau_{\rm A}$\\
        \hline
        I & 0.48  & 0.05 & 10 & 0.17 & 0.19 \\ 
        II & 0.66  & 0.09 & 0.31 & 0.46 & 0.34 \\ 
        III & 0.87  & 0.13 & 0.074 & 0.59 & 0.43 \\ 
        IV & 1.3  & 0.12 & 20 & 0.35 & 0.44  \\ 
        V & 0.63  & 0.1 & 28 & 0.25 & 0.26 \\ 
    \end{tabular}
    \caption{Inferred viscous and ideal-MHD timescales for Simulations I-V. See text for details.}
    \label{Table:tau}
\end{table}

To characterize the nature of the instabilities more quantitatively, we introduce the parameter $\mathfrak{R}_x$, which measures how close the unstable perturbations are to shellular rotation. 

To evaluate $\mathfrak{R}_x$, we divide the star radially into spherical shells of thickness $\sqrt{(\gamma_{\rm fr} - \gamma_{\rm eq}) E_{\rm grav}/E_{\rm mag}}\,  R / 5 \approx   (\omega_{\rm A}/\mathcal{N}) R/5$. Here $(\omega_{\rm A}/\mathcal{N}) R$ is an approximate characteristic radial scale of the instability (more precisely, it provides a lower estimate for the global instability and an upper estimate for the Tayler instability). The shells should be thinner than this scale. At the same time, choosing a thickness below the grid resolution is not meaningful. We therefore adopt $(\omega_{\rm A}/\mathcal{N}) R/5$ as a convenient shell thickness.
For each shell, we compute the following vector:
\begin{align}
   \m{\mathfrak{R}}\equiv \int_{\rm shell} \frac{\m{r} \times \m{u} \,dV}{r_{\rm shell} u_{\rm av} V_{\rm shell}},
\end{align}
where $\m r$ is the radius vector, $r_{\rm shell}$ is the characteristic radial coordinate of the 
shell, $V_{\rm shell}$ is the shell volume, $\m{u}$ is the fluid velocity, and $u_{\rm av}$ is the 
shell-averaged absolute value of the velocity, 
$u_{\rm av} \equiv 1/V_{\rm shell}  \int_{\rm shell} |\m u| dV$.
In the case of an ideal shellular rotation, this vector should be aligned in the same direction in all shells and should have magnitude $8/(3 \pi)$ (in the limit of an infinitesimally thin shell).
We then define the parameter $\mathfrak{R}_x$ as the projection of $\m{\mathfrak{R}}$ onto the 
$x$-axis. Since we choose the $x$- and $y$-axes so as to localize the motions mainly in the $y$-$z$ 
plane, we expect $\mathfrak{R}_x$ to approach $8/(3 \pi)$ in the case of global instability.
For the Tayler instability, by contrast, $\mathfrak{R}_x$ is expected to be smaller.

Figure~\ref{Fig:diagnostics} shows $\mathfrak{R}_x$ as a function of radial coordinate for 
simulations I--VII. 
Each panel corresponds to one particular simulation and contains several curves corresponding to 
different moments in time. These moments are also marked by circles in Figs.~\ref{Fig:d} and 
\ref{Fig:Energy}. Horizontal dotted lines indicate $\mathfrak{R}_x=\pm 8/(3 \pi)$. One can see that 
in the case of the Tayler instability $\mathfrak{R}_x$ remains close to zero 
(see 
panels,
which 
show Simulations I and VII, corresponding to the pure Tayler instability). By contrast, in the case 
of shellular rotation (see panels,
corresponding to Simulations IV and V), $\mathfrak{R}_x$ reaches the value $\pm 8/(3 \pi)$. In this 
case, the antinodes of the function $\mathfrak{R}_x$ correspond to the regions where the velocities 
are maximal, and in these regions $\mathfrak{R}_x$ approaches $\pm 8/(3 \pi)$. Meanwhile, the nodes 
of $\mathfrak{R}_x$ correspond to the nodes of the shellular rotation. In these regions, shellular 
rotation does not contribute to $\mathfrak{R}_x$, and the value of $\mathfrak{R}_x$ is determined 
by higher-order velocity corrections and numerical noise. 

\begin{figure}
	\center{\includegraphics[width=1\linewidth]{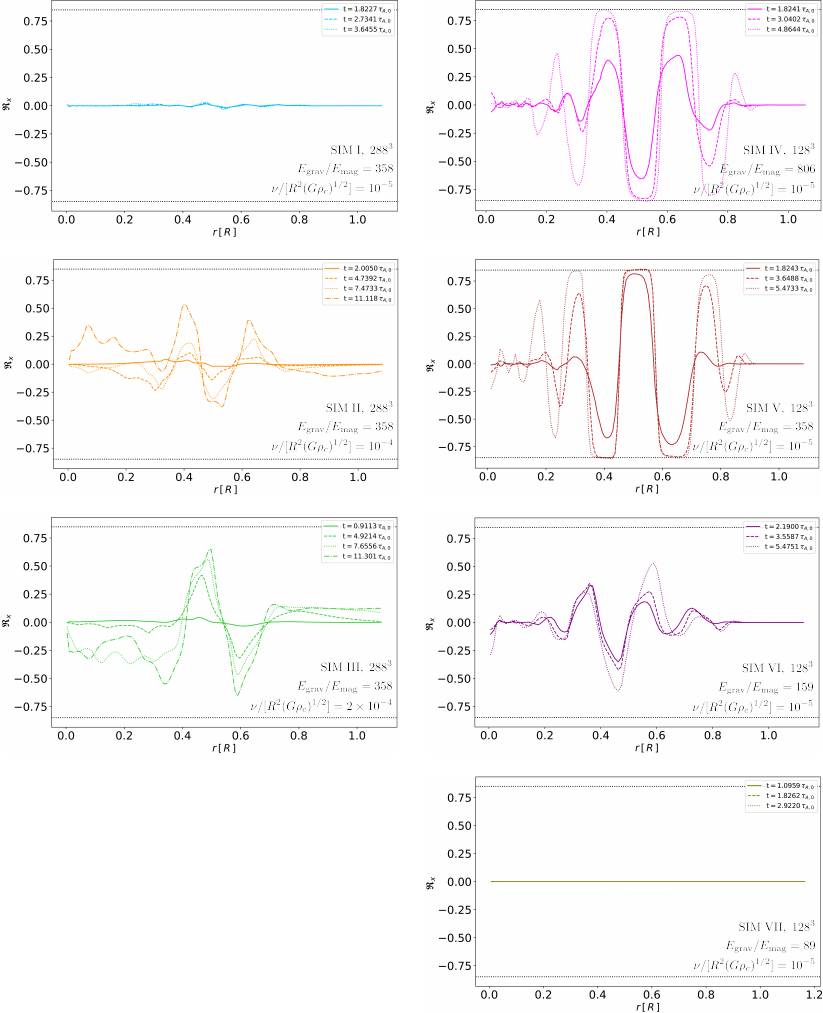}}
	\caption{
    $\mathfrak{R}_x$ as a function of radial coordinate for the simulations discussed in the paper. In each panel $\mathfrak{R}_x(r)$ is plotted for several snapshots (see labels in the Figure and the circles in Figs.~\ref{Fig:d} and \ref{Fig:Energy}).
	}
	\label{Fig:diagnostics}
\end{figure}

For Simulations II and III,
$\mathfrak{R}_x$ illustrates the transition from Tayler-type perturbations to the global 
instability discussed above. In both simulations, the instability initially develops through 
Tayler-like displacements, and the parameter $\mathfrak{R}_x$ remains small. The Tayler instability 
then saturates (probably due to the generation of a poloidal field component), after which the 
global instability, although initially slower, takes over, so that $\mathfrak{R}_x$ tends toward 
$\pm 8/(3 \pi)$. This interpretation is also supported by Fig.~\ref{Fig:Energy} (see the discussion 
above), where the initial steep slope associated with the Tayler instability is followed by a 
flatter one associated with the global instability.

\begin{figure}
	\center{\includegraphics[width=1\linewidth]{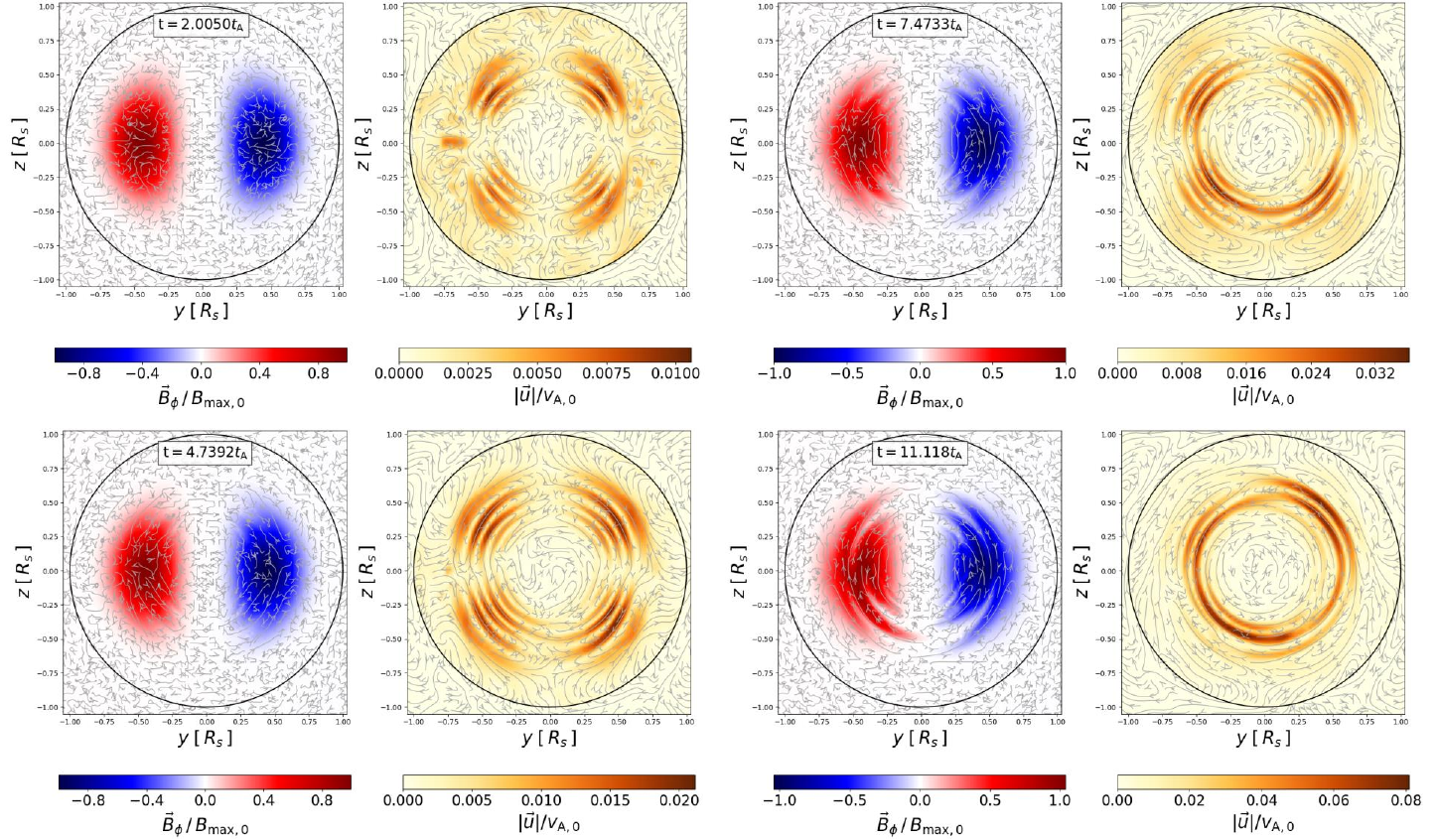}}
	\caption{
    Snapshots of Simulation II. Notations are the same as in Fig.~\ref{Fig:sim}.
	}
	\label{Fig:transition}
\end{figure}

To further illustrate the transition from one type of instability to the other, 
Fig.~\ref{Fig:transition} shows snapshots from Simulation II corresponding to the time moments 
indicated 
in the left middle panel
of Fig.~\ref{Fig:diagnostics}. One can see that the Tayler instability gradually gives way to the 
global instability, in agreement with the discussion above.

\section{Summary and conclusion}
\label{summary}

In this paper we analyzed the stability of toroidal magnetic fields in stably stratified nonrotating stars, using both analytical methods based on the energy principle of Bernstein et al.\ \cite{bernstein_etal58} and numerical simulations with the \texttt{Pencil Code} \cite{pencil21}.
Although most of the paper discusses stably stratified stellar matter, some of our results also apply to the barotropic case; see the comment after Eq.\ \eqref{gr} in Sec.~\ref{second}.

Analytically, we showed that, in addition to the well-known Tayler instability \cite{tayler73}, there exists a distinct class of unstable displacements that destabilize a toroidal field of essentially arbitrary (physically acceptable) form. Although both instabilities grow on the Alfv\'en timescale, they are qualitatively different. The Tayler instability is generally local in the $r$ and $\theta$ directions, i.e., it develops in specific regions of the star. 
In contrast to the Tayler instability, the instability we identify is global and, under certain 
conditions discussed in Sec.~\ref{pert} (see also Secs.~\ref{opt_diss} and \ref{sim}), may reveal 
itself by generating large-scale shellular differential rotation about an arbitrary axis 
perpendicular to the symmetry axis of the toroidal field. The corresponding Lagrangian displacement 
takes the form 
[see Eq.\ \eqref{lagany}]:
$\m\xi_{\rm rot}(r,\theta)=a(r)\,\hat{\m O}\times \m r$,
where $\hat{\m O}$ is 
a unit vector that sets the rotation-axis direction. 
In Sec.\ \ref{diffeq} we propose an approximate variational method to compute the function $a(r)$ 
that describes the differential rotation.

Another characteristic difference between the two instabilities is the scale of inhomogeneity in the radial coordinate $r$ that arises as the instability develops. The Tayler instability is characterized by small radial scales with characteristic wavenumbers $\mathcal{N}/(\omega_{\rm A}R) \lesssim k \lesssim \infty$ [see Eq.\ \eqref{Taylerk}]. In turn, the instability we find corresponds to smaller $k$, $1/R \lesssim k \lesssim \mathcal{N}/(\omega_{\rm A}R)$ [see Eq.\ \eqref{globalk}]. 
Consequently, the Tayler instability is more readily damped by dissipative effects (Sec.\ 
\ref{Tayler}).
Depending on the dissipation level, either instability can dominate. For example, according to our estimates, in young neutron stars where dissipation is governed by neutrino shear viscosity, the global instability is more likely to prevail.

Our general conclusions about the global instability were illustrated by (in)stability calculations for a star with magnetic field $\m B=\mathcal{L}\rho r\sin\theta \,\m e_\varphi$ within the Bernstein et al.\ energy principle (Sec.\ \ref{example}). For this field we numerically determined the function $a(r)$ and obtained the growth time of the instability, following the analytical approach developed in 
Sec.\ \ref{pert}.

In addition, we independently verified the existence of the global instability in a numerical experiment. 
To this end, we performed three-dimensional numerical simulations using the \texttt{Pencil Code} of the nonlinear, dissipative MHD equations that describe an initially equilibrated magnetized star with a toroidal field (Sec.~\ref{sim}). 
We showed that, as anticipated, 
depending on the dissipation strength, 
the simulations showed either Tayler-dominated or global-dominated behavior, with some runs 
exhibiting a mixture of the two.
For the chosen magnetic-field model, we found that the Tayler instability 
prevails for weak dissipation. However, for strong dissipation, we observed suppression of 
Tayler-type perturbations and the dominance of the new global instability, with the emergence and 
growth of the analytically predicted large-scale modes corresponding to shellular differential 
rotation.

We also estimated the strength of the poloidal magnetic field required to stabilize the system by opposing the growth of the unstable shellular differential rotation (Sec.\ \ref{poloidal1}). 
We showed that it constitutes only a small fraction of the toroidal field, and that the 
corresponding criterion is consistent with estimates reported in the literature 
\cite{braithwaite09,armm13,brvg22b}, although our criterion is more quantitative.

Frequently, numerical modeling of stellar magnetic-field evolution employs approximations that substantially simplify the governing MHD equations. We verify (see Appendix~\ref{approx}) that the global instability persists under four widely used approximations: the equilibrium Cowling approximation, the anelastic approximation, the LBR formulation \cite{Lantz92,br95,gtz19}, and the Boussinesq approximation.

Taken together, the results of this work indicate that, in addition to the classical Tayler 
instability, stars may be subject to a new current-driven $m = 1$ global instability that can 
become important and possibly dominant in some astrophysically relevant regimes.
This may call for a revision of theoretical models of stellar magnetism, in particular the Tayler-Spruit dynamo scenario, in which the presence of an additional class of unstable modes may significantly affect angular momentum transport and magnetic-field evolution in stellar interiors.
A natural next step is therefore to understand how these conclusions are modified by rotation.

It is clear that rotation should substantially affect the instability considered here once the 
Coriolis force in the equations of motion becomes comparable to the Lorentz force.
A simple estimate shows (see, e.g., Refs.\ \cite{pt85,spruit99}) that, in the absence of dissipative effects, this occurs when
$\Omega \gtrsim \omega_{\rm A}$.
Therefore, for objects that do not satisfy this inequality
(for example, magnetars),
the results obtained in the present work 
are likely to remain qualitatively applicable.

By contrast, for sufficiently rapidly rotating objects,
$\Omega \gtrsim \omega_{\rm A}$,
which include most stars,
a systematic treatment of rotation is required.
In this case, the analytical study of the global instability becomes considerably more complicated, although it is by no means hopeless.
By contrast, a numerical investigation of the problem including rotation presents no fundamental difficulties.
We have already made progress in both directions.
According to our preliminary results, rotation does not suppress the global instability, but it does significantly slow down its development:
at large $\Omega$, the characteristic growth time of the instability increases in proportion to $\Omega$.
In this respect, the situation resembles the effect of rotation on the Tayler instability \cite{pt85}.
A detailed study of the global instability in the presence of rotation will be presented elsewhere.

\acknowledgments
We thank the referee for their insightful comments and suggestions.
Preliminary results of this work were presented at the LUTh seminar (Paris Observatory, Meudon, 
February 2024), the workshop ``Extreme Physics of Neutron Star Interiors'' (Princeton University, 
14-16 May 2025), and the seminar of the theoretical astrophysics department at the Ioffe Institute 
(November 2025). The authors are grateful to the participants of these events for their questions 
and valuable comments.
The work of MEG and EMK is funded by the baseline project FFUG-2024-0002 of the Ioffe Institute. 
MEG acknowledges FONDECYT Project 11230837 for supporting his visit to UMCE (Santiago, Chile). JAV 
acknowledges the support of FONDECYT Project 1240697.
MEG and EMK dedicate their contribution to this paper to the memory of Yu.~A.~Shibanov.



%


\appendix

\section{\texorpdfstring{Accurate minimization of the ratio $W[\m\xi]/I[\m\xi]$ with respect to $\m\nabla \cdot \m\xi$}{Accurate minimization of the ratio W[mxi]/I[mxi] with respect to div(mxi)}}
\label{accurate}
In Sec.\ \ref{tor} we presented Eq.\ \eqref{div3}, which was used later in the text and which was obtained by minimizing the functional
$W[\m\xi]$ with respect to the $\varphi$-component of the Lagrangian displacement 
$\xi_{\varphi}$
or, equivalently,
with respect to the function $\m\nabla\cdot\m\xi$.
As discussed in footnote \ref{WI}, if we are interested in the growth time of the instability and in the optimal Lagrangian displacement, it would be more appropriate to minimize the ratio $W[\m\xi]/I[\m\xi]$.
The result of such a minimization with respect to $\m\nabla\cdot\m\xi$ is discussed in this Appendix.
As in the main text of the paper, we assume $m\neq 0$
and without any loss of generality present 
$\m\xi(r,\theta,\varphi)=[\xi_r(r,\theta,\varphi),\xi_\theta(r,\theta,\varphi),\xi_\varphi(r,\theta,\varphi)]$
in the form
\eqref{xiexpr2}.
We also introduce a new vector
$\overline{\m\xi}(r,\theta,\varphi) \equiv
[\xi_r(r,\theta,\varphi),\xi_\theta(r,\theta,\varphi),0]$.
Then minimization of $W[\m\xi]/I[\m\xi]$ with respect to $\m\nabla\cdot\m\xi$ at fixed
$\xi_r$ and $\xi_\theta$ yields
\begin{align}
&
\m\nabla \cdot\m\xi = -\frac{\rho m^2}{-m^2 p \gamma_{\rm fr}+\rho \Lambda \, r^2 \sin^2 \theta}
(\m\xi \cdot \m\nabla \Phi)
+\frac{\rho \Lambda \, r^2 \sin^2 \theta}{-m^2 p \gamma_{\rm fr}+\rho \Lambda \, r^2  \sin^2 \theta}
\m\nabla\cdot \overline{\m\xi}.
&
\label{div44}
\end{align}
We know (see Sec.\ \ref{timescale}) that for the global instability under consideration and for ideal MHD
\begin{align}
&
|\Lambda|\sim \omega_{\rm A}^2 \propto B^2.
&
\label{Lam}
\end{align}
Accordingly, $-m^2 p \gamma_{\rm fr}+\rho \Lambda \, r^2  \sin^2 \theta \approx -m^2 p
\gamma_{\rm fr}$ and Eq.\ \eqref{div44} simplifies:
\begin{align}
&
\m\nabla \cdot\m\xi = \frac{\rho}{p \gamma_{\rm fr}}
(\m\xi \cdot \m\nabla \Phi)
-\frac{\rho \Lambda \, r^2 \sin^2\theta}{m^2 p \gamma_{\rm fr}}
\m\nabla\cdot \overline{\m\xi}.
&
\label{div55}
\end{align}
This formula differs from expression \eqref{div3} by the presence of the second term on the right-hand side.
Let us estimate how large the contribution of this additional term is.
To this end, note that for the optimal Lagrangian displacement found in the paper, corresponding to the global instability, the following estimates hold:
$\m\nabla\cdot \overline{\m\xi} \sim a(r)$,
$\xi_r \sim \varepsilon r a(r) \sim (\varkappa^{1/2}/\sqrt{\gamma_{\rm fr, 0}-\gamma_{\rm eq, 
0}})\,r a(r) $
[see Eqs.\ \eqref{xir3}, \eqref{xitheta3}, \eqref{eps}, and \eqref{nopt}].
Using these estimates, it is easy to see that the ratio of the second to the first term on the right-hand side of \eqref{div55} is approximately
\begin{align}
&
\frac{\rm 2nd \,\, term \,\, in \,\, Eq.\ \eqref{div55}}{\rm 1st \,\, term\,\, in \,\, Eq.\ \eqref{div55}} \sim \left(\frac{\omega_{\rm A}}{\mathcal{N}}\right) \left(\frac{r
\mathcal{N}^2}{g}\right) \ll 1,
&
\label{2to1}
\end{align}
and is the product of two small dimensionless factors,
$\omega_{\rm A}/\mathcal{N}$ and $r \mathcal{N}^2/g$.
The latter factor is, in order of magnitude, equal to the ratio of the typical acceleration caused by the buoyancy force to the acceleration of gravity $g$.

The above estimate shows that performing a more accurate minimization of the ratio $W[\m\xi]/I[\m\xi]$, rather than of $W[\m\xi]$ alone, with respect to $\m\nabla \cdot \m\xi$, does not noticeably change Eq.\ \eqref{div3} and hence does not affect the optimal Lagrangian displacement or the characteristic growth time of the global instability.

\section{Dissipative timescale due to shear viscosity}
\label{shear1}
In this section we derive a more accurate expression for the dissipative timescale $\tau_{\nu}$.
We define $\tau_\nu$ as
\begin{align}
&
\tau_\nu = \frac{2 I[\m\xi]}{W_{\nu}[\m\xi]},
&
\label{taunu222}
\end{align}
where $I[\m\xi]$ is the kinetic energy functional \eqref{III}
and $W_{\nu}[\m\xi]$ is the dissipative integral,
\begin{align}
&
W_{\nu}[\m\xi] = \frac{1}{2} \int  \rho \nu \, \sigma_{ik}(\m\xi)\sigma^{ik}(\m\xi) \, dV.
&
\label{Wshear}
\end{align}
In Eq.\ \eqref{Wshear}, $\sigma_{ik}$ is the stress tensor (see, e.g., \cite{ll87}).
Since the kinetic energy functional $I[\m\xi]$ is expressed in terms of the Lagrangian displacement $\m\xi$ (rather than the velocity),
the argument of the stress tensor in \eqref{Wshear} should also be taken to be $\m\xi$.

For the global instability considered here, within the range of validity of the perturbation 
theory of Sec.\ \ref{pert}, the Lagrangian displacement is given by Eq.~\eqref{lagx} up to a small 
correction.
To compute $\tau_\nu$ to leading order, it is therefore sufficient to take $\m\xi=\m\xi_{\rm rot}(r,\theta,\varphi)$, where $\m\xi_{\rm rot}$ is given by Eq.~\eqref{lagx}.
Then the angular integrals in the expressions for $I[\m\xi]$ and $W_{\nu}[\m\xi]$ can be evaluated analytically and we obtain%
%
\footnote{An expression for $I[\m\xi]$ in this form has already been used, see Eq.\ \eqref{III4}.}
%
%
\begin{align}
&
I[\m\xi] \approx \frac{4 \pi}{3} \int_0^R r^4 \,\rho_0 \, a(r)^2 \, dr,
&
\label{IIIxx}\\
&
W_{\nu}[\m\xi]\approx \frac{8 \pi}{3} \int_0^R r^4 \,\rho_0 \nu \, [a'(r)]^2 \, dr.
&
\label{Wshear2}
\end{align}
Knowing the function $a(r)$, the integrals in these expressions are readily evaluated, allowing us to obtain $\tau_\nu$ from Eq.~\eqref{taunu222}.

\section{One-mode estimate for the viscous growth timescale and statement of a general variational theorem}
\label{proof1}

In this Appendix we first present a simple one-mode estimate leading to Eq.~\eqref{tau}. This 
derivation is intended only as a physically transparent argument. A more general variational 
theorem is stated at the end of the Appendix.

Consider a trial displacement $\m \xi_k(\m r)$ characterized by a radial wavenumber $k$. 
In the one-mode approximation, we represent the perturbation as
\begin{align}
&
\m \xi(\m r,t)=A(t)\,\m\xi_k(\m r),
&
\label{appC_one_mode_ansatz}
\end{align}
where $A(t)$ is a scalar amplitude and dots below denote derivatives with respect to time. 
Suppose that, in the absence of dissipation, this spatial pattern grows on the timescale $\tau_0(k)$. 
Then $A(t)$ satisfies
\begin{align}
&
\ddot A - \frac{1}{\tau_0^2(k)} A = 0.
&
\label{appC_A_ideal}
\end{align}
Now assume that shear viscosity does not substantially modify the spatial structure of the displacement and mainly affects its temporal evolution. Then the effect of viscosity can be modeled by adding a linear damping term with the characteristic timescale $\tau_\nu(k)$:
\begin{align}
&
\ddot A + \frac{1}{\tau_\nu(k)} \dot A - \frac{1}{\tau_0^2(k)} A = 0.
&
\label{appC_A_visc}
\end{align}
Looking for a growing solution in the form
\begin{align}
&
A(t)\propto e^{t/\tau(k)},
&
\label{appC_A_tau}
\end{align}
we obtain
\begin{align}
&
\frac{1}{\tau^2(k)}+\frac{1}{\tau_\nu(k)\tau(k)}-\frac{1}{\tau_0^2(k)}=0.
&
\label{appC_tau_quadratic}
\end{align}
Hence
\begin{align}
&
\tau(k)=
\frac{\tau_0^2(k)}{2\tau_\nu(k)}
+
\tau_0(k)\sqrt{1+\frac{\tau_0^2(k)}{4\tau_\nu^2(k)}}.
&
\label{appC_tau_final}
\end{align}
This is Eq.~\eqref{tau}.

The viscous timescale $\tau_\nu(k)$ entering Eqs.~(\ref{appC_A_visc})--(\ref{appC_tau_final}) should be evaluated for the same trial displacement $\m\xi_k$; in our case it is estimated from the dissipation functional derived in Appendix~\ref{shear1}.

Equation~(\ref{appC_tau_final}) reproduces the expected limiting cases:
\begin{align}
&
\tau(k)\approx \tau_0(k),
\qquad
\tau_0(k)\ll \tau_\nu(k),
&
\label{appC_weak_visc}
\end{align}
and
\begin{align}
&
\tau(k)\approx \frac{\tau_0^2(k)}{\tau_\nu(k)},
\qquad
\tau_0(k)\gg \tau_\nu(k).
&
\label{appC_strong_visc}
\end{align}
Thus, viscosity slows down the instability, and it slows down short-scale modes more efficiently because their viscous timescale is shorter.

The derivation above is a one-mode estimate. 
A more general variational theorem for linear dissipative systems explains why the resulting estimate has the form of Eq.~\eqref{tau}. 
We state the corresponding variational result below. Its proof, given by one of us (M.E.~Gusakov), is beyond the scope of the present paper and will be presented elsewhere.

Consider the abstract linear evolution equation
\begin{align}
&
\ddot{\xi}(t)+\hat D\,\dot{\xi}(t)+\hat L\,\xi(t)=0.
&
\label{appC_abstract}
\end{align}
Here $\xi(t)$ is a vector in a Hilbert space with scalar product $(\cdot,\cdot)$ and norm $\|\cdot\|$.
We assume that the operator $\hat L$ is self-adjoint and bounded from below, while the operator $\hat D$ is self-adjoint and non-negative:
\begin{align}
&
(\hat D z,z)\ge 0
\qquad {\rm for \,\, any \,\, admissible}\,\, z.
&
\label{appC_D_positive}
\end{align}
In the MHD problem considered in the main text, $\hat L$ is the force operator of the ideal system, 
while $\hat D$ represents viscous dissipation.
In this language, the operator $\hat L$ determines the quadratic potential-energy form of the ideal system, 
$W[z]=(\hat L z,z)/2$, while $\hat D$ determines the viscous 
dissipation functional
$(\hat D z,z)/2$. 
The kinetic-energy functional, in turn, is given by $I[z]=\|z\|^2/2$.

For a trial displacement $z$, one may define the corresponding ideal growth timescale $\tau_0[z]$ and viscous damping timescale $\tau_\nu[z]$ by
\begin{align}
&
\frac{1}{\tau_0^2[z]}=-\frac{(\hat L z,z)}{\|z\|^2},
\qquad
\frac{1}{\tau_\nu[z]}=\frac{(\hat D z,z)}{\|z\|^2}.
&
\end{align}
Thus, $\tau_0[z]$ and $\tau_\nu[z]$ characterize, respectively, the ideal growth and viscous damping of the same spatial pattern $z$.

For any $\lambda\ge 0$ and any admissible trial function $z$, let us define the quadratic form
\begin{align}
&
q_\lambda[z]\equiv (\hat L z,z)+\lambda (\hat D z,z)+\lambda^2 \|z\|^2.
&
\label{appC_q_lambda}
\end{align}
Suppose now that we have found a trial function $\eta$ such that
\begin{align}
&
(\hat L\eta,\eta)<0,
&
\label{appC_eta_negative}
\end{align}
i.e., the corresponding potential-energy variation is negative.
Then the system is unstable already in the absence of dissipation. 
Since $q_\lambda[\eta]$ is a convex quadratic polynomial in $\lambda$, negative at $\lambda=0$, and tends to $+\infty$ 
as $\lambda\to\infty$, it has one and only one positive root, $\lambda=\lambda_\eta$, given by
\begin{align}
&
\lambda_\eta=
\frac{
-(\hat D\eta,\eta)+
\sqrt{(\hat D\eta,\eta)^2-4(\hat L\eta,\eta)\|\eta\|^2}
}{
2\|\eta\|^2
}.
&
\label{appC_lambda_eta}
\end{align}

The quantity $\lambda_\eta$ provides a lower bound on the exact maximal growth rate of the dissipative system~(\ref{appC_abstract}). 
In other words, the true fastest-growing solution cannot grow slower than $e^{\lambda_\eta t}$.

Thus, solving the scalar equation
\begin{align}
&
q_\lambda[\eta]=0
\label{appC_scalar_eq}
&
\end{align}
for a given trial function $\eta$ yields a variational estimate of the growth rate from below. In the conservative limit, $\hat D=0$, Eq.~(\ref{appC_lambda_eta}) reduces to the standard variational formulation \eqref{III2}.

The one-mode derivation given at the beginning of this Appendix is recovered from this general 
variational estimate by restricting the trial functions to the form $\m \xi(\m 
r,t)=A(t)\,\m\xi_k(\m r)$.

\section{Global instability under various approximations}
\label{approx}

When investigating magnetohydrodynamic phenomena (both numerically and analytically), researchers often employ various simplifying approximations. However, the global instability discussed here may be sensitive to certain ones. In this section, we examine how these approximations affect the instability.

First, we emphasize that, as follows from the discussion in Secs.~\ref{general} and \ref{second}, the instability of a barotropic magnetized star should not necessarily exhibit the characteristic features described here; in particular, we do not expect to observe the suppression of the radial displacement in barotropic matter. Therefore, modeling a toroidal magnetic field instability under the barotropic approximation is not suitable. To observe the distinctive features of the described instability one should adopt a non-barotropic equation of state in the numerical simulations.

With this in mind, we begin by discussing the common approximation of neglecting magnetic 
field-induced perturbations to the background gravitational potential, which is often used to 
simplify calculations (in particular, we adopt it in the simulations of Sec.~\ref{sim}).
Such an approximation can be called the ``equilibrium Cowling approximation''. In this case, following the derivation of Sec.\ \ref{pert}, we find the same expression for $W$, but with vanishing $\partial_\theta \Phi_1$ in $W_2$ and $W_3 = 0$. The argumentation of Sec.\ \ref{pert} remains valid with the only modification: the expression (\ref{wxi3}) simplifies to
\begin{align}
	& 	
	W[\m\xi] \simeq
	\frac{\varkappa^2}{2} \int r^2 \sin\theta 
	\left(-\frac{W_2^2}{4 W_1} \right) \, 
	dr d\theta,
	&
	\label{Appwxi4}
\end{align}
implying that $W[\m\xi]$ is always non-positive and is negative for a wide class of 
$\mathfrak{a}(r,\theta)$ 
functions (namely, it is zero for solid-body rotation and negative for differential rotation). 
Notice that now we do not need $n \sim \partial_r\mathfrak{a}(r,\theta)R/\mathfrak{a}(r,\theta)$ to 
be $\gg 1$; however, it should not be $\ll 
1$ either, to ensure that the contribution of the next-order terms in $\varkappa$ and $\varepsilon$ 
is negligible. Nevertheless, since $W_2 \propto \partial_r\mathfrak{a}(r,\theta)$, the stronger 
radial gradients in the function 
$\mathfrak{a}(r,\theta)$ correspond to shorter instability timescales [see Eq.\ \eqref{III33}]. 
Thus, the displacements with larger $n$ should develop faster, and the $n$ value is again bounded from above by the third-order terms. 
Following the minimization procedure described in Sec.~\ref{second} and retaining the leading contribution from the third-order terms, 
one arrives at an approximate energy functional of the form~\eqref{wxi7}, but with $F_2=F_3=0$. Minimizing this functional yields the desired function $a(r)$. Since, in the general case (i.e., beyond the equilibrium Cowling approximation), the contributions of the terms $\propto F_2$ and $\propto F_3$ are negligible in the limit $n \gg 1$ [see the discussion preceding Eq.~\eqref{ar}], the optimal function $a(r)$ obtained within the equilibrium Cowling approximation practically coincides with that derived beyond this approximation.
In other words, the global instability analyzed in the equilibrium Cowling approximation should yield an $a(r)$ that is very close to the ``true'' one obtained in a self-consistent treatment.

Let us now restore magnetic field-induced perturbations to the background gravitational potential 
and consider one of the most frequently used approximations, the anelastic approximation (see, 
e.g., \cite{brown12} and references therein), which filters out fast variability in the equations 
(sound waves), replacing the continuity equation (\ref{cont13}) with $\m\nabla\cdot(\rho \pmb{\xi}) 
= 0$. Indeed, as discussed in Sec.~\ref{general}, for slow perturbations, $\m\nabla\cdot(\rho 
\pmb{\xi})$ is suppressed to ensure that the second term on the right-hand side of 
Eq.~(\ref{euler4}) becomes comparable to or smaller than the Lorentz force.

In addition to modifying the continuity equation, the Euler equation is often adjusted as well. We focus on the so called LBR formulation \cite{Lantz92,br95,gtz19}, widely used in numerical simulations. This formulation conserves energy \cite{brown12}, justifying the use of the energy functional, the approach adopted in this paper. Note, however, that the hydrodynamic part of the LBR energy functional differs from (\ref{Whydro}).  Before deriving it below, we examine the approximations used in the LBR formulation. To this end, let us rewrite Eq.\ \eqref{euler4} as
\begin{align}
&
	-\omega^2\m\xi=-{\pmb\nabla}\left(\frac{\delta p}{\rho}\right)-\left(\frac{\delta p}{\rho^2}{\m 
	\nabla}\rho+\frac{\delta p}{p\gamma_{\rm fr}}{\pmb\nabla}\Phi\right) -\frac{\partial \rho}{\partial x} 
	\delta x \frac{{\pmb\nabla}\Phi}{\rho}  +\frac{\delta \m F_{\rm L}}{\rho} 
    \label{LBR2}  
&
\end{align}
and estimate the terms on the right-hand side of this equation. When the magnetic field is small, one can approximately rewrite the second term as
\begin{align}
&	
    \frac{\delta p}{\rho^2}{\m \nabla}\rho+\frac{\delta p}{p\gamma_{\rm fr}}{\pmb\nabla}\Phi \approx 
	\frac{\delta p}{\rho^2}{\m \nabla}\rho-\frac{\delta p}{ p\gamma_{\rm fr,0}\rho}{\pmb\nabla}p\approx 
	\frac{\delta p}{\rho^2}\left(\frac{\gamma_{\rm fr,0}-\gamma_{{\rm eq},0} }{ \gamma_{\rm fr,0}}\right){\pmb\nabla}\rho.    
&
\end{align}
Matching two expressions for $\delta \rho$, $\delta \rho =\frac{\rho}{p\gamma_{\rm fr}} \delta 
p+\frac{\partial \rho}{\partial x}\delta x$ and $\delta \rho =\frac{\rho}{p\gamma_{\rm fr}} \delta 
p-\left[{\m \nabla}\rho \cdot {\m \xi}-\frac{\rho}{p\gamma_{\rm fr}}({\m \nabla}p\cdot {\m \xi})\right]$
[see Eq.\ \eqref{dp}], one can present the third term on the right-hand side of Eq.\ (\ref{LBR2}) as
\begin{align}
&	
    -\frac{\partial \rho}{\partial x} \delta x \frac{{\pmb\nabla}\Phi}{\rho} =\left[{\m \nabla}\rho 
	\cdot {\m \xi}-\frac{\rho}{p\gamma_{\rm fr}}({\m \nabla}p\cdot {\m \xi})\right]
	\frac{{\pmb\nabla}\Phi}{\rho}\sim -\frac{\gamma_{\rm fr,0}-\gamma_{{\rm eq},0} }{\gamma_{\rm fr,0}}\frac{{\m 
	\nabla}\rho}{\rho^2}({\m \nabla}p \cdot {\m \xi}),    
&
\end{align}

The results of Sec.\ \ref{general} imply that for perturbations of a general form caused by the Lorentz force, $\nabla p \cdot \pmb{\xi} \sim \frac{\gamma_{{\rm eq},0} }{\gamma_{{\rm fr,0}} - \gamma_{{\rm eq},0} } \delta p$ [see Eq.\ \eqref{dp} and the estimates in Sec.\ \ref{general}], and hence the third term is of the same order of magnitude as the first one. 
At the same time, the second term is smaller than the first term in Eq.\ (\ref{LBR2}) by a factor 
of $\frac{\gamma_{\rm fr} - \gamma_{{\rm eq},0} }{\gamma_{\rm fr}}$. Typically, in stellar matter, 
$\frac{\gamma_{\rm fr} - \gamma_{{\rm eq},0} }{\gamma_{\rm fr}} $ is small, indicating that the 
second term can be neglected%
%
\footnote{Note that for small-scale perturbations, the difference between the first and second terms becomes even larger (an additional factor, the ratio of the perturbation lengthscale to $R$, appears), while the third term remains of the same order of magnitude as the first one.} 
%
and we arrive at the LBR anelastic approximation of the Euler equation,
\begin{align}
&
    -\omega^2\m\xi=-{\pmb\nabla}\left(\frac{\delta p}{\rho}\right) -\frac{\partial \rho}{\partial x} \delta x 
	\frac{{\pmb\nabla}\Phi}{\rho}  +\frac{\delta \m F_{\rm L}}{\rho},    \label{LBR3}
&
\end{align}
where the second term stays for the buoyancy force.
In what follows we will use Eq.\ (\ref{LBR3}) in its equivalent form:
\begin{align}
&
    -\omega^2\m\xi=-{\pmb\nabla}\left(\frac{\delta p}{\rho}\right) +\left[{\m \nabla}\rho \cdot {\m 
	\xi}-\frac{\rho}{p\gamma_{\rm fr}}({\m \nabla}p\cdot {\m \xi})\right] \frac{{\pmb\nabla}\Phi}{\rho}  
	+\frac{\delta \m F_{\rm L}}{\rho}.    \label{LBR4}
&
\end{align}

Let us now derive the perturbation of the potential energy of the system under the displacement $\m 
\xi$. Following \cite{gtz19}, it can be calculated as the integral over the stellar volume
\begin{align}
&
	W[{\m \xi}]=-\frac{1}{2}\int {\m \xi}\cdot{\m F}({\m \xi})dV,
&
\end{align}
where ${\m F}({\m \xi})$ is a linear, self-adjoint operator defined by the equation
\begin{align}
&	
\rho \ddot{\m\xi}=	{\m F}({\m \xi}).
&
\end{align}
Then
\begin{align}
&	
W[{\m \xi}]=\frac{1}{2}\int \left[\rho{\m \xi}\cdot{\pmb\nabla}\left(\frac{\delta p}{\rho}\right)-\left({\m 
	\nabla}\rho \cdot {\m \xi}-\frac{\rho}{p\gamma_{\rm fr}}({\m \nabla}p\cdot {\m \xi})\right)\,\rho{\m 
	\xi}\cdot \frac{{\pmb\nabla} \Phi}{\rho}-\delta{\m F}_{\rm L}\cdot {\m \xi}\right]dV.
&
\end{align}
After integration by parts, the first term cancels out in the anelastic approximation for perturbations that obey the suitable boundary condition $({\m \nabla} p \cdot {\m \xi})\,{\m \xi} \cdot {\m n} = 0$ (where ${\m n}$ is the unit surface normal vector of the boundary), and we are left with
\begin{align}
&
W[{\m \xi}]=
	-\frac{1}{2}\int \left[ {\m \xi}\cdot\left({\m \nabla} \rho-\rho\frac{{\m F}_{\rm L}-\rho {\m 
	\nabla}\Phi}{p\gamma_{\rm fr}}\right)\,  ({\m \nabla} \Phi \cdot {\m \xi})+\delta{\m F}_{\rm L}\cdot {\m 
	\xi}\right] dV, 
&    
    \label{Wanel}
\end{align}
where the hydrostatic equilibrium condition has been used.
The magnetic part of $W[{\m \xi}]$, $W^{\rm mag}[{\pmb \xi}]$, remains unchanged and we present its 
derivation below for completeness.
\begin{align}
&
W^{\rm mag}[{\pmb \xi}]=
	-\frac{1}{2}\int {\m \xi}\cdot \delta{\m F}_{\rm L}\, dV=\frac{1}{8\pi}\int  {\m \xi}\cdot[{\m 
	B}\times ({\m \nabla} \times {\delta {\m B}})]\, dV+\frac{1}{8\pi}\int {\m \xi}\cdot[{\delta {\m B}}\times ({\m 
	\nabla} \times {\m B})]\, dV \equiv \nonumber
&
    \\ 
&
    \frac{1}{8\pi}\int {\rm I}\,dV+\frac{1}{8\pi}\int {\rm 
	II}\,dV,
&  
\label{magapp}
\end{align}
where
\begin{align}
&
    {\rm I}=({\m \nabla} \times {\delta {\m B}})\cdot({\m \xi} \times {\m B})={\delta {\m B}}\cdot ({\m \nabla} 
	\times ({\m \xi} \times {\m B}))-{\m \nabla}\cdot (({\m \xi} \times {\m B}) \times {\delta {\m B}})=\delta {\pmb B}\cdot\delta {\pmb B}-{\m \nabla}\cdot (({\m \xi} \times {\m B}) \times {\delta {\m B}}),
&    
    \\
&
    {\rm II}=({\m \nabla} \times {\m B})\cdot({\m \xi} \times {\delta {\m B}})=\frac{4\pi}{c}{\m J}\cdot 
	({\m \xi} \times {\delta {\m B}}).
\end{align}
Hence,
\begin{align}
&	
    W^{\rm mag}[{\pmb \xi}]=\frac{1}{2}\int\left(\frac{1}{4\pi}\delta {\pmb B}\cdot\delta {\pmb B}+\frac{1}{c}{\m J}\cdot ({\m \xi} 
	\times {\delta {\m B}})\right)\,dV+{\rm surface\, term}.
&    
\end{align}

Since only the magnetic part of $W[\pmb\xi]$, namely $W^{\rm mag}[\pmb\xi]$, contributes at leading order in $\varkappa$ and $\varepsilon$, the conclusions of Sec.~\ref{first} remain unchanged: for a toroidal field, the most unstable mode is the one with azimuthal number $m=1$.
To first order in the perturbation theory, the optimal Lagrangian displacement for this mode is given by shellular differential rotation of the star about an arbitrary axis in the $x$--$y$ plane, for example, the $x$-axis [see Eq.~\eqref{lagx}].
At the same time, this Lagrangian displacement is ``neutral'' in the sense that, in first-order 
perturbation theory, the perturbation-induced change in the energy functional vanishes, 
$W[\m\xi_{\rm rot}] \simeq \varkappa \alpha_{01} = 0$.

In the next order in $\varkappa$ and $\varepsilon$, using Eq.~\eqref{Wanel}, we find that 
$W[\m\xi]$ has the same structure as that discussed in Sec.~\ref{second}, with the functions $W_1$, 
$W_2$, and $W_3$ given by
\begin{align}
	& 	
	W_1 =  \frac{
		\pi 
		\, (\gamma_{{\rm fr}, 0}-\gamma_{{\rm eq},0} ) \, 
		[p_0'(r)]^2}{p_0 
		\gamma_{{\rm fr}, 0} \gamma_{{\rm eq},0} },
	&
	\label{AppW1}\\
	&
	W_2= \frac{2\pi
	}{p_0 \gamma_{{\rm fr}, 0} \gamma_{{\rm eq},0} }\left[
	- 
	2\mathcal{B}^2
	\, 
	{\rm cot}\theta
	\, p_0 \, \gamma_{{\rm fr}, 0} \, \gamma_{{\rm eq},0}  \, \partial_r\mathfrak{a}(r,\theta)
	+
	p_0'(r) 	\left(\rho_0\,\partial_\theta \phi_1 +\underline{\mathcal{B}^2 {\rm 
	cot}\theta+ \mathcal{B}	\partial_\theta 
		\mathcal{B}} \right) \, 
	(\gamma_{{\rm fr}, 0}-\gamma_{{\rm eq},0} )  \, \mathfrak{a}(r,\theta)
	\right],
	&
	\label{AppW2}\\
	&
	W_3=\frac{\pi
		\left(\rho_0\,\partial_\theta \phi_1  +\underline{2\mathcal{B}^2 {\rm cot}\theta+ 
		2\mathcal{B}	\partial_\theta 
			\mathcal{B} }\right)\, \mathfrak{a}(r,\theta)^2}{r p_0'(r) \, p_0\, \gamma_{{\rm fr}, 
			0}\, 
		\gamma_{{\rm eq},0} }
	\left[ 
	r \, p_0'(r)\, \partial_\theta \phi_1 \, 
	(\gamma_{{\rm fr}, 0}-\gamma_{{\rm eq},0} )  \, 
	\rho_0 
	+ 4 \mathcal{B} \, p_0 \, \gamma_{{\rm fr}, 0} \gamma_{{\rm eq},0}  \left( -
	\partial_\theta 
	\mathcal{B} +r 
	\,
	{\rm cot}\theta
	\, \partial_r \mathcal{B}\right)
	\right].
	&
	\label{AppW3}
\end{align}
As we see, the function $W_1$ is not affected by the anelastic LBR approximation, while $W_2$ and 
$W_3$ undergo minor modifications (additional underlined terms appear). Thus, the reasoning 
developed in Sec.~\ref{pert} applies to the LBR formulation of MHD and leads to the same 
conclusions. In other words, the approximations of the anelastic LBR 
do not eliminate the global instability.

Another widely used approach is the Boussinesq approximation (used, e.g., in the 
\href{https://magic-sph.github.io}{\texttt MagIC code}),
developed for MHD problems where the background lengthscale strongly exceeds the perturbation 
scale. Perturbations are assumed to be slow in the Boussinesq approximation, similarly to the 
anelastic one. Under this approximation, the density in the first term of the Euler equation 
(\ref{LBR4}) is factored out, while the continuity equation reduces to the incompressibility 
condition, $\m\nabla\cdot {\m \xi} = 0$. The energy functional $W[\m\xi]$ in the Boussinesq 
approximation then coincides with (\ref{Wanel}), indicating that the resulting equations also 
support the global toroidal instability.

\end{document}